\documentclass[preprint,superscriptaddress,nofootinbib]{revtex4-1}
\usepackage{graphicx}
\usepackage{float} 
\usepackage{subfigure}
\usepackage{dcolumn}
\usepackage{bm}
\usepackage{amsmath}
\usepackage{amsfonts} 
\usepackage{latexsym}
\usepackage{bbm}
\usepackage{color}
\usepackage{amssymb}
\usepackage{amsthm}
\usepackage{epsf}
\usepackage{epsfig}
\usepackage{caption3}
\usepackage{appendix}
\usepackage{cases}
\usepackage{subfigure}
\usepackage{multirow}
\usepackage{hyperref}
\makeatletter

\newcommand{\Rmnum}[1]{\expandafter\@slowromancap\romannumeral #1@}
\makeatother

\newcommand{\be}{\begin{equation}}
	\newcommand{\ee}{\end{equation}}
\newcommand{\ba}{\begin{array}}
	\newcommand{\ea}{\end{array}}
\newcommand{\bea}{\begin{eqnarray}}
	\newcommand{\eea}{\end{eqnarray}}

\usepackage{theorem}
{
	\theoremheaderfont{\bfseries}
	\theorembodyfont{\normalfont}
}

\begin{document}

\title{Light bending at two-loop order from scattering amplitudes}
\author{Jing Yang}

\author{Fa Peng Huang}
\email{ Corresponding Author. huangfp8@sysu.edu.cn}

\affiliation{MOE Key Laboratory of TianQin Mission, TianQin Research Center for
	Gravitational Physics \& School of Physics and Astronomy, Frontiers
	Science Center for TianQin, Gravitational Wave Research Center of
	CNSA, Sun Yat-sen University (Zhuhai Campus), Zhuhai 519082, China}

\bigskip
	
	
\begin{abstract}
We calculate the conservative deflection of a photon by a massive spinless source, such as a black hole, at third post-Minkowskian order using scattering amplitudes. We work in the heavy-source and long-range limits and adopt the geometric-optics approximation. The required five-point photon--graviton tree amplitude is constructed by on-shell recursion and sewn to three scalar--graviton vertices across a five-fold cut. Polynomial division organizes the cut information into sixteen contributions, including the residues with fewer propagators. We perform the source energy integrations and retain the near-on-shell photon regions in the remaining spatial integrals. Within the static region in which all exchanged momenta are soft, we relate a photon pole prescription to the subtraction of the cubic tree iteration from the conservative kernel. The sum of the sixteen contributions determines the classical logarithmic term at two loops. Its transverse Fourier transform gives the interaction radial action and yields the third-order deflection angle. The result agrees with the weak-deflection expansion of Schwarzschild null geodesics and demonstrates how nonlinear classical light bending is encoded in loop amplitudes around flat spacetime.
\end{abstract}

\maketitle
	
\section{Introduction}\label{sec:introduction}

Precision calculations of gravitational dynamics require control of nonlinear interactions beyond the leading graviton exchange. Scattering amplitudes provide a way to organize these interactions in quantum field theory and relate them to classical observables. The effective field theory description of gravity separates long-distance predictions from short-distance information, with nonanalytic momentum dependence encoding interactions that persist at large separation~\cite{Donoghue1994EFT}. Loop amplitudes contain classical terms as well as quantum corrections: propagation close to the mass shell can compensate the powers of $\hbar$ usually associated with a loop~\cite{Holstein2004Classical}. The extraction of these terms connects perturbation theory around flat spacetime to nonlinear motion in a gravitational field.

This connection has been developed systematically in the post-Minkowskian (PM) expansion, which orders the interaction in Newton's constant $G$ without requiring a small relative velocity. Generalized unitarity and effective field theory matching have been used to obtain classical two body potentials from on-shell amplitudes~\cite{Cheung2018Potentials}, while double copy constructions~\cite{Bern2008Relations,Bern2010DoubleCopy} have enabled calculations of binary dynamics at third PM order~\cite{Bern2019BlackHole}. Amplitude methods also determine the potential region contribution at fourth PM order~\cite{Bern2021G4}. Worldline quantum field theory supplies a related description in terms of particle trajectories and gravitational fields~\cite{Mogull2021WQFT}; recent worldline effective field theory calculations have obtained local-in-time conservative dynamics at fifth PM and first self-force orders~\cite{Dlapa2025G5}. In parallel, direct relations between amplitudes, momentum transfer, and the radial action have made it possible to extract scattering observables without first constructing a coordinate dependent potential~\cite{Kosower2019Observables,BjerrumBohr2022RadialAction,Damgaard2023ClassicalObservables}.The interaction radial action is obtained by integrating the radial momentum along the scattering orbit and subtracting the free-motion contribution at the same energy and angular momentum. Its negative derivative with respect to angular momentum at fixed energy gives the conservative scattering angle.

Light bending provided one of the early observational tests of general relativity: measurements of stellar deflection during the 1919 solar eclipse supported Einstein's prediction~\cite{Dyson1920LightDeflection}. It remains a basic observable in gravitational lensing and a direct probe of light propagation in curved spacetime~\cite{Iyer2007Bending}. Defined by the asymptotic photon directions at a fixed impact parameter, the deflection angle permits a comparison between quantum field theory calculations and classical geometry without introducing a coordinate dependent potential. Its higher order terms encode the corrections beyond linearized gravity. Deriving these terms from photon scattering therefore tests how graviton interactions and propagation near the mass shell combine to reproduce nonlinear classical motion for a massless probe.

Earlier amplitude calculations of photon scattering from a massive source obtained the leading classical deflection and its one loop correction at order $G^2$, together with distinct long-range quantum terms~\cite{BjerrumBohr2015LightBending,BjerrumBohr2016Lightlike}. Eikonal methods organize repeated exchanges and provide a direct connection to the scattering angle~\cite{KoemansCollado2019Eikonal}. A worldline treatment of the photon has also reproduced light bending through second PM order in the geometric-optics limit~\cite{Bastianelli2022LightBending}, while the massless classical limit can be formulated using localized wave packets and classical radiation states~\cite{Cristofoli2022Waveforms}. Related work on gravitational Compton scattering has recently connected a third order PM graviton--source amplitude to black hole perturbation theory~\cite{BjerrumBohr2026ComptonBHPT}. The Schwarzschild weak deflection expansion is already known beyond this order~\cite{Iyer2007Bending}; it provides a benchmark for extending the photon amplitude calculation from the one loop classical correction to the two loop contribution at order $G^3$.

The extension to two loops introduces difficulties in both the on-shell reconstruction and the classical integration. The required photon tree amplitude contains three graviton legs, and its factorization channels generate momentum-dependent residues with different propagator structures. Contact terms must be assigned consistently between these structures, since cuts with the largest number of propagators do not determine all residues. In the integration, the nearly on shell source propagators select small exchange energies, while two internal photon propagators can couple the remaining loop integrations. Near a photon pole, the longitudinal momentum is parametrically smaller than a generic soft momentum. Quadratic terms in the photon denominators can then contribute at the retained classical order, so a uniform linearization before integration is insufficient. The relevant regions and their correlated limits must be resolved before truncating the soft expansion~\cite{Beneke1998Regions,Jantzen2011Regions}.

A further difficulty is that the two-loop amplitude contains repeated lower order interactions as well as the third-order interaction to be extracted. In particular, the cubic tree iteration contributes to the Hermitian part of the two loop transition operator. Taking the real part of the phase reduced amplitude therefore does not by itself isolate the interaction radial action~\cite{Damgaard2021Exponential,Damgaard2023ClassicalObservables}. An amplitude derivation of the two-loop bending angle must relate the treatment of the photon poles to the removal of this iteration while retaining the numerator terms that contribute to the classical logarithm.

In this work, we calculate the third PM conservative deflection of a photon by a neutral spinless source with minimal gravitational coupling. We take the source mass $m$ to be much larger than the photon energy $E$ and work at momentum transfer $Q\ll E$. Our on shell construction starts from the five-point tree amplitude with two photons and three gravitons, obtained by Britto--Cachazo--Feng--Witten (BCFW) recursion~\cite{Britto2005BCFW,Cheung2010Recursion}. We sew this amplitude to three scalar--graviton vertices across a cut through the three exchanged gravitons and two source propagators. This choice exposes the source vertices and pole residues responsible for the leading heavy source contribution, following the use of multiple cuts to isolate classical information~\cite{Cachazo2020Singularities,BjerrumBohr2022RadialAction}. We then use the multiloop extension of Ossola--Papadopoulos--Pittau (OPP) integrand reduction, implemented through polynomial division with Gr\"obner bases~\cite{Ossola2007Reduction,Mastrolia2012Division,Zhang2012Integrand}, to reconstruct sixteen contributions: eight with seven propagators, seven with six, and one with five. This construction fixes the irreducible numerator residues and the assignment of contact terms within the cut sector. The numerators of all sixteen contributions enter the subsequent classical integration.

We evaluate the loop energy integrals at the source poles and sum the attachment orderings on the source line before performing the remaining spatial integrations. Within the static contribution in which all exchanged momenta are soft, we relate the chosen photon prescription to the subtraction of the cubic tree iteration in the Hermitian logarithm of the scattering operator. This relation is established through the factorization of the double photon residue and the associated intermediate state phase space convolution; the conservative restriction is imposed separately. For the spatial integration, successive tensor bubble reductions treat the topologies admitting a nested integration, while a joint Schwinger-parameter representation retains the coupled poles of the two photon topologies. Numerator insertions and the quadratic photon terms are kept through the soft order required for the logarithm. The sum of the sixteen contributions determines the $Q^0\log Q^2$ term of the third order interaction kernel. A transverse Fourier transform converts this kernel into the interaction radial action, whose derivative at fixed energy determines the deflection angle.

The calculation applies for $E/m\ll1$, $Eb\gg1$, and $Gm/b\ll1$, and the comparison uses the same asymptotic impact parameter in both descriptions. The known geodesic coefficient tests the combined reconstruction and integration. The methodological contribution is an explicit photon amplitude derivation in which the contact residues, coupled photon-pole regions, and subtraction of repeated scattering are treated consistently. It identifies how the two loop integrand produces the classical observable and provides a concrete basis for extending amplitude calculations of massless probes beyond the one loop approximation.

The paper is organized as follows. Sec.~\ref{sec:kinematics} specifies the kinematics and classical probe limit. Sec.~\ref{sec:tree_amplitudes_cut} constructs the tree amplitudes and the five-fold cut, and Sec.~\ref{sec:integrand_reconstruction} reconstructs the integrand by polynomial reduction. Sec.~\ref{sec:classical_integration} develops the classical integration and extracts the conservative kernel. Sec.~\ref{sec:radial_action_angle} derives the radial action and deflection angle at two-loop level, followed by conclusions in Sec.~\ref{sec:conclusions}. The appendices give the double-copy representation of the five-point amplitude, the reconstruction coefficients, the spatial integral identities, and the mixed and contact terms entering the iteration subtraction.

\section{Kinematics and the classical probe limit}\label{sec:kinematics}

\subsection{Photon--scalar scattering }\label{subsec:scattering_conventions}

We consider the gravitational scattering of a photon by a massive spinless particle,
\be
 \gamma(k,\sigma)+\Phi(p)\longrightarrow\gamma(k',\sigma')+\Phi(p'),
 \label{eq:scattering_process}
\ee
where $\sigma,\sigma'=\pm1$ are the photon helicities and $\Phi$ is a neutral real scalar field with mass $m$, which serves as the heavy object, i.e. black hole. Throughout this work we use natural units $\hbar=c=1$ and  the mostly-minus metric $\eta_{\mu\nu}=\operatorname{diag}(1,-1,-1,-1)$.

The photon and scalar fields are minimally coupled to Einstein gravity. The action is
\be
 \begin{aligned}
 S=\int d^4x\sqrt{-g}\bigg[&\frac{2R}{\kappa^2}
 -\frac14g^{\mu\rho}g^{\nu\sigma}F_{\mu\nu}F_{\rho\sigma}\\
 &+\frac12g^{\mu\nu}\partial_\mu\Phi\partial_\nu\Phi
 -\frac12m^2\Phi^2\bigg],
 \label{eq:einstein_maxwell_scalar_action}
 \end{aligned}
\ee
where $\kappa^2=32\pi G$ and $F_{\mu\nu}=\partial_\mu A_\nu-\partial_\nu A_\mu$. We expand the metric as $g_{\mu\nu}=\eta_{\mu\nu}+\kappa h_{\mu\nu}$, with $h_{\mu\nu}$ the graviton field. This expansion gives the scalar--graviton and photon--graviton interactions, while the Einstein–Hilbert term generates the graviton self interactions. Since $\Phi$ is electrically neutral, the scattering process in Eq.~\eqref{eq:scattering_process} proceeds purely through gravitational exchange.

All external momenta are taken physical, with positive energies, and satisfy
\be
 p+k=p'+k',\qquad p^2=p'^2=m^2,\qquad k^2=k'^2=0.
 \label{eq:external_onshell}
\ee
We define the momentum transfer $q$ as that gained by the source,
\be
 q=p'-p=k-k',\qquad t=q^2=-Q^2<0,
 \label{eq:momentum_transfer}
\ee
where $Q>0$. In the rest frame of the incoming source, we write $p^\mu=(m,\boldsymbol{0})$ and $k^\mu=E(1,\hat{\boldsymbol n})$, where $E$ and $\hat{\boldsymbol n}$ the photon energy and direction of incidence. We use the Mandelstam invariants and the dimensionless energy variable
\be
 \begin{aligned}
 s&=(p+k)^2=m^2+2mE,\qquad u=(p-k')^2,\\
 s+t+u&=2m^2,\qquad \xi=\frac{s-m^2}{m^2}=\frac{2E}{m}.
 \end{aligned}
 \label{eq:mandelstam_variables}
\ee
Using the on-shell conditions for the outgoing state, we find
\be
 2p\cdot q=-t,\qquad 2k\cdot q=t,\qquad
 q^0=\frac{Q^2}{2m}.
 \label{eq:recoil_kinematics}
\ee
Thus the outgoing photon energy is $E'=E-Q^2/(2m)$. The exchanged energy vanishes only in the static limit; accordingly, we retain the exact on-shell relations when constructing the amplitudes, and expand in the heavy-source limit only when extracting the classical contribution

For the tree amplitudes and generalized cuts, we use the all-outgoing convention
\be
 p_1=-p,\qquad p_2=-k,\qquad p_3=k',\qquad p_4=p',
 \qquad \sum_{i=1}^4p_i=0.
 \label{eq:all_outgoing}
\ee
with legs $1, 4 ~(2, 3)$ the scalars (photons). In this convention, $s=(p_1+p_2)^2$ and $t=(p_2+p_3)^2$. We denote the transfer entering the cut momenta by 
\be
 q_c=p_2+p_3=-q,\qquad q_c^2=t,\qquad (p_1+q_c)^2=m^2.
 \label{eq:cut_transfer}
\ee
This fixes the relative sign between the transfer used in the cuts and the physical momentum gained by the source.

We normalize one-particle states as $\langle\boldsymbol p'|\boldsymbol p\rangle=2E_{\boldsymbol p}(2\pi)^3\delta^{(3)}(\boldsymbol p'-\boldsymbol p)$, with a Kronecker delta for photon helicities left implicit. The invariant amplitude is defined by
\be
 \langle p',k',\sigma'|S-1|p,k,\sigma\rangle
 =i(2\pi)^4\delta^{(4)}(p+k-p'-k')\mathcal M_{\sigma'\sigma}.
 \label{eq:amplitude_normalization}
\ee
Loop integrals are dimensionally regulated in $D=4-2\epsilon$ dimensions, with the measure
\be
 \int_\ell\equiv
 \left(\frac{e^{\gamma_E}\bar\mu^2}{4\pi}\right)^\epsilon
 \int\frac{d^{4-2\epsilon}\ell}{(2\pi)^{4-2\epsilon}},
 \label{eq:loop_measure}
\ee
where $\bar\mu$ is the dimensional regularization scale and $\gamma_E$ is Euler's constant. The tree amplitudes entering the cuts are evaluated in the four-dimensional helicity (FDH) scheme~\cite{Bern2002FDH}.

\subsection{Scale hierarchy and the target classical contribution}\label{subsec:classical_probe_limit}

We work in the probe limit $E/m\ll1$, in which the recoil of the heavy source is negligible at leading order. Let $b$ be the asymptotic impact parameter of the photon and $J$ its orbital angular momentum relative to the source. In this limit, $J=Eb$. A classical ray description and a weak-deflection expansion require
\be
 \frac{1}{Eb}\ll1,\qquad \frac{Gm}{b}\ll1.
 \label{eq:geometric_optics_weak_field}
\ee
The first condition states that the photon wavelength is much smaller than the impact parameter. Restoring $\hbar$, the same condition reads $J/\hbar=Eb/\hbar\gg1$, so that a wave packet can be localized on a scale intermediate between the wavelength and the orbital scale~\cite{Kosower2019Observables,Cristofoli2022Waveforms}. The second condition controls the weak field expansion of the deflection angle in powers of $Gm/b$. The two conditions parametrically suppress finite-wavelength and strong field corrections, respectively. 

The long-range interaction is obtained by Fourier transforming the amplitude kernel to impact-parameter space. At a fixed order in $G$, the inverse impact parameter sets the soft scale of the exchanged momentum. We therefore expand the integrands according to the hierarchy 
\be
 Q\ll E\ll m,\qquad
 \lambda=\frac{Q}{E}\ll1,\qquad \eta=\frac{E}{m}\ll1.
 \label{eq:scale_hierarchy}
\ee
Here $\lambda$ controls the soft expansion relative to the photon energy, while $\eta$ controls the recoil of the source. These two expansions are parametrically distinct. For example, Eq.~\eqref{eq:recoil_kinematics} gives $q^0\sim Q^2/m$, which is smaller than the spatial momentum transfer of order $Q$. Near a massive propagator pole, the loop energy scales in the same way, $l^0\sim Q^2/m$. These scales determine the classical regions of the loop integrals.

The PM expansion is an expansion in $G$ at fixed asymptotic energy and impact parameter. In the probe limit, we write  the conservative  part of the deflection angle as
\be
 \chi(b)=c_1\frac{Gm}{b}
 +c_2\left(\frac{Gm}{b}\right)^2
 +c_3\left(\frac{Gm}{b}\right)^3
 +\mathcal O\!\left(\frac{G^4m^4}{b^4}\right).
 \label{eq:pm_deflection}
\ee
Our aim is to determine the third-order contribution from scattering amplitudes. For the minimally coupled theory, the amplitude has the expansion
\be
 \mathcal M=\mathcal M_1+\mathcal M_2+\mathcal M_3+\cdots,
 \qquad \mathcal M_n=\mathcal O(G^n),
 \label{eq:pm_amplitude}
\ee
where $\mathcal M_1$, $\mathcal M_2$, and $\mathcal M_3$ are the tree, one-loop, and two-loop terms, respectively. Loop amplitudes contain classical contributions when the exchanged massless momenta are soft and the massive propagators become nearly on shell~\cite{Holstein2004Classical,Cheung2018Potentials}. The two-loop term also contains iterations of lower orders, which must be subtracted when extracting the conservative radial action from the amplitude~\cite{BjerrumBohr2022RadialAction}.

We denote by $\mathcal K_{3,\mathrm{cl}}$ the 3PM classical kernel entering the radial action. In the amplitude normalization of Eq.~\eqref{eq:amplitude_normalization}, the long-range term relevant to $c_3$ has the form
\be
 \left.\mathcal K_{3,\mathrm{cl}}(Q^2)\right|_{Q^0\log Q^2}
 =C_3G^3m^4E^2\log\frac{Q^2}{\bar\mu^2},
 \label{eq:target_classical_log}
\ee
where $C_3$ is a dimensionless coefficient. Its value is determined by the two-loop integrand and the separation of lower-order iterations. The factor $G^3m^4E^2$ is dimensionless in four dimensions, as required for the amplitude kernel. A logarithm in a loop amplitude is not by itself sufficient to identify a classical contribution; its coefficient must also exhibit the correct mass, energy, and $\hbar$ scaling~\cite{Holstein2004Classical,BjerrumBohr2016Lightlike}.

\begin{figure}[H]
 \centering
 \includegraphics[width=\textwidth]{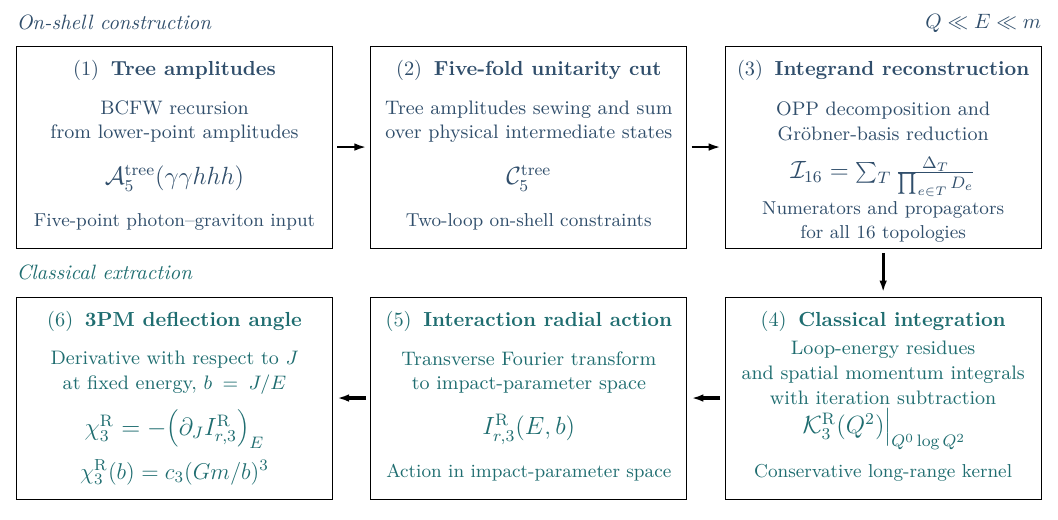}
 \caption{Outline of the calculation procedure for the 3PM contribution to the conservative light deflection angle.}
 \label{fig:calculation_workflow}
\end{figure}

For nonzero $b$, terms polynomial in the transverse momentum transfer yield only contact terms $\delta(b)$ (and derivatives thereof) and do not contribute to the long-range deflection. In contrast, the logarithm in Eq.~\eqref{eq:target_classical_log} gives a $1/b^2$ term in the radial action after the transverse Fourier transform. Differentiating with respect to $J=Eb$ then produces the $1/b^3$ contribution in Eq.~\eqref{eq:pm_deflection}. We summarize the calculation from the on-shell tree amplitudes to the 3PM contribution of the conservative deflection angle in Fig.~\ref{fig:calculation_workflow}. In what follows we extract this logarithm by combining the on-shell construction of the two-loop integrand (Secs.~\ref{sec:tree_amplitudes_cut} and \ref{sec:integrand_reconstruction}) with its classical expansion (Sec.~\ref{sec:classical_integration}).

\section{Tree amplitudes and the five-fold unitarity cut}\label{sec:tree_amplitudes_cut}

We construct the on-shell input for the two-loop calculation from tree amplitudes. The photon side of the cut requires the five-point amplitude with two photons and three gravitons, while the heavy-source side reduces to three scalar--graviton three-point amplitudes on the five-fold cut introduced in Sec.~\ref{subsec:five_fold_cut}. We first obtain the five-point amplitude from its factorization channels and then sew these tree amplitudes across the five cut lines.

\subsection{Tree-level building blocks and BCFW recursion}\label{subsec:tree_bcfw}

We use four-dimensional spinor-helicity variables for the massless external and cut momenta~\cite{Xu1987Helicity}. For a null momentum $K_i$, our conventions are
\be
 K_{i,\alpha\dot\alpha}=\lambda_{i,\alpha}\widetilde\lambda_{i,\dot\alpha},
 \qquad s_{ij}=(K_i+K_j)^2=\langle ij\rangle[ji],
 \qquad [i|K_a|j\rangle=[ia]\langle aj\rangle .
 \label{eq:tree_spinor_conventions}
\ee
A subscript $h$ denotes a graviton and a subscript $\gamma$ denotes a photon; their superscripts $\pm$ denote helicities $\pm2$ for gravitons and $\pm1$ for photons. We write the tree amplitude, including its Feynman phase, as
\be
 \mathcal A_n^{\mathrm{tree}}=-ig^{n-2}H_n,
 \qquad g=\frac{\kappa}{2}=\sqrt{8\pi G}.
 \label{eq:tree_phase_normalization}
\ee
Thus $\mathcal A^{\mathrm{tree}}=i\mathcal M^{\mathrm{tree}}$ in the $S$-matrix convention of Eq.~\eqref{eq:amplitude_normalization}. The functions $H_n$ are stripped of the coupling and the overall phase. With this definition, sewing two tree amplitudes with $i/P^2$ gives a positive factorization relation for $H_n$.

The three-point photon--graviton amplitudes needed here are
\be
 \begin{aligned}
 H_3(1_h^-,2_\gamma^+,3_\gamma^-)
 &=\frac{\langle13\rangle^4}{\langle23\rangle^2},&
 H_3(1_h^+,2_\gamma^+,3_\gamma^-)
 &=\frac{[12]^4}{[23]^2}.
 \end{aligned}
 \label{eq:photon_graviton_three}
\ee
The graviton self-interaction enters through
\be
 \begin{aligned}
 H_3(1_h^-,2_h^+,3_h^-)
 &=\frac{\langle13\rangle^6}{\langle12\rangle^2\langle23\rangle^2},&
 H_3(1_h^+,2_h^+,3_h^-)
 &=\frac{[12]^6}{[13]^2[23]^2}.
 \end{aligned}
 \label{eq:graviton_three}
\ee
Other nonzero components follow by permuting the legs or reversing all helicities. These expressions are evaluated on complexified three-point kinematics. Momentum conservation implies $\langle ij\rangle[ji]=0$, but the two brackets do not vanish simultaneously in general. The angle-bracket (square-bracket) amplitudes are supported on the branch where the square (angle) brackets vanish.

\begin{figure}[!htb]
 \centering
 \includegraphics[width=0.96\linewidth]{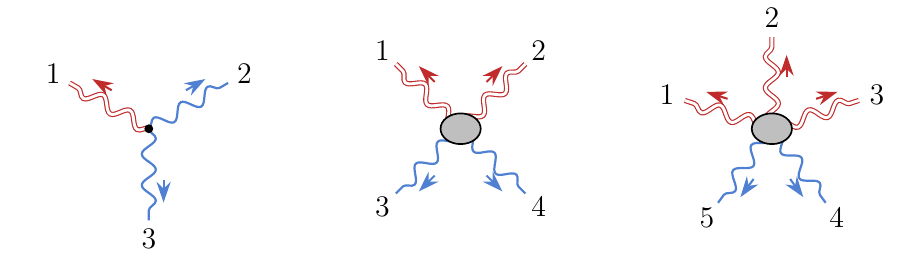}
 \caption{Photon--graviton tree amplitudes. Blue single and red double wavy lines denote photons and gravitons; gray blobs denote complete tree amplitudes.}
 \label{fig:tree_amplitudes}
\end{figure}

For the four-point amplitude we define $s_4=s_{12}$, $t_4=s_{13}$, and $u_4=s_{14}$, with $s_4+t_4+u_4=0$. Sewing pairs of three-point amplitudes across their exchanged propagator gives~\cite{BjerrumBohr2015GravitonPhoton}
\be
 H_4(1_h^+,2_h^-,3_\gamma^-,4_\gamma^+)
 =-\frac{\bigl(\langle23\rangle[14]\bigr)^2
 [1|K_3-K_4|2\rangle^2}{4s_4t_4u_4}.
 \label{eq:photon_graviton_four}
\ee
The component with graviton helicities $(-,+)$ is obtained by exchanging the graviton legs $1$ and $2$. Momentum conservation gives $[1|K_3-K_4|2\rangle=2[13]\langle32\rangle$. The helicity configuration $H_4(h^\pm,h^\pm,\gamma^-,\gamma^+)$ vanish. Fig.~\ref{fig:tree_amplitudes} shows the external states of the photon--graviton amplitudes.

On the heavy-source side, we parametrize the outgoing momenta as $(P,-P-K,K)$, with $P^2=(P+K)^2=m^2$ and $K^2=0$. The on-shell conditions imply $P\cdot K=0$. Contracting the scalar energy-momentum tensor with the graviton polarization gives
\be
 H_3(P,-P-K,K_h^\pm)=2\bigl[P\cdot\varepsilon^\pm(K)\bigr]^2,
 \qquad
 \mathcal A_3=-2ig\bigl[P\cdot\varepsilon^\pm(K)\bigr]^2,
 \label{eq:scalar_graviton_three}
\ee
where the graviton polarization is $\varepsilon^\pm_\mu\varepsilon^\pm_\nu$. The condition $P\cdot K=0$ ensures that this result is independent of the choice of reference vector in $\varepsilon^\pm_\mu$.

For the explicit cut calculation, we use the spinor-helicity representation of this source vertex. With outgoing momenta $K_1=P$, $K_2=-P-K$, and $K_3=K$, the auxiliary gauge-theory input consists of the scalar--gluon three-point amplitudes
\be
\begin{aligned}
	A_3^{\mathrm{tree}}(1_m,2_m,3^+)
	&=-i\,\frac{\langle\zeta|K_1|3]}{\langle\zeta3\rangle},&
	A_3^{\mathrm{tree}}(1_m,2_m,3^-)
	&=-i\,\frac{\langle3|K_1|\zeta]}{[3\zeta]}.
\end{aligned}
\label{eq:scalar_gluon_three_spinor}
\ee
Here $1_m$ and $2_m$ are the massive scalar legs, while $3^\pm$ denotes a gluon of helicity $\pm1$. The color factor and gauge coupling are omitted. The reference spinor $\zeta$ is chosen so that the indicated denominator is nonzero, and $K_1$ between the angle and square bracket is the massive momentum matrix. Removing the common factor $-i$ and taking the double copy of two identical gauge-theory factors gives
\be
\begin{aligned}
	H_3(1_m,2_m,3_h^+)
	&=\left(\frac{\langle\zeta|K_1|3]}{\langle\zeta3\rangle}\right)^2,&
	H_3(1_m,2_m,3_h^-)
	&=\left(\frac{\langle3|K_1|\zeta]}{[3\zeta]}\right)^2.
\end{aligned}
\label{eq:scalar_graviton_three_spinor}
\ee
 Restoring the gravitational coupling and phase gives $\mathcal A_3^{\mathrm{tree}}=-igH_3$, as in Eq.~\eqref{eq:tree_phase_normalization}. These are the scalar--graviton factors used in the cut product of Eq.~\eqref{eq:five_cut_tree_product}.

We now construct the five-point photon--graviton amplitude using BCFW recursion~\cite{Britto2005BCFW,Cheung2010Recursion}. We label the gravitons by $1,2,3$ and fix the photon helicities to $4_\gamma^-,5_\gamma^+$. We deform the spinors of legs $1$ and $2$ as
\be
 |\widehat1\rangle=|1\rangle-z|2\rangle,\qquad
 |\widehat2]=|2]+z|1],
 \label{eq:bcfw_shift}
\ee
with the other spinors unchanged. This $[1,2\rangle$ shift is chosen so that the deformed amplitude vanishes at large $z$ for the graviton helicities used below. The shifted momenta are $\widehat K_1=K_1-zr$ and $\widehat K_2=K_2+zr$, where $r_{\alpha\dot\alpha}=\lambda_{2,\alpha}\widetilde\lambda_{1,\dot\alpha}$. Since $r^2=r\cdot K_1=r\cdot K_2=0$, the deformation preserves both on-shellness and total momentum conservation.

The deformed tree amplitude is a rational function of $z$. Its finite poles occur at values of $z$ where an internal momentum goes on shell. For the graviton helicities used below, the amplitude vanishes at large $z$~\cite{ArkaniHamed2008Tree,Cheung2010Recursion}, so the contour integral of $H_5(z)/z$ gives
\be
 H_5(0)=\sum_{I,\sigma}
 H_L(z_I;\widehat P_I^\lambda)\,
 \frac{1}{P_I^2}\,
 H_R(z_I;-\widehat P_I^{-\lambda}).
 \label{eq:bcfw_recursion}
\ee
Here the two shifted legs lie on opposite sides of the channel, and $\sigma$ runs over the physical helicities of the internal particle. The tree factors are evaluated at $\widehat P_I^2(z_I)=0$, whereas $P_I^2$ in the denominator is the undeformed invariant. This denominator comes from the residue of the shifted propagator. We use $|-P\rangle=-|P\rangle$ and $|-P]=|P]$ for the spinors of a reversed null momentum.

\begin{figure}[!htb]
 \centering
 \includegraphics[width=0.96\linewidth]{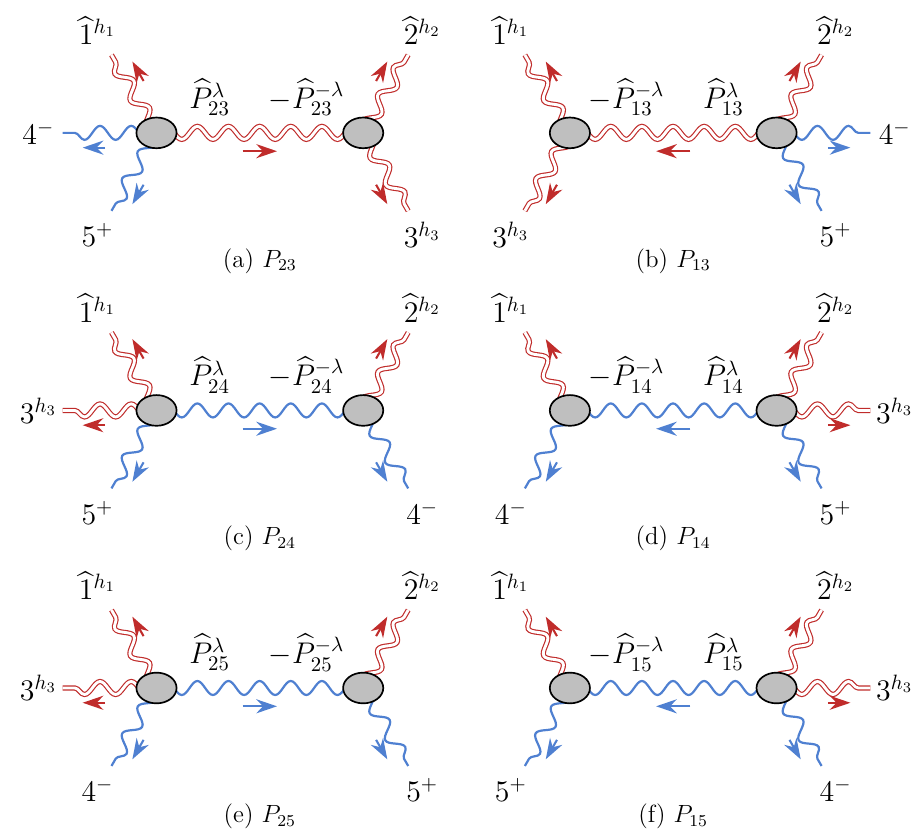}
 \caption{Six BCFW factorization channels. Blue single and red double wavy lines denote photons and gravitons; blobs denote on-shell three- and four-point amplitudes.}
 \label{fig:bcfw_channels}
\end{figure}

At five points, each channel contains one three-point and one four-point amplitude. There are six possible partitions,
\be
 P_{23},\quad P_{24},\quad P_{25},\qquad
 P_{13},\quad P_{14},\quad P_{15},
 \qquad P_{ij}=K_i+K_j,
 \label{eq:bcfw_six_channels}
\ee
shown in Fig.~\ref{fig:bcfw_channels}. The $P_{23}$ and $P_{13}$ channels exchange a graviton, and the other four exchange a photon. Their pole locations are
\be
 \begin{aligned}
 \widehat P_{2j}^2(z)
 &=\langle2j\rangle\bigl([j2]+z[j1]\bigr),
 &z_{2j}&=\frac{[2j]}{[j1]},\\
 \widehat P_{1j}^2(z)
 &=\bigl(\langle1j\rangle-z\langle2j\rangle\bigr)[j1],
 &z_{1j}&=\frac{\langle1j\rangle}{\langle2j\rangle},
 \qquad j=3,4,5 .
 \end{aligned}
 \label{eq:bcfw_poles}
\ee
The first line selects the angle-bracket three-point amplitudes because $[j\widehat2]=0$. The second selects the square-bracket amplitudes because $\langle\widehat1j\rangle=0$. Together with the vanishing four-point components, this determines which partitions contribute for each external helicity assignment.

For $(1_h^+,2_h^-,3_h^-)$, the nonzero channels are $P_{23}$, $P_{25}$, and $P_{24}$. For example, the graviton-exchange contribution is
\be
 \mathcal R_{23}=
 \left.
 H_4(\widehat1_h^+,\widehat P_{23,h}^-,4_\gamma^-,5_\gamma^+)
 \frac{1}{P_{23}^2}
 H_3((-\widehat P_{23})_h^+,\widehat2_h^-,3_h^-)
 \right|_{z=z_{23}} .
 \label{eq:bcfw_channel_23}
\ee
At $z_{23}=[23]/[31]$, a convenient choice of the internal spinors is
\be
 [\widehat2|=\frac{[12]}{[13]}[3|,\qquad
 |\widehat P_{23}\rangle=|3\rangle+\frac{[12]}{[13]}|2\rangle,\qquad
 [\widehat P_{23}|=[3|.
 \label{eq:bcfw_internal_spinors}
\ee
Using Eqs.~\eqref{eq:graviton_three} and \eqref{eq:photon_graviton_four}, we obtain
\be
 H_3((-\widehat P_{23})_h^+,\widehat2_h^-,3_h^-)
 =\frac{\langle23\rangle^2[13]^2}{[12]^2},\qquad
 \mathcal R_{23}
 =\frac{[15]^5[14]\langle45\rangle\langle23\rangle}
 {[35][34][45][12]^2[23]} .
 \label{eq:bcfw_evaluated_channel}
\ee
The arbitrary little-group normalization of the internal spinors cancels between the two tree factors. Evaluating the two photon-exchange channels in the same way gives
\be
 \mathcal A_5(1_h^+,2_h^-,3_h^-,4_\gamma^-,5_\gamma^+)=ig^3U_5,
 \qquad H_5(1_h^+,2_h^-,3_h^-,4_\gamma^-,5_\gamma^+)=-U_5,
 \label{eq:five_point_u_normalization}
\ee
where
\be
 \begin{aligned}
 U_5={}&\frac{[15]^5}{[35][34][45][12]^2}
 \bigg[-\frac{\langle45\rangle[14]\langle23\rangle}{[23]}
 -\frac{\langle34\rangle[14][15]\langle25\rangle}{[13][25]}\\
 &\hspace{43mm}
 +\frac{\langle35\rangle[14]^2\langle24\rangle}{[13][24]}\bigg].
 \end{aligned}
 \label{eq:five_point_u}
\ee
The three terms in brackets correspond to $P_{23}$, $P_{25}$, and $P_{24}$, respectively. The common minus sign between $H_5$ and $U_5$ follows from the phase convention in Eq.~\eqref{eq:tree_phase_normalization}.

For $(1_h^+,2_h^+,3_h^-)$, the nonzero channels are instead $P_{13}$, $P_{14}$, and $P_{15}$. Their sum gives
\be
 \mathcal A_5(1_h^+,2_h^+,3_h^-,4_\gamma^-,5_\gamma^+)=ig^3V_5,
 \qquad H_5(1_h^+,2_h^+,3_h^-,4_\gamma^-,5_\gamma^+)=-V_5,
 \label{eq:five_point_v_normalization}
\ee
with
\be
 \begin{aligned}
 V_5={}&\frac{\langle34\rangle^5\langle35\rangle}
 {\langle12\rangle^2\langle45\rangle}
 \bigg[-\frac{[45][13]}{\langle13\rangle\langle24\rangle\langle25\rangle}
 +\frac{[35][14]}{\langle14\rangle\langle23\rangle\langle25\rangle}\\
 &\hspace{43mm}
 -\frac{[15][34]}{\langle15\rangle\langle23\rangle\langle24\rangle}\bigg].
 \end{aligned}
 \label{eq:five_point_v}
\ee
These terms follow the channel order $P_{13}$, $P_{14}$, and $P_{15}$. With the photon helicities fixed, the components with three positive-helicity or three negative-helicity gravitons vanish. The six nonzero components are obtained by permuting the graviton legs in $U_5$ and $V_5$. Reversing all helicities interchanges angle and square brackets in $H_5$, while the common phase in $\mathcal A_5$ remains fixed. The same tree amplitude can also be obtained through a double-copy construction from gauge-theory trees using the field-theory limit of the Kawai--Lewellen--Tye relations~\cite{Kawai1986KLT}. We use the BCFW form here to keep the internal factorization channels explicit; the double-copy representation is given in Appendix~\ref{app:five_point_double_copy}.

\subsection{Five-fold cut and physical-state sums}\label{subsec:five_fold_cut}

\begin{figure}[!htb]
 \centering
 \includegraphics[width=0.65\linewidth]{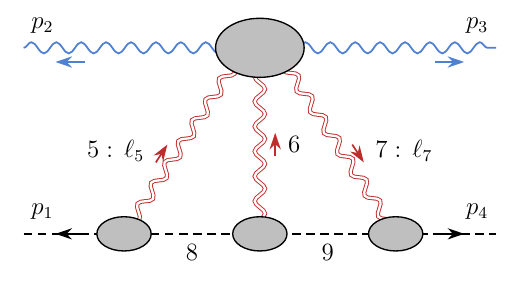}
 \caption{The five-fold generalized cut with three graviton and two scalar cutting lines. Labeled lines $5,6,7$ are gravitons, $8,9$ are massive scalars, and $2,3$ are photons; gray blobs denote tree amplitudes.}
 \label{fig:five_fold_cut}
\end{figure}

Generalized unitarity constrains a loop integrand by placing selected internal propagators on shell and matching the resulting residue to a product of tree amplitudes~\cite{Bern1995Fusing,Britto2005Generalized}. For the classical contribution considered here, we cut the three exchanged gravitons and the two internal scalar lines on the source side, as shown in Fig.~\ref{fig:five_fold_cut}. Cutting only the gravitons would leave a five-point amplitude on each side. The additional scalar cuts factorize the source amplitude into three scalar--graviton vertices.

This cut choice makes both the coupling order and the leading power of the source mass explicit. The five-point photon amplitude contributes $g^3$, and the three source amplitudes contribute an additional $g^3$, giving $g^6=(8\pi G)^3$. To count the powers of $m$, we work in the heavy-source limit at fixed $E$ and $Q$. Writing a source momentum as $P^\mu=mv^\mu+\mathcal O(Q)$, with $v^2=1$, Eq.~\eqref{eq:scalar_graviton_three} gives
\be
 \mathcal A_3
 =-2ig m^2\bigl[v\cdot\varepsilon^\pm\bigr]^2
 +\mathcal O(gmQ).
 \label{eq:source_vertex_mass_counting}
\ee
Thus each scalar--graviton vertex supplies two powers of the source mass, and the three vertices give $m^6$.

The two internal scalar lines supply the compensating inverse powers of $m$ through their energy-pole residues. For $p=(m,\boldsymbol0)$, a source propagator carrying momentum $p+L$ has denominator
\be
 (p+L)^2-m^2+i0
 =2mL^0+(L^0)^2-\boldsymbol L^2+i0 .
 \label{eq:source_pole_denominator}
\ee
The pole near zero lies at
\be
 L_*^0=-m+\sqrt{m^2+\boldsymbol L^2}
 =\frac{\boldsymbol L^2}{2m}
 +\mathcal O\!\left(\frac{|\boldsymbol L|^4}{m^3}\right).
 \label{eq:source_near_shell_root}
\ee
For $|\boldsymbol L|\sim Q$, this is the $Q^2/m$ energy scale anticipated in Sec.~\ref{subsec:classical_probe_limit}. At this pole, the energy residue is
\be
 \mathop{\mathrm{Res}}_{L^0=L_*^0}
 \frac{i}{(p+L)^2-m^2}
 =\frac{i}{2(m+L_*^0)}
 =\frac{i}{2m}\left[1+\mathcal O\!\left(\frac{Q^2}{m^2}\right)\right].
 \label{eq:source_pole_mass_counting}
\ee
Each of the two source poles therefore gives a factor $1/(2m)$ at leading order in the energy integrations~\cite{BjerrumBohr2022RadialAction}. Combining the three vertices with these two residues, the source contribution scales as
\be
 \underbrace{(gm^2)^3}_{\text{three source vertices}}\,
 \underbrace{\left(\frac{1}{2m}\right)^2}_{\text{two energy residues}}
 \ \propto\ g^3m^4 .
 \label{eq:source_combined_mass_counting}
\ee
Including the coupling $g^3$ from the photon amplitude gives $g^6m^4\propto G^3m^4$, precisely the power of the source mass appearing in the target term $G^3m^4E^2\log(Q^2/\bar\mu^2)$ of Eq.~\eqref{eq:target_classical_log}. The inverse powers of $m$ in this argument come from the energy residues rather than from a pointwise power counting of the unintegrated propagators. In particular, the algebraic cut product contains the three source vertices; the two residue factors enter when this pole information is used in the classical energy integrals.

This mass counting explains why the chosen cut isolates the required source-pole structure. The $E^2\log Q^2$ dependence and its coefficient are determined by the photon amplitude, the remaining integrations, and the separation of lower-order iterations. The photon side remains a complete five-point amplitude and therefore retains photon exchange, graviton self-interactions, and the contact interactions required by the minimally coupled theory.

We use the external momenta of Eq.~\eqref{eq:all_outgoing} and take $\ell_5,\ell_7$ as independent loop momenta. Their directions of flow are indicated in Fig.~\ref{fig:five_fold_cut}: $\ell_5$ points from the source to the photon side, whereas $\ell_7$ points toward the source. It is convenient to define the three outgoing graviton momenta on the source side by
\be
 r_1=\ell_5,\qquad r_2=-\ell_7,\qquad
 r_3=q_c+\ell_7-\ell_5,\qquad r_1+r_2+r_3=q_c .
 \label{eq:five_cut_routing}
\ee
The common cut conditions are
\be
 \begin{aligned}
 g_1&=\ell_5^2=0,\qquad
 g_2=\ell_7^2=0,\qquad
 g_3=(\ell_5-\ell_7-q_c)^2=0,\\
 h_1&=(p_1+\ell_5)^2-m^2=0,\qquad
 h_2=(p_1+q_c+\ell_7)^2-m^2=0 .
 \end{aligned}
 \label{eq:five_cut_denominators}
\ee
Here $g_i$ and $h_j$ denote propagator denominators. The first three equations put the exchanged gravitons on shell, and the last two put the source propagators on shell. These equations define an algebraic generalized cut, on which the loop momenta are taken complex. They do not impose a positive-energy phase-space integration.

To display the sewing explicitly, we define
\be
 P_c=p_1+q_c=-p_4,\qquad
 P_L=p_1+\ell_5,\qquad P_R=P_c+\ell_7 .
 \label{eq:source_cut_momenta}
\ee
The outgoing momenta at the left, middle, and right source vertices are then $(p_1,-P_L,r_1)$, $(P_L,-P_R,r_3)$, and $(-P_c,P_R,r_2)$, respectively. Each triplet momenta sums to zero, and all scalar momenta are on shell with mass $m$ on the cut. Expressing the incoming photon to the all-outgoing convention reverses its helicity label, so we take the external photon states to be $p_{2,\gamma}^{-\sigma}$ and $p_{3,\gamma}^{\sigma'}$. The tree product is
\be
 \begin{aligned}
 \mathcal C_5^{\mathrm{tree}}
 =\sum_{\sigma_1,\sigma_2,\sigma_3=\pm2}
 &\mathcal A_5\bigl((-r_1)_h^{-\sigma_1},
 (-r_2)_h^{-\sigma_2},(-r_3)_h^{-\sigma_3},
 p_{2,\gamma}^{-\sigma},p_{3,\gamma}^{\sigma'}\bigr)\\
 {}\times&
 \mathcal A_3(p_1,-P_L,r_1^{\sigma_1})
 \mathcal A_3(-P_c,P_R,r_2^{\sigma_2})
 \mathcal A_3(P_L,-P_R,r_3^{\sigma_3}) .
 \end{aligned}
 \label{eq:five_cut_tree_product}
\ee
The scalar nature of the first two legs of each $\mathcal A_3$ is left implicit. Opposite momentum and helicity labels on the two sides of a graviton cut refer to the same internal state. With the phase convention of Eq.~\eqref{eq:tree_phase_normalization}, the four tree phases multiply to $(-i)^4=1$, so $\mathcal C_5^{\mathrm{tree}}/g^6$ is the corresponding sum of amplitude product structure $H_5H_3H_3H_3$. For helicity-preserving photon scattering, the photons have opposite all-outgoing helicities, and six of the eight graviton-helicity assignments contribute, as the components with three like-sign graviton helicities vanish (Sec.~\ref{subsec:tree_bcfw}).

The sum runs over physical graviton states only. Equivalently, in four dimensions the state sum is the symmetric-traceless transverse projector
\be
 \begin{aligned}
 \sum_{\lambda=\pm2}\varepsilon_{\mu\nu}^{\lambda}(K)
 \varepsilon_{\rho\sigma}^{\lambda *}(K)
 &=\frac12\bigl(\Pi_{\mu\rho}\Pi_{\nu\sigma}
 +\Pi_{\mu\sigma}\Pi_{\nu\rho}
 -\Pi_{\mu\nu}\Pi_{\rho\sigma}\bigr),\\
 \Pi_{\mu\nu}
 &=-\eta_{\mu\nu}
 +\frac{K_\mu n_\nu+n_\mu K_\nu}{K\cdot n},
 \qquad K^2=n^2=0 .
 \end{aligned}
 \label{eq:four_dimensional_graviton_sum}
\ee
Here $n$ is an auxiliary null reference vector with $K\cdot n\ne0$. The reference-vector terms cancel when contracted with the gauge-invariant tree amplitudes. For complex cut momenta, Eq.~\eqref{eq:five_cut_tree_product} implements the analytic continuation of this state sum through opposite helicities on the two sides of each cut. No additional scalar or antisymmetric tensor states are included, consistent with the use of four-dimensional helicity states in Sec.~\ref{subsec:scattering_conventions}.

For completeness, let $\mathcal I_F$ denote the loop integrand with Feynman phases included, so that its integral contributes to $i\mathcal M_3$. Since each cut propagator carries a factor $i/D_e$, removing these factors gives the matching equation
\be
 \left.
 \left[\prod_{D_e\in\{g_1,g_2,g_3,h_1,h_2\}}\frac{D_e}{i}\right]
 \mathcal I_F
 \right|_{g_i=h_j=0}
 =\mathcal C_5^{\mathrm{tree}} .
 \label{eq:five_cut_matching}
\ee
This equation determines the residue on the common cut. In the loop integral the propagators are restored with their $i0$ prescriptions; the algebraic cut does not by itself perform the energy integrations.

The five-point tree amplitude has at most two internal propagators. Resolving these poles gives rise to terms carrying two, one, or no additional denominators beyond the five common cut lines. In the integrand basis used below, these form eight seven-propagator terms, seven six-propagator terms, and one five-propagator term. We denote their numerators by $N_{7,1},\ldots,N_{7,8}$, $N_{6,1},\ldots,N_{6,7}$, and $N_5$. All sixteen terms enter the same cut constraint: the seven-propagator contributions must be fixed before the six-propagator residues, and both must be subtracted before determining the remaining five-propagator term. This provides the on-shell input for the polynomial reduction of Sec.~\ref{sec:integrand_reconstruction}.

\section{Integrand reconstruction by polynomial reduction}\label{sec:integrand_reconstruction}

The cuts in Sec.~\ref{sec:tree_amplitudes_cut} determine the numerator of each topology on the cut surfaces. To perform the loop integrals, we recast this information in terms of polynomials in independent loop-momentum variables. We use the integrand decomposition of OPP method~\cite{Ossola2007Reduction} and its multiloop formulation through polynomial division~\cite{Mastrolia2011TwoLoop,Mastrolia2012Division,Zhang2012Integrand}. The propagators specify the allowed residues, while the products of tree amplitudes determine their coefficients. We carry out the reconstruction proceeding from the seven-propagator terms down to the common five-propagator structure.

\subsection{Topologies and irreducible scalar products}\label{subsec:topologies_isps}

All terms share the five denominators in Eq.~\eqref{eq:five_cut_denominators}. For the algebraic reduction, we define their product stripped of the $i0$ prescriptions, since the reduction is insensitive to them,
\be
 \mathcal D_5=g_1g_2g_3h_1h_2.
 \label{eq:common_denominator_product}
\ee
The additional propagators from the photon-side tree amplitude are
\be
 \begin{aligned}
 D_A&=(\ell_5-p_2)^2,&
 D_B&=(\ell_7+p_3)^2,\\
 D_C&=(\ell_5-\ell_7-p_3)^2,&
 D_D&=(\ell_5-\ell_7-p_2)^2,\\
 D_E&=(\ell_7+p_2)^2,&
 D_F&=(\ell_5-p_3)^2,\\
 D_H&=(\ell_5-\ell_7)^2.&&
 \end{aligned}
 \label{eq:additional_denominators}
\ee
The denominators $D_A,\ldots,D_F$ correspond to internal photon lines, and $D_H$ to an additional graviton line. The loop momenta follow the momentum routing of Fig.~\ref{fig:five_fold_cut} for every topology.

\begin{figure}[!htbp]
\centering
\begin{minipage}{0.45\linewidth}\centering
  \includegraphics[width=\linewidth,height=2.8cm,keepaspectratio]{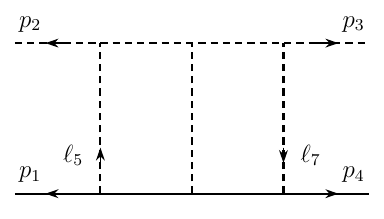}\\[-1mm]
  {\small $N_{7,1}$}
\end{minipage}
\hfill
\begin{minipage}{0.45\linewidth}\centering
  \includegraphics[width=\linewidth,height=2.8cm,keepaspectratio]{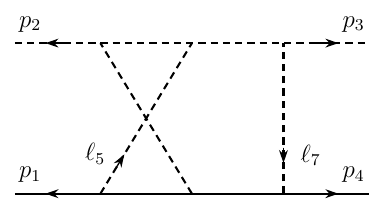}\\[-1mm]
  {\small $N_{7,2}$}
\end{minipage}
\\[2mm]
\begin{minipage}{0.45\linewidth}\centering
  \includegraphics[width=\linewidth,height=2.8cm,keepaspectratio]{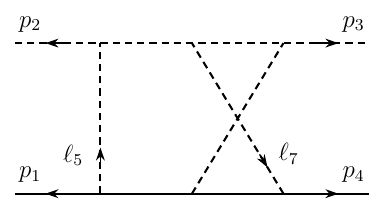}\\[-1mm]
  {\small $N_{7,3}$}
\end{minipage}
\hfill
\begin{minipage}{0.45\linewidth}\centering
  \includegraphics[width=\linewidth,height=2.8cm,keepaspectratio]{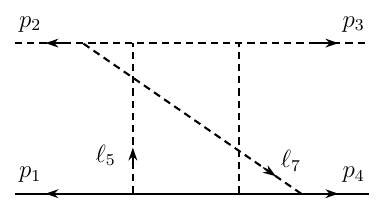}\\[-1mm]
  {\small $N_{7,4}$}
\end{minipage}
\\[2mm]
\begin{minipage}{0.45\linewidth}\centering
  \includegraphics[width=\linewidth,height=2.8cm,keepaspectratio]{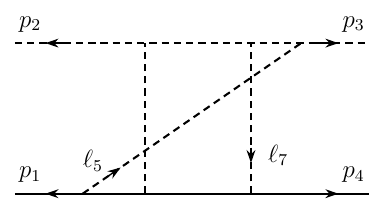}\\[-1mm]
  {\small $N_{7,5}$}
\end{minipage}
\hfill
\begin{minipage}{0.45\linewidth}\centering
  \includegraphics[width=\linewidth,height=2.8cm,keepaspectratio]{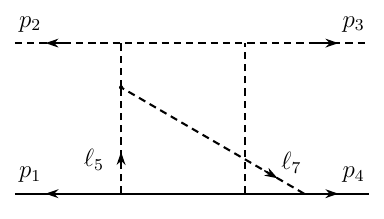}\\[-1mm]
  {\small $N_{7,6}$}
\end{minipage}
\\[2mm]
\begin{minipage}{0.45\linewidth}\centering
  \includegraphics[width=\linewidth,height=2.8cm,keepaspectratio]{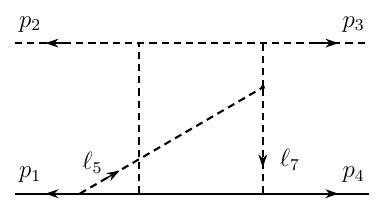}\\[-1mm]
  {\small $N_{7,7}$}
\end{minipage}
\hfill
\begin{minipage}{0.45\linewidth}\centering
  \includegraphics[width=\linewidth,height=2.8cm,keepaspectratio]{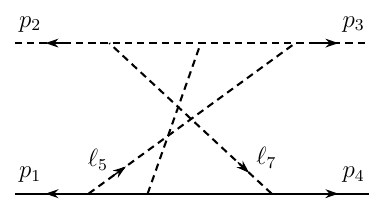}\\[-1mm]
  {\small $N_{7,8}$}
\end{minipage}
\caption{Seven-propagator topologies. Solid lines denote the massive scalar, dashed lines denote massless bosons, and arrows indicate momentum flow. Unmarked crossings are not vertices.}
\label{fig:topologies_seven}
\end{figure}

\begin{figure}[!htbp]
\centering
\begin{minipage}{0.45\linewidth}\centering
  \includegraphics[width=\linewidth,height=2.8cm,keepaspectratio]{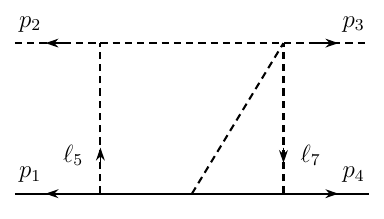}\\[-1mm]
  {\small $N_{6,1}$}
\end{minipage}
\hfill
\begin{minipage}{0.45\linewidth}\centering
  \includegraphics[width=\linewidth,height=2.8cm,keepaspectratio]{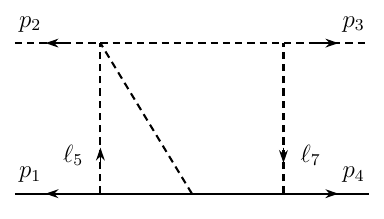}\\[-1mm]
  {\small $N_{6,2}$}
\end{minipage}
\\[2mm]
\begin{minipage}{0.45\linewidth}\centering
  \includegraphics[width=\linewidth,height=2.8cm,keepaspectratio]{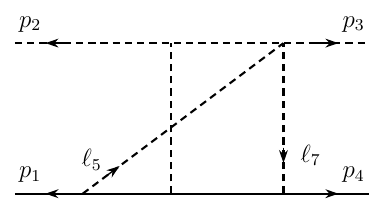}\\[-1mm]
  {\small $N_{6,3}$}
\end{minipage}
\hfill
\begin{minipage}{0.45\linewidth}\centering
  \includegraphics[width=\linewidth,height=2.8cm,keepaspectratio]{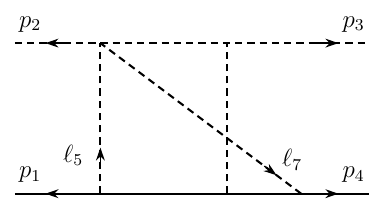}\\[-1mm]
  {\small $N_{6,4}$}
\end{minipage}
\\[2mm]
\begin{minipage}{0.45\linewidth}\centering
  \includegraphics[width=\linewidth,height=2.8cm,keepaspectratio]{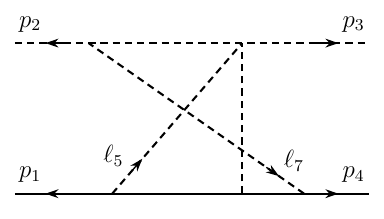}\\[-1mm]
  {\small $N_{6,5}$}
\end{minipage}
\hfill
\begin{minipage}{0.45\linewidth}\centering
  \includegraphics[width=\linewidth,height=2.8cm,keepaspectratio]{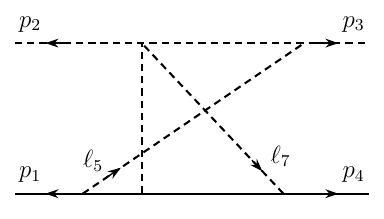}\\[-1mm]
  {\small $N_{6,6}$}
\end{minipage}
\\[2mm]
\begin{minipage}{0.45\linewidth}\centering
  \includegraphics[width=\linewidth,height=2.8cm,keepaspectratio]{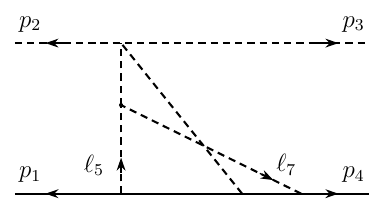}\\[-1mm]
  {\small $N_{6,7}$}
\end{minipage}
\caption{Six-propagator topologies. The conventions for line styles and momentum flow are the same as  Fig.~\ref{fig:topologies_seven}.}
\label{fig:topologies_six}
\end{figure}

\begin{figure}[htbp]
\centering
\begin{minipage}{0.45\linewidth}\centering
  \includegraphics[width=\linewidth,height=2.8cm,keepaspectratio]{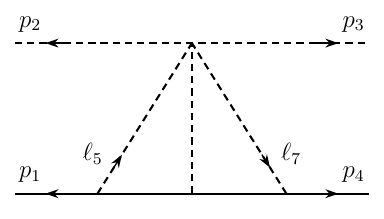}\\[-1mm]
  {\small $N_5$}
\end{minipage}
\caption{The five-propagator topology. The conventions for line styles and momentum flow are the same as Fig.~\ref{fig:topologies_seven}.}
\label{fig:topologies_five}
\end{figure}

Figures~\ref{fig:topologies_seven}--\ref{fig:topologies_five} show the sixteen terms, grouped by their number of propagators. Their extra denominators are listed in Table~\ref{tab:topology_denominators}. Six of the seven-propagator terms contain two internal photons, and the other two contain one photon and one additional graviton. The terms with fewer propagators collect the residues remaining after the higher-propagator contributions are subtracted.

\begin{table}[!htb]
 \centering
 \caption{Additional denominators of the sixteen terms. Every entry also contains the common factor $\mathcal D_5$.}
 \label{tab:topology_denominators}
 \begin{tabular}{cc@{\qquad\qquad}cc}
 \hline
 Numerator & Extra denominators & Numerator & Extra denominator\\
 \hline
 $N_{7,1}$ & $D_AD_B$ & $N_{6,1}$ & $D_A$\\
 $N_{7,2}$ & $D_BD_C$ & $N_{6,2}$ & $D_B$\\
 $N_{7,3}$ & $D_AD_D$ & $N_{6,3}$ & $D_C$\\
 $N_{7,4}$ & $D_DD_E$ & $N_{6,4}$ & $D_D$\\
 $N_{7,5}$ & $D_CD_F$ & $N_{6,5}$ & $D_E$\\
 $N_{7,6}$ & $D_DD_H$ & $N_{6,6}$ & $D_F$\\
 $N_{7,7}$ & $D_CD_H$ & $N_{6,7}$ & $D_H$\\
 $N_{7,8}$ & $D_ED_F$ & $N_5$ & $1$\\
 \hline
 \end{tabular}
\end{table}

We extract the common external spinor factor for the all-outgoing photon helicities $(2_\gamma^-,3_\gamma^+)$ by writing
\be
 S_{312}=\langle3|p_1|2],\qquad
 \mathcal I_F^{(\mathcal C_5)}
 =\frac{ig^6}{S_{312}^{\,2}}\mathcal I_{16},\qquad
 \mathcal I_{16}=\sum_T\frac{\Delta_T}{\prod_{e\in T}D_e}.
 \label{eq:reduced_integrand_normalization}
\ee
The superscript $(\mathcal C_5)$ specifies the sector reconstructed from the common cut and the associated six- and seven-fold cuts. Multiplication by $S_{312}^2$ removes the external little-group weights, allowing the numerators to be written in scalar-product variables. We use the residue conventions
\be
 \Delta_{7,u}=-N_{7,u},\qquad
 \Delta_{6,r}=iN_{6,r},\qquad
 \Delta_5=N_5.
 \label{eq:residue_numerator_phases}
\ee
The relative signs and phases in Eq.~\eqref{eq:residue_numerator_phases} account for the additional internal-propagator phases. To keep these factors explicit in the matching, for a cut $T$ consisting of $r$ propagators we define the normalized tree data by
\be
 \widetilde{\mathcal C}_T
 =i^{r-1}\frac{S_{312}^{\,2}}{g^6}\mathcal C_T^{\mathrm{tree}}
 =\left.\left[\prod_{e\in T}D_e\right]\mathcal I_{16}\right|_{D_e=0\ (e\in T)}.
 \label{eq:normalized_general_cut}
\ee
This follows from removing the propagator phases $i/D_e$ in Eq.~\eqref{eq:five_cut_matching}. The phase multiplying the tree data is $1$, $i$, or $-1$ for five-, six-, or seven-fold cuts, respectively.

The polynomial decomposition is based on a simple numerator identity. For a topology $T$, we write
\be
 P=\Delta_T+\sum_{e\in T}D_eQ_e,\qquad
 \frac{P}{\prod_{e\in T}D_e}
 =\frac{\Delta_T}{\prod_{e\in T}D_e}
 +\sum_{e\in T}\frac{Q_e}{\prod_{f\in T\setminus\{e\}}D_f}.
 \label{eq:polynomial_contact_identity}
\ee
Here $\Delta_T$ is the residue, and each term $D_eQ_e$ cancels one propagator and contributes to a lower-propagator topology. We refer to such contributions as contact terms. They may still contain massless exchange poles, so ``contact'' here refers to the algebraic reduction and does not imply a local interaction in impact-parameter space.

To identify independent numerator structures, we use the four-dimensional basis $e_A=(p_1,p_2,p_3,\omega)$, where $\omega$ is orthogonal to $p_1$, $p_2$, and $p_3$. Define
\be
 \begin{aligned}
 G_{AB}&=e_A\cdot e_B,\qquad
 X_A=\ell_5\cdot e_A,\qquad Y_A=\ell_7\cdot e_A,\\
 \ell_5^2&=X^{\mathrm T}G^{-1}X,\qquad
 \ell_7^2=Y^{\mathrm T}G^{-1}Y,\qquad
 \ell_5\cdot\ell_7=X^{\mathrm T}G^{-1}Y .
 \end{aligned}
 \label{eq:scalar_product_coordinates}
\ee
A scalar product that can be eliminated in favor of propagator denominators and external invariants is reducible. The remaining variables are irreducible scalar products (ISPs). Products of ISPs can themselves obey polynomial relations on a cut, so choosing independent variables does not by itself yield an independent monomial basis~\cite{Badger2012Hepta,Zhang2012Integrand}.

We work in the polynomial ring $\mathbb F[X,Y]$ over the function field $\mathbb F=\mathbb C(s,t,m^2)$. The propagator ideal is
\be
 I_T=\langle D_e:e\in T\rangle
 =\left\{\sum_{e\in T}f_e(X,Y)D_e(X,Y):f_e\in\mathbb F[X,Y]\right\}.
 \label{eq:propagator_ideal}
\ee
Two polynomials whose difference belongs to $I_T$ agree on the cut. A Gr\"obner basis generates the same ideal and, for a fixed monomial ordering, gives a unique remainder under polynomial division~\cite{Mastrolia2012Division,Zhang2012Integrand}. Monomials that are not divisible by any leading monomial of this basis form the allowed remainder structures. After imposing the numerator degree bounds, the residue takes the finite form
\be
 \Delta_T=\sum_{\alpha\in\mathcal B_T}
 c_{T,\alpha}(s,t,m^2)\,M_\alpha(X,Y).
 \label{eq:residue_polynomial_ansatz}
\ee
The basis $\mathcal B_T$ is determined by the propagators and the maximal numerator rank compatible with the tree input. The coefficients $c_{T,\alpha}$ are then fixed by the tree data. This separates the algebraic reduction from the dynamical input carried by the amplitudes.

\subsection{A seven-propagator reconstruction}\label{subsec:seven_propagator_example}

We illustrate the procedure with the topology of $N_{7,1}$, whose additional denominators are $D_A$ and $D_B$. The two additional cut conditions impose $\ell_5\cdot p_2=0$ and $\ell_7\cdot p_3=0$. Together with the massive cut conditions, they imply
\be
 X_1=X_2=Y_3=0,\qquad Y_1=-Y_2.
 \label{eq:seven_cut_linear_relations}
\ee
These linear relations follow from identities such as $2X_1=h_1-g_1$, $2X_2=g_1-D_A$, $2Y_3=D_B-g_2$, and $2(Y_1+Y_2+Y_3)=h_2-g_2$. Thus the eight loop coordinates reduce to $X_3,X_4,Y_2,Y_4$ before imposing the remaining quadratic equations.

We normalize the transverse basis vector and introduce two external invariants as
\be
 \omega_\mu=\frac{2i}{s}\epsilon_{\mu\nu\rho\sigma}
 p_1^\nu p_2^\rho p_3^\sigma,\qquad
 \rho_s=s-m^2,\qquad
 \mathcal S=\rho_s^2+st,\qquad
 \omega^2=-\frac{t\mathcal S}{s^2}.
 \label{eq:transverse_vector_normalization}
\ee
The reduction is performed at generic kinematics with $s\,t\,\rho_s\,\mathcal S\ne0$. Substituting Eq.~\eqref{eq:seven_cut_linear_relations} into the three massless cut conditions gives
\be
 \begin{aligned}
 s^2X_4^2&=\rho_s^2X_3^2,\qquad
 s^2Y_4^2=\rho_s^2Y_2^2,\\
 s^2X_4Y_4&=(\rho_s^2+2st)X_3Y_2
 +t\mathcal S\left(X_3-Y_2-\frac t2\right).
 \end{aligned}
 \label{eq:seven_cut_quadratics}
\ee
The first two equations fix the squares of the transverse components in terms of the longitudinal ones. The third relates their product to the scalar products along the external momenta.

These equations also induce relations among higher powers of the ISPs. Squaring the last equation and using the first two gives
\be
 \left[(\rho_s^2+2st)X_3Y_2
 +t\mathcal S\left(X_3-Y_2-\frac t2\right)\right]^2
 -\rho_s^4X_3^2Y_2^2=0 .
 \label{eq:seven_cut_quartic}
\ee
Since $(\rho_s^2+2st)^2-\rho_s^4=4st\mathcal S$, this yields the explicit reduction
\be
 \begin{aligned}
 X_3^2Y_2^2\equiv{}&
 -\frac{\rho_s^2+2st}{2s}X_3Y_2
 \left(X_3-Y_2-\frac t2\right)\\
 &-\frac{t\mathcal S}{4s}
 \left(X_3-Y_2-\frac t2\right)^2
 \pmod{I_7},
 \end{aligned}
 \label{eq:quartic_monomial_reduction}
\ee
where $I_7=\langle g_1,g_2,g_3,h_1,h_2,D_A,D_B\rangle$. The right-hand side contains monomials of lower total degree. The equality holds modulo the propagator ideal: away from the cut, the difference is absorbed into contact terms through Eq.~\eqref{eq:polynomial_contact_identity}.

With lexicographic order $X_4\succ Y_4\succ Y_2\succ X_3$, the leading monomials of the Gr\"obner basis after the linear relations \eqref{eq:seven_cut_linear_relations} have been used to eliminate $X_1,X_2,Y_1,Y_3$ are
\be
 \{X_4^2,\ Y_4^2,\ X_4Y_4,\ X_4Y_2,\ X_3^2X_4,\ X_3^2Y_2^2\}.
 \label{eq:seven_cut_leading_monomials}
\ee
For this reconstruction, we retain numerator degrees
\be
 \deg_X\le8,\qquad \deg_Y\le8,\qquad \deg_{X,Y}\le12,
 \label{eq:seven_numerator_degree_bounds}
\ee
which follow from the mass dimension and the pole structure of the tree input.
Removing all monomials divisible by those in Eq.~\eqref{eq:seven_cut_leading_monomials} gives
\be
 \begin{aligned}
 \mathcal B_7={}&
 \{X_3^aY_2^bY_4^c:a=0,1;\ c=0,1;\ 0\le b\le8-c\}\\
 &\cup\{X_3^aY_2^bY_4^c:2\le a\le8;\ b,c=0,1\}\\
 &\cup\{X_4,X_3X_4\}.
 \end{aligned}
 \label{eq:seven_basis_64}
\ee
The three sets contain $34$, $28$, and $2$ monomials, respectively. Thus this seven-propagator residue has 64 candidate coefficients. This count is specific to the present topology and rank bounds, rather than a universal basis size for two-loop gravity amplitudes.

To determine the coefficients, we parameterize all branches of the cut. The first two equations in Eq.~\eqref{eq:seven_cut_quadratics} give
\be
 X_4=\sigma_5\frac{\rho_s}{s}X_3,\qquad
 Y_4=\sigma_7\frac{\rho_s}{s}Y_2,\qquad
 \sigma_5,\sigma_7=\pm1 .
 \label{eq:seven_branch_signs}
\ee
The signs $\sigma_5,\sigma_7$ label the algebraic roots and carry no helicity information. Substituting them into the mixed equation yields
\be
 \begin{cases}
 2sX_3Y_2+\mathcal S(X_3-Y_2-t/2)=0,
 &\sigma_5=\sigma_7,\\
 (2X_3-t)(Y_2+t/2)=0,
 &\sigma_5=-\sigma_7 .
 \end{cases}
 \label{eq:seven_branch_equations}
\ee
There are six one-parameter branches. With $\sigma=\pm1$, they can be written as
\be
 \begin{array}{c|c|c|c}
 \text{Branch}&(\sigma_5,\sigma_7)&X_3&Y_2\\ \hline
 \mathcal V^{(1)}_\sigma&(\sigma,\sigma)&\zeta&
 \displaystyle\frac{\mathcal S(\zeta-t/2)}{\mathcal S-2s\zeta}\\[2mm]
 \mathcal V^{(2)}_\sigma&(\sigma,-\sigma)&\zeta&-t/2\\[1mm]
 \mathcal V^{(3)}_\sigma&(\sigma,-\sigma)&t/2&\zeta
 \end{array}.
 \label{eq:seven_cut_branches}
\ee
The other scalar products follow from Eqs.~\eqref{eq:seven_cut_linear_relations} and \eqref{eq:seven_branch_signs}. In particular, the seven-fold cut does not consist of six isolated loop-momentum points; each branch is parametrized by a complex variable $\zeta$.

Let $F_\sigma^{(j)}(\zeta)$ denote the normalized tree data of Eq.~\eqref{eq:normalized_general_cut} evaluated on these branches. Sewing the three-point amplitudes gives
\be
 \begin{aligned}
 F_+^{(1)}(\zeta)&=\rho_s^4(\mathcal S-2s\zeta)^2,&
 F_-^{(1)}(\zeta)&=\frac{\rho_s^8\mathcal S^2}{(\mathcal S-2s\zeta)^2},\\
 F_+^{(2)}(\zeta)=F_+^{(3)}(\zeta)&=\rho_s^8,&
 F_-^{(2)}(\zeta)=F_-^{(3)}(\zeta)&=\rho_s^4\mathcal S^2 .
 \end{aligned}
 \label{eq:seven_cut_tree_values}
\ee
All terms have mass dimension 16, consistent with the mass dimension of the seven-propagator numerator once the external spinor factor is removed. The rational denominator in $F_-^{(1)}$ originates from the rational parametrization of the branch and cancels against the corresponding factor in the basis evaluation below.

The coefficients satisfy
\be
 \left.\sum_{\alpha=1}^{64}c_\alpha M_\alpha
 \right|_{\mathcal V_\sigma^{(j)}(\zeta)}
 =F_\sigma^{(j)}(\zeta),\qquad j=1,2,3,\quad\sigma=\pm1 .
 \label{eq:seven_coefficient_matching}
\ee
For example, the first branch gives the polynomial
$\rho_s^4\mathcal S^2-4\rho_s^4s\mathcal S\,\zeta+4\rho_s^4s^2\zeta^2$ on the right-hand side. On the left-hand side, the basis monomials are evaluated using the rational expression for $Y_2$ in Eq.~\eqref{eq:seven_cut_branches}. Clearing the common denominator and comparing powers of $\zeta$ gives linear equations for $c_\alpha$. Applying this procedure to all six branches determines the residue uniquely. Equivalently, one can sample nonsingular values of $\zeta$ and solve the resulting linear system, which has full column rank.

The matching leaves 16 nonzero coefficients, listed in Appendix~\ref{app:seven_reconstruction}. Two simple examples are
\be
 [X_3^2Y_2]\Delta_{7,1}=-4\rho_s^2s^3,\qquad
 [X_3Y_2^2]\Delta_{7,1}=4\rho_s^2s^3,
 \qquad \Delta_{7,1}=-N_{7,1},
 \label{eq:seven_example_coefficients}
\ee
where the square brackets extract the corresponding monomial coefficient. Terms involving the transverse variables must be retained during this matching because they distinguish the cut branches. Under the simultaneous reflection $(X_4,Y_4)\mapsto(-X_4,-Y_4)$, all propagators and the integration measure are invariant. Terms odd under this reflection therefore integrate to zero, even though they are indispensable for reproducing the full four-dimensional cut.

For the subsequent integration, the part of $N_{7,1}$ even under this reflection has the compact form
\be
 \begin{aligned}
 \left.N_{7,1}\right|_{\mathrm{even}}=-\rho_s^2\bigg[
 &\mathcal S\left(\rho_s^4-\frac32s^2t^2\right)
 +4s^2t\mathcal S(X_3-Y_2)\\
 &+2s^2(4\rho_s^2+5st)X_3Y_2
 -2s^3t(X_3^2+Y_2^2)\\
 &-4s^3X_3Y_2(X_3-Y_2)\bigg].
 \end{aligned}
 \label{eq:seven_numerator_even}
\ee
This expression contains eight monomials. It can replace the full numerator under the loop integral; the odd terms remain part of the full residue used in Eq.~\eqref{eq:seven_coefficient_matching}.

\subsection{Lower cuts and contact terms}\label{subsec:lower_cuts_contacts}

A maximal (seven-fold) cut isolates the residue of the corresponding topology. On a six-fold cut, seven-propagator terms still contribute through their single uncut denominator. Their contribution must be subtracted before fitting the six-propagator residue~\cite{Mastrolia2011TwoLoop,Mastrolia2012Division}. More generally, for a cut $T$ we form the subtracted tree data
\be
 R_T=\widetilde{\mathcal C}_T
 -\left.
 \sum_{T'\supsetneq T}
 \frac{\Delta_{T'}}{\prod_{f\in T'\setminus T}D_f}
 \right|_{D_e=0\ (e\in T)} .
 \label{eq:recursive_cut_subtraction}
\ee
The residue $\Delta_T$ is determined by matching $R_T$ on this cut. Expanding $\Delta_T$ in the basis of Eq.~\eqref{eq:residue_polynomial_ansatz} and sampling at cut solutions $z^{(a)}$ gives
\be
 \sum_{\alpha\in\mathcal B_T}
 M_\alpha(z^{(a)})c_{T,\alpha}=R_T(z^{(a)}).
 \label{eq:residue_linear_system}
\ee
The samples must resolve all branches and yield a linear system of full column rank. A six-propagator topology uses its own propagator ideal and monomial basis while the 64-term basis of the preceding example does not carry over to a different cut.

As a concrete example, we consider the six-fold cut with $D_C=0$ in addition to the five common conditions. Table~\ref{tab:topology_denominators} shows that only three seven-propagator terms are compatible with this cut: $N_{7,2}$, $N_{7,5}$, and $N_{7,7}$. Their remaining uncut denominators are $D_B$, $D_F$, and $D_H$, respectively. The cut equation is
\be
 \widetilde{\mathcal C}_{6,C}
 =\left.
 \left(iN_{6,3}
 -\frac{N_{7,2}}{D_B}
 -\frac{N_{7,5}}{D_F}
 -\frac{N_{7,7}}{D_H}\right)
 \right|_{g_i=h_j=D_C=0}.
 \label{eq:six_cut_content}
\ee
Consequently, the polynomial to be matched to the six-propagator residue is
\be
 \left.iN_{6,3}\right|_{g_i=h_j=D_C=0}
 =\left.
 \left(\widetilde{\mathcal C}_{6,C}
 +\frac{N_{7,2}}{D_B}
 +\frac{N_{7,5}}{D_F}
 +\frac{N_{7,7}}{D_H}\right)
 \right|_{g_i=h_j=D_C=0}.
 \label{eq:six_cut_subtraction}
\ee
The plus signs on the right follow from the minus sign in the seven-propagator residue convention of Eq.~\eqref{eq:residue_numerator_phases}. This subtraction removes the remaining poles of these terms before determining $N_{6,3}$.

Polynomial division also dictates how contact terms are assigned to the lower-propagator topologies. For the topology of $N_{6,4}$, the extra denominator obeys
\be
 D_D=g_3+2X_3-2Y_3-t.
 \label{eq:six_cut_extra_relation}
\ee
Thus its six-fold cut imposes $X_3=Y_3+t/2$, in addition to the common conditions. Let $P_6$ be the numerator polynomial obtained after subtracting the seven-propagator contributions and bringing the result to the common denominator $\mathcal D_5D_D$. Division by this six-propagator ideal gives
\be
 P_6=iN_{6,4}+D_DQ_6+\sum_{D_j\in\{g_1,g_2,g_3,h_1,h_2\}}D_jQ_j.
 \label{eq:six_polynomial_division}
\ee
The numerator $N_{6,4}$ is already the remainder in its irreducible basis. The division by the propagators exposes the role of each quotient:
\be
 \frac{P_6}{\mathcal D_5D_D}
 =\frac{iN_{6,4}}{\mathcal D_5D_D}
 +\frac{Q_6}{\mathcal D_5}
 +\sum_{D_j\in\{g_1,g_2,g_3,h_1,h_2\}}
 \frac{Q_j}{D_D\displaystyle\prod_{D_k\ne D_j}D_k}.
 \label{eq:contact_term_assignment}
\ee
The product in the last term runs over the remaining common denominators. The term $Q_6/\mathcal D_5$ carries no $D_D$ pole and therefore belongs to the five-propagator residue. Each term in the final sum lacks at least one common propagator and has zero residue on the common five-fold cut. Thus the assignment of contact terms follows from the same polynomial division that defines the numerators.

After this subtraction has been repeated for all seven six-propagator terms, the remaining five-propagator numerator is fixed by
\be
 \left.N_5\right|_{\mathcal C_5}
 =\left.
 \left[\widetilde{\mathcal C}_5
 -\sum_{r=1}^{7}\frac{iN_{6,r}}{d_r}
 +\sum_{u=1}^{8}\frac{N_{7,u}}{d_{u1}d_{u2}}\right]
 \right|_{\mathcal C_5},
 \label{eq:five_residue_subtraction}
\ee
where $\mathcal C_5$ denotes $g_i=h_j=0$. The denominators $d_r$ and $(d_{u1},d_{u2})$ are the single and paired extra denominators listed in Table~\ref{tab:topology_denominators}. The five-propagator term therefore contains both the contact interactions in the tree input and the contact contributions assigned to it by the reduction of the higher-propagator terms.

The resulting representation is
\be
 \begin{aligned}
 \mathcal I_{16}
 &=\frac{1}{\mathcal D_5}\left[
 N_5+\sum_{r=1}^{7}\frac{iN_{6,r}}{d_r}
 -\sum_{u=1}^{8}\frac{N_{7,u}}{d_{u1}d_{u2}}
 \right],\\
 \mathcal M_3^{(\mathcal C_5)}
 &=\frac{g^6}{S_{312}^{\,2}}
 \int_{\ell_5}\int_{\ell_7}\mathcal I_{16}.
 \end{aligned}
 \label{eq:assembled_two_loop_integrand}
\ee
The loop measure is the one defined in Eq.~\eqref{eq:loop_measure}; the $i0$ prescriptions are restored when the integrals are evaluated. In this normalization, $N_5$, $N_{6,r}$, and $N_{7,u}$ have mass dimensions $12$, $14$, and $16$, respectively, so that every term carries the mass dimension of a four-point amplitude in four dimensions. By construction,
\be
 \left.\mathcal D_5\mathcal I_{16}\right|_{\mathcal C_5}
 =\widetilde{\mathcal C}_5.
 \label{eq:reconstructed_common_cut}
\ee
All sixteen numerators are defined in the same momentum routing and with a uniform assignment of contact terms. They can be found in the ancillary
file~\cite{numeratorfile}. The reconstruction uses four-dimensional helicity states and loop coordinates, while the subsequent integration is performed in dimensional regularization. We expand this representation in the classical probe regime in Sec.~\ref{sec:classical_integration}, evaluating the three topology groups in turn.

\section{Classical integration and the conservative kernel}\label{sec:classical_integration}

In this section, we evaluate the classical contribution of Eq.~\eqref{eq:assembled_two_loop_integrand} in the hierarchy $Q\ll E\ll m$. Our objective is the coefficient $C_3$ of $G^3m^4E^2\log(Q^2/\bar\mu^2)$ in Eq.~\eqref{eq:target_classical_log}, which determines the third-order deflection angle. The expansion in $E/m$ selects the leading heavy-source contribution, while the expansion in $Q/E$ determines which terms in the photon propagators and numerator polynomials contribute to the logarithm.

The calculation proceeds from the loop-energy integrations to the remaining spatial integrals. At leading order in the heavy-source expansion, the sum over attachment orderings localizes the exchanged energies. We then define a photon pole prescription that separates the cubic tree iteration from the genuine third-order interaction, and evaluate the spatial integrals with all sixteen reconstructed numerators. We denote the resulting classical kernel by $\mathcal K_3^{\mathrm R}$; its relation to iteration subtraction is derived in Sec.~\ref{subsec:prescription_iteration_proof}.

\subsection{Classical regions and pole prescriptions}\label{subsec:classical_regions}

The heavy-source contribution arises from the region where the spatial exchange momenta are of order $Q$ and their energies of order $Q^2/m$. A uniform Taylor expansion of the propagators before integration is insufficient because their denominators become small in different momentum regions. We therefore use expansion by regions: the integrand is expanded according to the scaling of each relevant region, and each expanded term is integrated over the full loop-momentum domain~\cite{Beneke1998Regions,Jantzen2011Regions}. Dimensional regularization, supplemented by analytic regularization when required, defines these integrals and their overlaps; scaleless overlap integrals vanish. Below we first identify the heavy and photon poles and then apply their respective expansions.

We consider a heavy propagator carrying momentum $p+\ell$, where $p=(m,\mathbf0)$ and $\ell=(\omega,\boldsymbol\ell)$. We define
\be
 E_\ell=\sqrt{m^2+\boldsymbol\ell^2},\qquad
 \delta_\ell=E_\ell-m.
 \label{eq:integration_heavy_energy}
\ee
The exact partial-fraction decomposition separates the positive- and negative-energy poles of the internal heavy line,
\be
 \frac1{(p+\ell)^2-m^2+i0}
 =\frac1{2E_\ell}
 \left[\frac1{\omega-\delta_\ell+i0}
 -\frac1{\omega+E_\ell+m-i0}\right].
 \label{eq:integration_heavy_partial_fractions}
\ee
The first pole lies in the classical (near-on-shell) energy region. Its position and residue have the heavy-mass expansions
\be
 \delta_\ell=\frac{\boldsymbol\ell^2}{2m}
 +\mathcal O\!\left(\frac{Q^4}{m^3}\right),\qquad
 \frac1{2E_\ell}=\frac1{2m}
 \left[1-\frac{\boldsymbol\ell^2}{2m^2}
 +\mathcal O\!\left(\frac{Q^4}{m^4}\right)\right].
 \label{eq:integration_heavy_expansion}
\ee
The second pole, at $\omega=-m-E_\ell+i0\simeq-2m+i0$, lies outside the classical region. The corresponding partial fraction is regular for $\omega\sim Q^2/m$ and expands as
\[
 -\frac1{2E_\ell(\omega+E_\ell+m-i0)}
 =-\frac1{4m^2}+\mathcal O\!\left(\frac{Q^2}{m^4}\right).
\]
This regular term receives no enhancement from the near-on-shell heavy pole. The leading source contribution therefore retains only the residue $1/(2m)$ of the first pole. The pole location must be determined keeping both $2m\omega$ and $\boldsymbol\ell^2$ in the denominator, since the two are of the same order near the pole. In the remaining photon subgraph, the recoil shift $\delta_\ell\sim Q^2/m$ is parametrically smaller than the photon-pole scale $Q^2/E$ by a factor $E/m$. This hierarchy justifies setting the exchanged energies to zero at leading order once the heavy poles have been isolated.

\begin{figure}[!htbp]
 \centering
 \subfigure[\enspace Ordering $(1,2)$.\label{fig:source_ordering_12}]{%
  \includegraphics[trim=0mm 3mm 80mm 9mm,clip,width=0.43\textwidth]{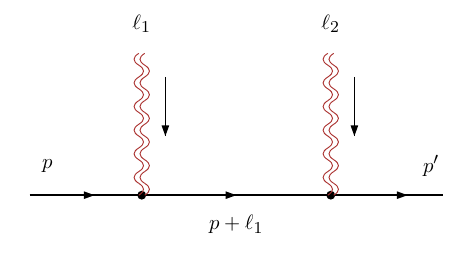}%
 }\hspace{0.04\textwidth}%
 \subfigure[\enspace Ordering $(2,1)$.\label{fig:source_ordering_21}]{%
  \includegraphics[trim=80mm 3mm 0mm 9mm,clip,width=0.43\textwidth]{heavy_source_orderings.pdf}%
 }
 \caption{Two attachment orderings on the scalar line. Black solid lines denote the scalar and red double-wavy lines denote gravitons.}
 \label{fig:source_orderings}
\end{figure}

The sum over the attachment orderings on the heavy line implements this static limit. Fig.~\ref{fig:source_orderings} illustrates the two orderings for two exchanged gravitons, with $\ell_1+\ell_2=q$ and $\omega=\ell_1^0$. At leading order, each scalar--graviton vertex is proportional to the same source tensor $p^\mu p^\nu$. The two orders therefore have identical source-vertex factors. With the remaining subgraph expressed in a common momentum routing, the linearized heavy propagators satisfy
\be
 \frac1{2m\omega+i0}+\frac1{-2m\omega+i0}
 =-2\pi i\,\delta(2m\omega)
 =-\frac{\pi i}{m}\delta(\omega).
 \label{eq:integration_two_orderings}
\ee
The principal-value terms cancel between the two orderings, leaving a delta function that fixes the exchange energy. Eq.~\eqref{eq:integration_two_orderings} omits the recoil shifts, the subleading residue corrections, and the distant negative-energy pole. For the full integrand, the common factor must be identified before applying this propagator identity. In particular, a permutation of the exchanged momenta acts on the numerator and the photon subgraph together with the heavy propagators.

For three exchanges with $\omega_1+\omega_2+\omega_3=0$, the two heavy propagators in each ordering give the identity which underlies the classical two-loop source-pole reduction,
\be
 \sum_{\sigma\in S_3}
 \frac1{(\omega_{\sigma(1)}+i0)
 (\omega_{\sigma(1)}+\omega_{\sigma(2)}+i0)}
 =(-2\pi i)^2\delta(\omega_1)\delta(\omega_2).
 \label{eq:integration_three_orderings}
\ee
The sum runs over all six permutations of the three attachments. In a time-integral representation of the linear propagators, these orders cover the full domain of the two independent relative times. Integrating over that domain gives the two delta functions in Eq.~\eqref{eq:integration_three_orderings}, which localize both independent exchange energies~\cite{Bern2019BlackHole}.

For the source-ordering argument, we consider the complete photon subgraph with three labeled graviton attachments, contracted with the leading source currents. Its on-shell limit is the five-point tree input of Sec.~\ref{sec:tree_amplitudes_cut}. Bose symmetry of the three graviton legs, together with the identical source tensors, makes this complete factor symmetric under simultaneous permutations of the exchanged momenta and graviton indices. This symmetry applies to the full subgraph, including its internal propagators and contact terms. By contrast, the individual terms in its sixteen-topology decomposition need not be symmetric.

To make the permutation average explicit, let $\mathcal H_\sigma$ denote the summand in Eq.~\eqref{eq:integration_three_orderings}, and let $\mathcal F_0$ contain the remaining subgraph contraction and the three exchange propagators with their $+i0$ prescriptions. For the reference ordering $(1,2,3)$, relabeling the integration variables gives
\[
 \int\mathcal F_0\mathcal H_{(1,2,3)}
 =\frac1{3!}\sum_{\sigma\in S_3}
   \int\mathcal F_0(r_\sigma)\mathcal H_\sigma
 =\frac1{3!}\int\mathcal F_0(r)
   \sum_{\sigma\in S_3}\mathcal H_\sigma,
\]
where $r_\sigma=(r_{\sigma(1)},r_{\sigma(2)},r_{\sigma(3)})$ and the second equality uses the symmetry of the complete subgraph. For the reconstructed representation, we apply this equality to the retained classical contribution, with common-propagator contact terms treated as in Appendix~\ref{app:contact_corrections}. The energy reduction is thus performed on the complete source-pole contribution, after which linearity permits separate spatial integrations of the sixteen terms. The factor $1/3!$ is the weight of this average, including the two loop-energy measures gives
\be
 w_E=\frac1{3!}\frac{(-2\pi i)^2}{(2\pi)^2}
 =-\frac16.
 \label{eq:integration_energy_weight}
\ee
The factors $1/(2m)$ from the two linearized heavy propagators supply an additional factor $1/(2m)^2$. The delta functions in Eq.~\eqref{eq:integration_three_orderings}, together with energy conservation, enforce $r_1^0=r_2^0=r_3^0=0$. The three graviton denominators thus reduce to $-\mathbf r_i^2+i0$, giving an overall minus sign. For a constant numerator over the five common propagators, the resulting normalization is
\be
 w_E\left(-\frac1{4m^2}\right)=\frac1{24m^2}.
 \label{eq:integration_common_prefactor}
\ee
We keep the overall factor $w_E$ explicit when tabulating the sixteen spatial integrals in Sec.~\ref{subsec:classical_kernel_sum}.

After the heavy-pole reduction, the photon propagators retain a near-on-shell region in the remaining spatial integration. For $k=(E,0,0,E)$ and a static exchange $\ell=(0,\boldsymbol\ell_\perp,\ell_\parallel)$, a representative photon denominator reads
\be
 (k+\ell)^2+i0
 =-2E\ell_\parallel-\boldsymbol\ell_\perp^2-\ell_\parallel^2+i0.
 \label{eq:integration_photon_region}
\ee
For generic soft momenta this denominator is of  $\mathcal{O}(EQ)$. Near the photon pole, however, $\ell_\parallel\simeq-\boldsymbol\ell_\perp^2/(2E)$ is of $\mathcal{O}( Q^2/E)$. The enhancement of the propagator compensates for the smaller longitudinal integration range. Contributions from this region must therefore be kept in the expansion of the spatial integrand.

Let $z$ denote a longitudinal variable for which the photon denominator takes the local form $2Ez-\Delta+i0$, with $\Delta$ quadratic in soft momenta. The following distributional expansion retains the displaced photon pole:
\be
 \begin{aligned}
 \frac1{2Ez-\Delta+i0}
 &\sim\sum_{n=0}^{\infty}
 \frac{\Delta^n}{(2E)^{n+1}(z+i0)^{n+1}},\\
 \frac1{(z+i0)^{n+1}}
 &=\operatorname{FP}\frac1{z^{n+1}}
 -i\pi\frac{(-1)^n}{n!}\delta^{(n)}(z).
 \end{aligned}
 \label{eq:integration_photon_distribution}
\ee
Here $\operatorname{FP}$ denotes the Hadamard finite-part distribution. For a simple pole it coincides with the Cauchy principal value, $\operatorname{FP}(1/z)=\operatorname{PV}(1/z)$. For higher poles, the finite part is defined by subtracting the divergent pieces from a symmetric exclusion of the pole; for example, for a smooth test function $f$,
\[
 \left\langle\operatorname{FP}\frac1{z^2},f\right\rangle
 =\lim_{\varepsilon\to0^+}
 \left[\int_{|z|>\varepsilon}\frac{f(z)}{z^2}\,dz
 -\frac{2f(0)}{\varepsilon}\right].
\]
The delta-function derivatives in Eq.~\eqref{eq:integration_photon_distribution} retain the corresponding on-shell contribution. For a $z$-independent $\Delta$, their convolution with a test function $f$ reproduces the expansion of $-\pi f(\Delta/(2E))/(2E)$, the imaginary part obtained from the shifted pole. If $\Delta$ depends on $z$, the same expansion applies with $\Delta^n$ multiplied into the test function. Both the finite-part and on-shell terms are therefore required in the soft expansion.

We implement the subtraction of repeated scattering in the static kernel through the following boundary values for the extra propagators of Eq.~\eqref{eq:additional_denominators}:
\be
 \left.\frac1{D_j}\right|_{\mathrm R}
 =\begin{cases}
 (D_j-i0)^{-1},&j=B,C,F,\\
 (D_j+i0)^{-1},&j=A,D,E,H.
 \end{cases}
 \label{eq:integration_R_prescription}
\ee
The common propagators retain $+i0$. Eq.~\eqref{eq:integration_R_prescription} assigns a boundary value to each labeled denominator in every topology, including terms with only one internal photon. The labels refer to the common momentum routing of Sec.~\ref{subsec:topologies_isps}. This prescription is applied to the static kernel after the source-energy reduction; its relation to the velocity-cut organization of classical terms~\cite{BjerrumBohr2022RadialAction} and to the cubic tree iteration is established in Sec.~\ref{subsec:prescription_iteration_proof}.

The real part is taken after the propagators have been multiplied together with the numerator. In particular, the product of two imaginary on-shell terms yields a real contribution when a topology contains two internal photon lines. Replacing each photon propagator separately by its principal value would discard this contribution and alter the iteration subtraction.

\subsection{Removal of the cubic tree iteration}\label{subsec:prescription_iteration_proof}

The prescription in Eq.~\eqref{eq:integration_R_prescription} removes the cubic tree iteration from the classical kernel. To establish this relation, we compare the change of propagator boundary values with the lower-order terms generated by the exponential representation of the scattering operator. We work with the leading heavy-source contribution where all three exchanged momenta are soft, and retain the expansion through the order contributing to $Q^0\log Q^2$. We denote the soft expansion, conservative restriction, Hermitian projection, and dimensionally regulated integration by $\Pi_{\rm cl}$.

The full scattering operator is unitary and admits the exponential representation~\cite{Damgaard2021Exponential}

\be
 S=1+iT=e^{i\mathcal N},\qquad
 \mathcal N^\dagger=\mathcal N,\qquad
 \operatorname{Herm}X=\frac{X+X^\dagger}{2}.
 \label{eq:prescription_operator_definition}
\ee

Here $\mathcal N=-i\log S$ is an operator. Its Hermiticity follows from $S^\dagger=S^{-1}$: on the branch of the logarithm continuously connected to $S=1$, $(\log S)^\dagger=\log S^{-1}=-\log S$. This statement refers to the full space of physical scattering states. Elastic matrix elements are taken only after $\mathcal N$ is defined; an elastic block of $S$ need not itself be unitary when radiation channels are open.

Let $T_r$ and $\mathcal N_r$ denote the elastic contributions at order $G^r$, with the coupling factors included. The elastic terms in the exponential expansion are\footnote{In the full state space, the lowest radiative channel also contributes to the third-order expansion through $i/2$ times the square of a Hermitian radiation operator. This term is anti-Hermitian and does not change Eq.~\eqref{eq:prescription_iteration_operator}. The conservative static projection is still imposed separately.}

\be
 \begin{aligned}
 T_1&=\mathcal N_1,\\
 T_2&=\mathcal N_2+\frac i2\mathcal N_1^2,\\
 T_3&=\mathcal N_3+\frac i2\{\mathcal N_1,\mathcal N_2\}
 -\frac16\mathcal N_1^3 ,
 \end{aligned}
 \label{eq:prescription_operator_expansion}
\ee
where $\{A,B\}=AB+BA$. Since $\mathcal N_1$ and $\mathcal N_2$ are Hermitian, their anticommutator is Hermitian and the term carrying the factor $i/2$ is anti-Hermitian. By contrast, the cubic term $-\mathcal N_1^3/6$ is Hermitian. Taking the Hermitian part and using $T_1=\mathcal N_1$ therefore gives

\be
 \mathcal N_3=\operatorname{Herm}T_3+\frac16T_1^3.
 \label{eq:prescription_iteration_operator}
\ee

Thus $\operatorname{Herm}T_3$ still contains the iterated tree contribution $-T_1^3/6$, and extracting $\mathcal N_3$ therefore requires adding back $T_1^3/6$. The imaginary one-loop iteration in Eq.~\eqref{eq:prescription_operator_expansion} also illustrates that an imaginary part of the amplitude does not necessarily signal dissipation.

For elastic matrix elements, the Hermitian projection is
\be
 (\operatorname{Herm}T)_{fi}
 =\frac12\bigl(T_{fi}+T_{if}^{*}\bigr).
 \label{eq:prescription_hermitian_matrix_element}
\ee
Scattering reciprocity relates an amplitude to that of the reversed process, with the external momenta and polarizations transformed consistently~\cite{Deak2012Reciprocity}. For helicity-preserving scattering from the static spinless source, this relation reduces to symmetry under interchange of the initial and final states after removal of the external helicity phase. We separate this phase from the integrated amplitude according to
\be
 \mathcal M_{fi}=U_{fi}\widehat{\mathcal M}_{fi},\qquad
 U_{fi}=\frac{\mathcal S}{S_{312}^{\,2}},\qquad
 |U_{fi}|=1,\quad U_{if}=U_{fi}^{*}.
 \label{eq:prescription_external_phase_factor}
\ee
Here $\mathcal S$ is defined in Eq.~\eqref{eq:transverse_vector_normalization}, and $\widehat{\mathcal M}_{fi}$ retains the real normalization factor $1/\mathcal S$. The reverse-channel phase follows from the complex conjugation of the external photon polarizations under initial--final interchange.

Terms odd under the transverse reflection of Sec.~\ref{subsec:seven_propagator_example} vanish upon integration. At the retained static order, $|\mathbf k|=|\mathbf k'|=E$ and $Q^2=|\mathbf k-\mathbf k'|^2$. Rotational invariance therefore restricts the reduced amplitude to a scalar function of $m$, $E$, and $Q^2$, which are unchanged by interchange of the initial and final states. Consequently,
\be
 \widehat{\mathcal M}_{fi}=F_h(m,E,Q^2)
 =\widehat{\mathcal M}_{if}.
 \label{eq:prescription_reduced_reciprocity}
\ee
Eq.~\eqref{eq:prescription_reduced_reciprocity} imposes no reality condition on $F_h$. Substituting Eqs.~\eqref{eq:prescription_external_phase_factor} and \eqref{eq:prescription_reduced_reciprocity} into Eq.~\eqref{eq:prescription_hermitian_matrix_element}, with the common momentum-conservation delta function removed, gives
\be
 U_{fi}^{-1}\frac{\mathcal M_{fi}+\mathcal M_{if}^{*}}2
 =\frac{\widehat{\mathcal M}_{fi}
       +\widehat{\mathcal M}_{if}^{*}}2
 =\operatorname{Re}\widehat{\mathcal M}_{fi}.
 \label{eq:prescription_hermitian_real_part}
\ee
Eqs.~\eqref{eq:prescription_reduced_reciprocity} and \eqref{eq:prescription_hermitian_real_part} apply to the summed, integrated kernel at the retained static order. The phase convention $S_{312}^{\,2}=\mathcal S$ in Eq.~\eqref{eq:integration_external_phase} sets $U_{fi}=1$, so the Hermitian projection is implemented by taking the real part of $\mathcal M$ in the normalization of Eq.~\eqref{eq:amplitude_normalization}, with the common factor $i$ removed. The restriction to conservative scattering is imposed independently and excludes dissipative radiation-reaction effects.

The product $T_1^3$ chains three successive tree interactions through intermediate physical states. Inserting two complete sets of physical photon--scalar states and removing the overall momentum-conservation delta function gives

\be
 \mathcal C_{fi}\equiv[T_1^3]_{fi}
 =\sum_{\lambda_1,\lambda_2=\pm1}
 \int d\Phi_2^{(1)}\,d\Phi_2^{(2)}
 \mathcal M_{1;f2}\mathcal M_{1;21}\mathcal M_{1;1i}.
 \label{eq:prescription_tree_convolution}
\ee

The labels $1,2$ denote the two intermediate states, and $\mathcal M_1$ is the tree amplitude defined in Eq.~\eqref{eq:pm_amplitude}. For the total momentum $P_{\rm tot}=p+k$, each intermediate state is assigned the two-body phase-space measure

\be
 d\Phi_2(P_{\rm tot};K,P_h)
 =(2\pi)^D\delta^{(D)}(P_{\rm tot}-K-P_h)
 \frac{d^d\mathbf K}{(2\pi)^d\,2|\mathbf K|}
 \frac{d^d\mathbf P_h}{(2\pi)^d\,2E_h},
 \qquad d=D-1.
 \label{eq:prescription_two_body_phase_space}
\ee

Both intermediate particles have positive energy. The photon-helicity sums link adjacent tree amplitudes, and their relative phases are retained throughout the convolution. Consequently, $\mathcal C_{fi}$ is a convolution over the two intermediate states, rather than the cube of an amplitude at fixed $s$ and $t$.

We now express the change of boundary prescription in terms of on-shell distributions. For a real inverse propagator $x$, we define

\be
 G_\sigma(x)=\frac1{x+i\sigma0}
 =\operatorname{PV}\frac1x-\frac{i\sigma}{2}\varrho(x),
 \qquad \varrho(x)=2\pi\delta(x),\qquad \sigma=\pm1.
 \label{eq:prescription_spectral_identity}
\ee

This decomposition applies to each extra denominator appearing in Eq.~\eqref{eq:integration_R_prescription}, including the unchanged graviton denominator $D_H$. In particular, the real correction can involve one photon and one graviton spectral density. For two inverse propagators $x_1,x_2$, we set $H_j=\operatorname{PV}(1/x_j)$ and $\varrho_j=\varrho(x_j)$. Distributional multiplication gives

\be
 G_{\sigma_1}(x_1)G_{\sigma_2}(x_2)
 =H_1H_2-\frac i2(\sigma_1\varrho_1H_2+\sigma_2H_1\varrho_2)
 -\frac{\sigma_1\sigma_2}{4}\varrho_1\varrho_2.
 \label{eq:prescription_spectral_product}
\ee

Relative to the standard $+i0$ prescription, the real part of the product changes by

\be
 \operatorname{Re}\!\left[
 G_{\sigma_1}(x_1)G_{\sigma_2}(x_2)
 -G_+(x_1)G_+(x_2)\right]
 =\frac{1-\sigma_1\sigma_2}{4}\varrho_1\varrho_2.
 \label{eq:prescription_double_spectral_change}
\ee

Flipping exactly one of the two boundary signs therefore contributes $+\varrho_1\varrho_2/2$; flipping neither or both signs gives no real correction. These statements concern the product of propagators with the retained real numerator. Taking the real part of each propagator separately would discard the spectral product and miss this contribution. Eq.~\eqref{eq:prescription_double_spectral_change} holds for the quadratic denominators before expansion. For the integrated kernel with its external helicity phase removed, the real part implements the Hermitian projection in Eq.~\eqref{eq:prescription_hermitian_real_part}.

We first consider the six topologies with two internal photon propagators. We use the exchange labels $r_1,r_2,r_3$ of Eq.~\eqref{eq:five_cut_routing}, where $r_3$ is attached at the middle source vertex. Table~\ref{tab:prescription_photon_orderings} lists the ordering of the attachments along the photon line, read from the incident momentum $k$ to the outgoing momentum $k'$. The ordering on the source line is held fixed in the table. The signs assigned by Eq.~\eqref{eq:integration_R_prescription} are obtained by reversing the boundary sign of each internal photon propagator that follows the $r_3$ attachment. If $r_3$ is last, neither photon sign changes; if it is first, both signs change; if it lies between the other two exchanges, only one sign changes. Hence only the topologies $N_{7,1}$ and $N_{7,8}$ receive nonzero double-photon spectral corrections.

\begin{table}[!htbp]
 \centering
 \caption{Photon-line attachment orderings and boundary signs. The digits label $r_1,r_2,r_3$ and are read from the incident photon to the outgoing photon.}
 \label{tab:prescription_photon_orderings}
 \renewcommand{\arraystretch}{1.15}
 \begin{tabular}{clcc}
 \hline\hline
 Ordering & Numerator & Photon denominators & R signs\\
 \hline
 $123$ & $N_{7,3}$ & $D_A,D_D$ & $+,+$\\
 $213$ & $N_{7,4}$ & $D_E,D_D$ & $+,+$\\
 $132$ & $N_{7,1}$ & $D_A,D_B$ & $+,-$\\
 $231$ & $N_{7,8}$ & $D_E,D_F$ & $+,-$\\
 $312$ & $N_{7,2}$ & $D_C,D_B$ & $-,-$\\
 $321$ & $N_{7,5}$ & $D_C,D_F$ & $-,-$\\
 \hline\hline
 \end{tabular}
\end{table}

Let $\Delta_{\gamma\gamma}$ denote the change of the retained real kernel induced by these two double-photon terms. Combining the spectral coefficient of Eq.~\eqref{eq:prescription_double_spectral_change} with the source-energy reduction gives the complete spatial integral
\be
 \begin{aligned}
 \Delta_{\gamma\gamma}
 ={}&\frac{w_Eg^6}{2(2m)^2}\,
 \Pi_{\rm cl}\int
 \frac{d^d\boldsymbol\ell_5\,d^d\boldsymbol\ell_7}{(2\pi)^{2d}}
 \\
 &\times\left.
 \frac{(2\pi)^2\bigl[
 N_{7,1}\delta(D_A)\delta(D_B)
 +N_{7,8}\delta(D_E)\delta(D_F)\bigr]}
 {S_{312}^{\,2}\boldsymbol\ell_5^2\boldsymbol\ell_7^2
 (\boldsymbol q_c+\boldsymbol\ell_7-\boldsymbol\ell_5)^2}
 \right|_{\ell_5^0=\ell_7^0=0}.
 \end{aligned}
 \label{eq:prescription_complete_photon_integral}
\ee
The common dimensional-regularization factor $(e^{\gamma_E}\bar\mu^2/4\pi)^{2\epsilon}$ from Eq.~\eqref{eq:loop_measure} is suppressed here and in the intermediate-state convolution below. The external helicity phase and the reflection-odd numerator terms are treated as described above. Eq.~\eqref{eq:prescription_complete_photon_integral} retains all photon numerators and exchange propagators.

Every factor in this expression originates from the reconstructed integrand. The two heavy residues supply the factor $1/(2m)^2$. The product of photon spectral densities in Eq.~\eqref{eq:prescription_double_spectral_change} gives the two delta functions and the coefficient $(2\pi)^2/2$. The three static graviton denominators have a negative product; this minus sign cancels the one multiplying $N_{7,a}$ in Eq.~\eqref{eq:assembled_two_loop_integrand}. The remaining energy weight is $w_E=-1/6$. Together with the external spinor normalization, these factors give $-g^6/(48m^2S_{312}^{\,2})$ before the soft expansion. The transfer-dependent factor $1/S_{312}^{\,2}$ is included in the action of $\Pi_{\rm cl}$. We now identify the intermediate states and the tree amplitudes within this integral.

For the two terms, we choose positive-energy intermediate photon momenta as
\be
 \begin{aligned}
 N_{7,1}:&\qquad K_1=k+\ell_5,\qquad K_2=k'+\ell_7,
 &D_A=K_1^2,\quad D_B=K_2^2,\\
 N_{7,8}:&\qquad K_1=k-\ell_7,\qquad K_2=k'-\ell_5,
 &D_E=K_1^2,\quad D_F=K_2^2.
 \end{aligned}
 \label{eq:prescription_intermediate_photon_momenta}
\ee
The source-energy integration fixes $\ell_5^0=\ell_7^0=0$. With $k^0=k'^0=E$, the intermediate photon energies are therefore $K_1^0=K_2^0=E>0$. Both changes of variables have a Jacobian of unit absolute value. The delta functions then reduce to $\delta(E^2-\mathbf K_1^2)\delta(E^2-\mathbf K_2^2)$. The three momentum transfers along the photon line are $K_1-k$, $K_2-K_1$, and $k'-K_2$, whose sum is $q_c=k'-k$. For either ordering, the product of exchange denominators in Eq.~\eqref{eq:prescription_complete_photon_integral} becomes
\[
 |\mathbf K_1-\mathbf k|^2
 |\mathbf K_2-\mathbf K_1|^2
 |\mathbf k'-\mathbf K_2|^2.
\]

The numerator on the double photon shell is fixed by tree factorization. The residue of the complete five-point tree amplitude factorizes into three photon--graviton vertices. With the normalization of Eq.~\eqref{eq:tree_phase_normalization},
\be
 \left.x_1x_2 H_5\right|_{x_1=x_2=0}
 =\sum_{\lambda_1,\lambda_2=\pm1}
 H_3^{(1)}H_3^{(2)}H_3^{(3)}.
 \label{eq:prescription_double_tree_factorization}
\ee
This follows by taking the photon residues successively.  Sewing each graviton leg to its scalar vertex and restoring the corresponding exchange propagator gives a photon--scalar tree amplitude. In the variables of Eq.~\eqref{eq:prescription_intermediate_photon_momenta}, the resulting relation at the retained source order is
\be
 \begin{aligned}
 &\left.
 -\frac{g^6}{S_{312}^{\,2}}
 \frac{N_{7,a}}
 {|\mathbf K_1-\mathbf k|^2
  |\mathbf K_2-\mathbf K_1|^2
  |\mathbf k'-\mathbf K_2|^2}
 \right|_{K_1^2=K_2^2=0}
 \\
 &\hspace{12mm}=
 \sum_{\lambda_1,\lambda_2=\pm1}
 \mathcal M_{1;f2}\mathcal M_{1;21}\mathcal M_{1;1i}
 +\mathcal R_{\rm ct}^{(a)},\qquad a=1,8.
 \end{aligned}
 \label{eq:prescription_sewn_numerator}
\ee
Here $N_{7,a}$ is evaluated through its own inverse change of variables in Eq.~\eqref{eq:prescription_intermediate_photon_momenta}. The three amplitudes describe $k\to K_1$, $K_1\to K_2$, and $K_2\to k'$ respectively, with the associated scalar momenta fixed by two-body momentum conservation before taking the static limit. The coupling factors, helicity phases, and single-graviton propagators are included in each amplitude. The sign follows from $\mathcal A_n=-ig^{n-2}H_n=i\mathcal M_n$ and $\Delta_{7,a}=-N_{7,a}$: the seven-fold residue gives $-N_{7,a}/S_{312}^{\,2}$ as the product of the six stripped three-point vertices.

The terms $\mathcal R_{\rm ct}^{(a)}$ account for common-graviton contact terms when this residue relation is extended to the static spatial integral. They cancel at least one of the three exchange denominators. We keep their integrated contribution, denoted by $\Delta_{\rm ct}$, explicitly; its vanishing at the retained order is shown in Appendix~\ref{app:contact_corrections}. 

The remaining measure in Eq.~\eqref{eq:prescription_complete_photon_integral} admits a direct intermediate-state interpretation. Integrating over the intermediate heavy momentum in Eq.~\eqref{eq:prescription_two_body_phase_space} gives
\be
 \begin{aligned}
 d\Phi_2
 &=\frac{d^d\mathbf K}{(2\pi)^d\,2|\mathbf K|}
   \frac{2\pi\delta(m+E-|\mathbf K|-E_h)}{2E_h},
 \qquad E_h=\sqrt{m^2+|\mathbf k-\mathbf K|^2},\\
 d\Phi_2\big|_{\rm static}
 &=\frac1{2m}\frac{d^d\mathbf K}{(2\pi)^d}\,
 2\pi\delta(E^2-\mathbf K^2),\qquad K^0=E>0.
 \end{aligned}
 \label{eq:prescription_spectral_phase_space}
\ee
The second line follows from expanding $E_h=m+\mathcal O(Q^2/m)$ and using $\delta(E-|\mathbf K|)/(2|\mathbf K|)=\delta(E^2-\mathbf K^2)$ for $E,|\mathbf K|>0$. Thus each source residue $1/(2m)$ combines with one photon delta function and one spatial integration to form a complete two-body phase-space measure. Two such factors in Eq.~\eqref{eq:prescription_complete_photon_integral} supply precisely the two intermediate-state integrations of Eq.~\eqref{eq:prescription_tree_convolution}.

This identification retains the quadratic static photon shell. In the forward soft region, we write $\mathbf K=(\mathbf r,K_\parallel)$ and $d_\perp=d-1=2-2\epsilon$. Solving the shell condition gives
\be
 \begin{aligned}
 K_\parallel&=+\sqrt{E^2-\mathbf r^2},\\
 d\Phi_2\big|_{\rm forward}
 &=\frac{d^{d_\perp}\mathbf r}{(2\pi)^{d_\perp}\,4mE}
 \left[1+\frac{\mathbf r^2}{2E^2}
 +\mathcal O\!\left(\frac{\mathbf r^4}{E^4}\right)\right].
 \end{aligned}
 \label{eq:prescription_forward_phase_space}
\ee
The backward longitudinal root lies outside the region in which all three exchanged momenta are soft. Within the forward region, the pole displacement, the phase-space Jacobian, the exchange denominators, and the helicity numerators must be expanded through the same soft order. Their quadratic corrections can multiply the leading soft singularity and contribute to the classical logarithm. Replacing the photon denominators by their linear terms before taking the residues would miss some of these contributions.

The two terms in Eq.~\eqref{eq:prescription_complete_photon_integral} give the same ordered product of tree amplitudes after factorization. The explicit exchange of variables is
\be
 (r_1,r_2,r_3)\mapsto(r_2,r_1,r_3),\qquad
 (\ell_5,\ell_7)\mapsto(-\ell_7,-\ell_5).
 \label{eq:prescription_ordering_change_variables}
\ee
It maps the photon ordering $132$ to $231$, preserves $r_1+r_2+r_3=q_c$, and has unit absolute Jacobian. The numerator and the intermediate-state labels are transformed together; the physical $K_1,K_2$ in Eq.~\eqref{eq:prescription_intermediate_photon_momenta} are then unchanged. Substituting the sewn numerators and the two phase-space factors into the complete integral gives
\be
 \begin{aligned}
 \Delta_{\gamma\gamma}
 ={}&\frac{1}{6(2m)^2}\Pi_{\rm cl}
 \sum_{\lambda_1,\lambda_2=\pm1}
 \int\frac{d^d\mathbf K_1\,d^d\mathbf K_2}{(2\pi)^{2d}}
 \\
 &\times(2\pi)^2\delta(E^2-\mathbf K_1^2)
 \delta(E^2-\mathbf K_2^2)
 \mathcal M_{1;f2}\mathcal M_{1;21}\mathcal M_{1;1i}
 +\Delta_{\rm ct}
 \\
 ={}&\frac16\Pi_{\rm cl}
 \sum_{\lambda_1,\lambda_2=\pm1}
 \int d\Phi_2^{(1)}\,d\Phi_2^{(2)}\,
 \mathcal M_{1;f2}\mathcal M_{1;21}\mathcal M_{1;1i}
 +\Delta_{\rm ct}
 \\
 ={}&\frac16\Pi_{\rm cl}[T_1^3]_{fi}+\Delta_{\rm ct}.
 \end{aligned}
 \label{eq:prescription_combinatorial_factor}
\ee
Both the integration measure and the full tree product are displayed in the first equality. The coefficient is $2(-w_E/2)=2\times(1/3!)\times(1/2)=1/6$: the factor $2$ counts the two topologies $N_{7,1}$ and $N_{7,8}$, $1/2$ comes from the double spectral product, and the $1/3!$ source-ordering weight from $w_E$ is included once. The remaining four double-photon orderings give no real prescription change.

The topology $N_{7,6}$ retains the standard boundary values, while the five- and six-propagator terms have no real prescription correction in the retained kernel. The contact term $\Delta_{\rm ct}$ and the mixed photon--graviton prescription correction from $N_{7,7}$ vanish at the retained static and soft orders. The complete integrals and derivations are collected in Appendix~\ref{app:mixed_contact_corrections}.

Let $T_3^{(\mathcal C_5)}$ be the operator kernel obtained from Eq.~\eqref{eq:assembled_two_loop_integrand} with the standard $+i0$ prescription, and let $T_3^{\mathrm R}$ denote the same representation with Eq.~\eqref{eq:integration_R_prescription}. With the mixed and contact terms vanishing under $\Pi_{\rm cl}$, the double-photon spectral terms reproduce the tree convolution. Including the phase-space and ordering factors therefore yields

\be
 \Pi_{\rm cl}\operatorname{Herm}T_3^{\mathrm R}
 =\Pi_{\rm cl}\left[
 \operatorname{Herm}T_3^{(\mathcal C_5)}+\frac16T_1^3\right].
 \label{eq:prescription_iteration_subtraction}
\ee

Comparison with Eq.~\eqref{eq:prescription_iteration_operator} identifies the added convolution with the subtraction of the cubic tree iteration. Eq.~\eqref{eq:prescription_iteration_subtraction} relates the integrated kernels after the static, all-soft, and conservative projections defined at the beginning of this subsection. The intermediate states in the subtraction are on shell, as required by the operator product in the Hermitian logarithm of $S$. The new third-order interaction is then obtained by evaluating the remaining numerator-dependent spatial integrals of all sixteen topologies, which determine $\mathcal K_3^{\mathrm R}$.

\subsection{Spatial integrals and representative topologies}\label{subsec:classical_spatial_integrals}

The spatial integrals determine the $Q^0\log Q^2$ term of the kernel at leading order in the heavy-source expansion. All topologies contain the same three common graviton denominators; they differ in their numerator polynomials and additional propagators. We first extract the common dependence on $m$ and $E$, then reduce the resulting dimensionless integrals. With the momentum routing of Fig.~\ref{fig:five_fold_cut} and the definition $m\xi=2E$, we introduce the rescaled momenta
\be
 \mathbf L=\frac{\boldsymbol\ell_5}{m\xi},\qquad
 \mathbf R=-\frac{\boldsymbol\ell_7}{m\xi},\qquad
 \mathbf Q=\frac{\mathbf q_c}{m\xi},\qquad
 \mathbf K=\mathbf Q-\mathbf L-\mathbf R.
 \label{eq:integration_spatial_momenta}
\ee
The vectors $\mathbf L$, $\mathbf R$, and $\mathbf K$ are the momenta transferred from the source to the photon subgraph by the left, right, and middle gravitons, respectively. The sign of $\mathbf R$ follows from the orientation of $\ell_7$ in Fig.~\ref{fig:five_fold_cut}; momentum conservation fixes $\mathbf K$ in terms of the two independent integration variables $\mathbf L$ and $\mathbf R$. Throughout this subsection the bold $\mathbf Q$ denotes the dimensionless vector defined above, whereas $Q=\sqrt{-t}$ remains the physical momentum-transfer scale.

We also introduce the average photon momentum and the squared momentum transfer in the same normalization,
\be
 \mathbf W=\frac{\mathbf k+\mathbf k'}{2m\xi},\qquad
 \rho=\mathbf Q^2=\frac{Q^2}{m^2\xi^2},\qquad
 \tau=\frac{t}{m^2\xi^2}=-\rho .
 \label{eq:integration_average_photon}
\ee
The static kinematics, $|\mathbf k|=|\mathbf k'|=E=m\xi/2$, imply
\be
 \mathbf W\cdot\mathbf Q=0,\qquad
 W^2=\frac{1-\rho}{4}.
 \label{eq:integration_static_kinematics}
\ee
Thus $\mathbf W$ is directed along the bisector of the incident and outgoing photon momenta, with its magnitude fixed by $\rho$. The scalar products entering the numerators and photon denominators are abbreviated as
\be
 u=\mathbf W\cdot\mathbf L,\quad v=\mathbf Q\cdot\mathbf L,\qquad
 w=\mathbf W\cdot\mathbf R,\quad z=\mathbf Q\cdot\mathbf R,\qquad
 u_K=-u-w,\quad v_K=\rho-v-z .
 \label{eq:integration_spatial_products}
\ee

To combine the numerator residues with the external-state normalization, we use the phase convention described in Sec.~\ref{subsec:prescription_iteration_proof}, in which
\be
 S_{312}^{\,2}=\mathcal S,\qquad
 \mathcal S=(s-m^2)^2+st
 =m^4\xi^2[1-(1+\xi)\rho].
 \label{eq:integration_external_phase}
\ee
The invariant $\mathcal S$ was defined in Eq.~\eqref{eq:transverse_vector_normalization}. For each topology with $5+k$ propagators, we extract the powers of $m$ and $\xi$ from the residue in Eq.~\eqref{eq:residue_numerator_phases} and define
\be
 \left.\Delta_a\right|_{\text{required probe order}}
 =m^{12+2k}\xi^{4+2k}n_a,\qquad k=0,1,2.
 \label{eq:integration_probe_polynomials}
\ee
The probe order indicated in Eq.~\eqref{eq:integration_probe_polynomials} retains the powers of $\xi$ that produce the overall scaling $G^3m^4E^2$ after restoring the common factors. The subsequent soft expansion selects the $Q^0\log Q^2$ term within this contribution. The topology-dependent phases of Eq.~\eqref{eq:residue_numerator_phases} are included in this definition: $n_{7,u}$, $n_{6,r}$, and $n_5$ are obtained from $-N_{7,u}$, $iN_{6,r}$, and $N_5$, respectively. In the source-pole region, we implement the probe expansion by the rescalings $t=m^2\xi^2\tau$, $X_i=m^2\xi^2\bar X_i$, and $Y_i=m^2\xi^2\bar Y_i$, with $i=1,2,3$. Terms odd under the transverse reflection of Sec.~\ref{subsec:seven_propagator_example} vanish upon integration. The polynomials used here therefore follow directly from the numerators and irreducible bases constructed in Sec.~\ref{sec:integrand_reconstruction}.

For the examples below, the required scalar products take the form
\be
 \bar X_2=u-\frac v2,\qquad \bar X_3=-u-\frac v2,\qquad
 \bar Y_2=-w+\frac z2,\qquad \bar Y_3=w+\frac z2.
 \label{eq:integration_isp_spatial_map}
\ee
These relations follow from the all-outgoing convention of Eq.~\eqref{eq:all_outgoing} and the on-shell conditions of Sec.~\ref{subsec:scattering_conventions}.

Each of the $k$ extra denominators supplies a factor $m^2\xi^2$, so that removing them leaves the same overall factor $m^{12}\xi^4$ for all three topology groups. At zero exchanged energy, the dimensionless denominators $\widehat D_j=D_j/(m^2\xi^2)$ are
\be
 \begin{aligned}
 \widehat D_A&=-\mathbf L^2-2u+v,&
 \widehat D_F&=-\mathbf L^2+2u+v,\\
 \widehat D_B&=-\mathbf R^2+2w+z,&
 \widehat D_E&=-\mathbf R^2-2w+z,\\
 \widehat D_C&=-\mathbf K^2-2u_K+v_K,&
 \widehat D_D&=-\mathbf K^2+2u_K+v_K,\\
 \widehat D_H&=-(\mathbf K-\mathbf Q)^2.&&
 \end{aligned}
 \label{eq:integration_spatial_denominators}
\ee
Their $i0$ prescriptions are assigned according to Eq.~\eqref{eq:integration_R_prescription}. The remaining spatial integral for topology $a$ is then
\be
 \mathcal J_a=
 \int\frac{d^d\mathbf L\,d^d\mathbf R}{(2\pi)^{2d}}\,
 \frac{n_a(\mathbf L,\mathbf R)}
 {\mathbf L^2\mathbf R^2\mathbf K^2
 \displaystyle\prod_{j\in a}\widehat D_j},
 \qquad d=3-2\epsilon .
 \label{eq:integration_spatial_topology}
\ee
The product contains zero, one, or two extra denominators, depending on the topology. The momentum dependence of the extra denominators dictates the order in which the integrations are performed. For $N_5$, the seven six-propagator topologies $N_{6,r}$, and $N_{7,6},N_{7,7}$, an outer momentum can be chosen such that the inner integral is a tensor bubble. The other six topologies contain two photon denominators associated with different exchange momenta. We treat them in a joint Schwinger-parameter representation so as to retain their coupled pole structure.

The successive integrations in the first group are based on the scalar bubble
\be
 \begin{aligned}
 B_{\alpha\beta}(P^2)
 &=\int\frac{d^d\mathbf L}{(2\pi)^d}
 \frac1{(\mathbf L^2)^\alpha[(\mathbf L+\mathbf P)^2]^\beta}\\
 &=\frac{(P^2)^{d/2-\alpha-\beta}}{(4\pi)^{d/2}}
 \frac{\Gamma(\alpha+\beta-d/2)\Gamma(d/2-\alpha)
 \Gamma(d/2-\beta)}
 {\Gamma(\alpha)\Gamma(\beta)\Gamma(d-\alpha-\beta)}.
 \end{aligned}
 \label{eq:integration_bubble}
\ee
Combining the denominators with a Feynman parameter and shifting $\mathbf L$ gives the second line, understood via analytic continuation in $d$, $\alpha$, and $\beta$. Polynomial numerators are treated by the corresponding tensor moments. For $\alpha=\beta=1$, dividing by $B_{11}$ defines the normalized moments, of which the first two are
\be
 \langle L^i\rangle_B=-\frac{P^i}{2},\qquad
 \langle L^iL^j\rangle_B
 =\frac{dP^iP^j-P^2\delta^{ij}}{4(d-1)}.
 \label{eq:integration_bubble_moments}
\ee
The first relation follows from the interchange $\mathbf L\mapsto-\mathbf L-\mathbf P$ of the two propagators. In the second relation, contraction with $\delta^{ij}$ cancels one propagator, leaving a scaleless integral, so that the trace vanishes. These identities reduce the inner-loop numerator to scalar products of the outer momentum. Their derivations via Feynman parameters, together with those of the master integrals, are collected in Appendix~\ref{app:spatial_integrals}.

The product of the three common graviton denominators defines the two-loop master integral
\be
 \begin{aligned}
 J_2(\rho)
 &=\int\frac{d^d\mathbf L\,d^d\mathbf R}{(2\pi)^{2d}}
 \frac1{\mathbf L^2\mathbf R^2(\mathbf L+\mathbf R-\mathbf Q)^2}\\
 &=\frac{\rho^{d-3}}{(4\pi)^d}
 \frac{\Gamma(3-d)\Gamma(d/2-1)^3}
 {\Gamma(3d/2-3)}.
 \end{aligned}
 \label{eq:integration_master}
\ee
Applying Eq.~\eqref{eq:integration_bubble} twice gives this result: the first integration produces a bubble factor and an outer denominator with exponent $\beta_B=2-d/2$. At $d=3-2\epsilon$, the simple pole of $\Gamma(3-d)=\Gamma(2\epsilon)$ multiplies $\rho^{d-3}=\rho^{-2\epsilon}$ and generates the logarithm of the momentum transfer. We use $J_2$ to normalize the logarithmic terms from the individual topologies.

The topology $N_{6,4}$ illustrates how tensor reduction combines with the soft expansion of a photon propagator. Its extra denominator, $\widehat D_D$, depends only on $\mathbf K$. We choose $\mathbf K$ as the outer integration variable and set $\mathbf R=\mathbf Q-\mathbf K-\mathbf L$, giving
\be
 \mathcal J_{6,4}
 =\int\frac{d^d\mathbf K}{(2\pi)^d}
 \frac1{h(2u_K+v_K-h+i0)}
 \int\frac{d^d\mathbf L}{(2\pi)^d}
 \frac{n_{6,4}(\mathbf L,\mathbf Q-\mathbf K-\mathbf L)}
 {\mathbf L^2(\mathbf L+\mathbf K-\mathbf Q)^2},
 \label{eq:integration_six_four_nested}
\ee
where $h=\mathbf K^2$, $r=(\mathbf K-\mathbf Q)^2$, and $v_K=(h+\rho-r)/2$. To identify the terms producing $Q^0\log Q^2$, we assign soft degrees under a simultaneous scaling of $\mathbf L$, $\mathbf K$, and $\mathbf Q$. The scalar products $u$ and $u_K$ carry degree one, while $h,r,v_K,\rho$ have degree two. The relevant expansion of the numerator is
\be
 n_{6,4}=-4+4(u_K+u)+\mathcal O(\lambda^2).
 \label{eq:integration_six_four_numerator}
\ee
The inner bubble has $\mathbf P=\mathbf K-\mathbf Q$. Eq.~\eqref{eq:integration_bubble_moments} therefore gives $\langle u\rangle_B=-u_K/2$, reducing the integral to
\be
 \begin{aligned}
 \mathcal J_{6,4}
 &=\mathcal B_d\int\frac{d^d\mathbf K}{(2\pi)^d}
 \frac{-4+2u_K+\mathcal O(\lambda^2)}
 {h\,r^{\beta_B}(2u_K+v_K-h+i0)},\\
 \mathcal B_d&=
 \frac{\Gamma(2-d/2)\Gamma(d/2-1)^2}
 {(4\pi)^{d/2}\Gamma(d-2)},\qquad
 \beta_B=2-\frac d2 .
 \end{aligned}
 \label{eq:integration_six_four_bubble}
\ee
We keep the $d$-dependence of both the bubble factor $\mathcal B_d$ and the exponent $\beta_B$ until the outer integration has been performed. The photon denominator is expanded through the corresponding soft order,
\be
 \frac1{2u_K+v_K-h+i0}
 =\frac1{2u_K+i0}
 +\frac{h-v_K}{(2u_K+i0)^2}+\cdots .
 \label{eq:integration_six_four_photon}
\ee
The leading and subleading propagator terms have soft degrees $-1$ and zero, respectively. The degree-zero part of their product with the numerator receives two contributions: the linear numerator term multiplying the leading propagator term and the constant term multiplying the subleading propagator term. This degree is selected because the measure and three graviton denominators together scale as $\lambda^{-4\epsilon}$. Taking the real part of the degree-zero product gives
\be
 1-\frac{h-v_K}{u_K^2}
 =1-\frac{h}{2u_K^2}-\frac{r}{2u_K^2}
 +\frac{\rho}{2u_K^2}.
 \label{eq:integration_six_four_target}
\ee
The inverse squares are understood in the finite-part sense, $\operatorname{FP}(1/u_K^2)$. All four terms have the same soft degree, including the term proportional to $\rho/u_K^2$, and all of them must be integrated before the $\rho$-independent coefficient is extracted.

Eq.~\eqref{eq:integration_six_four_target} gives only the soft degree-zero part of the numerator divided by the photon denominator. The inner $\mathbf L$ integration has already been performed in Eq.~\eqref{eq:integration_six_four_bubble}; the remaining task is to integrate this expression over $\mathbf K$, including the factor $\mathcal B_d/(h r^{\beta_B})$. Substituting it back into Eq.~\eqref{eq:integration_six_four_bubble} gives the complete contribution to be evaluated,
\be
 \begin{aligned}
 \left.\operatorname{Re}\mathcal J_{6,4}\right|_{\rm target}
 ={}&\mathcal B_d\int\frac{d^d\mathbf K}{(2\pi)^d}
 \frac1{h\,r^{\beta_B}}
 \left[1-\frac{h}{2u_K^2}-\frac{r}{2u_K^2}
 +\frac{\rho}{2u_K^2}\right]\\
 ={}&\mathcal B_d\int\frac{d^d\mathbf K}{(2\pi)^d}
 \left[\frac1{h\,r^{\beta_B}}
 -\frac1{2r^{\beta_B}u_K^2}
 -\frac1{2h\,r^{\beta_B-1}u_K^2}
 +\frac{\rho}{2h\,r^{\beta_B}u_K^2}\right].
 \end{aligned}
 \label{eq:integration_six_four_outer_integral}
\ee
Here ${\rm target}$ denotes the real contribution of soft degree zero that supplies the $Q^0\log Q^2$ term, and $h=\mathbf K^2$, $r=(\mathbf K-\mathbf Q)^2$, and $u_K=\mathbf W\cdot\mathbf K$. The second line makes the remaining integrations explicit: the four terms differ only in the powers of these three factors. We therefore introduce a common moment $\mathcal M_{\alpha\beta n}$, where $\alpha$ and $\beta$ are the powers of the two quadratic denominators and $2n$ is the power of the longitudinal factor $u_K$. The first term has $n=0$, while the other three have $n=-1$. Their respective pairs $(\alpha,\beta)$ are $(1,\beta_B)$, $(0,\beta_B)$, $(1,\beta_B-1)$, and $(1,\beta_B)$. All four integrals can thus be evaluated using
\be
 \begin{aligned}
 \mathcal M_{\alpha\beta n}
 &=\int\frac{d^d\mathbf K}{(2\pi)^d}
 \frac{(\mathbf W\cdot\mathbf K)^{2n}}
 {(\mathbf K^2)^\alpha[(\mathbf K-\mathbf Q)^2]^\beta}\\
 &=\frac{(W^2)^n\rho^{d/2+n-\alpha-\beta}}{(4\pi)^{d/2}}
 \frac{\Gamma(n+1/2)}{\sqrt\pi}\\
 &\quad\times
 \frac{\Gamma(\alpha+\beta-d/2-n)
 \Gamma(d/2+n-\alpha)\Gamma(d/2+n-\beta)}
 {\Gamma(\alpha)\Gamma(\beta)\Gamma(d+2n-\alpha-\beta)} .
 \end{aligned}
 \label{eq:integration_outer_moments}
\ee
Since $\mathbf W\cdot\mathbf Q=0$, the shift used in Feynman parametrization leaves the longitudinal numerator unchanged. Analytic continuation to negative integer $n$ defines the finite-part terms appearing in Eq.~\eqref{eq:integration_photon_distribution}. In this notation, the complete integral in Eq.~\eqref{eq:integration_six_four_outer_integral} becomes
\be
 \begin{aligned}
 \left.\operatorname{Re}\mathcal J_{6,4}\right|_{\rm target}
 =\mathcal B_d\bigg[
 &\mathcal M_{1,\beta_B,0}
 -\frac12\mathcal M_{0,\beta_B,-1}\\
 &-\frac12\mathcal M_{1,\beta_B-1,-1}
 +\frac{\rho}{2}\mathcal M_{1,\beta_B,-1}\bigg].
 \end{aligned}
 \label{eq:integration_six_four_moments}
\ee
The second moment is scaleless: setting $\mathbf k=\mathbf K-\mathbf Q$ removes $\mathbf Q$ from the remaining quadratic denominator without changing $u_K=\mathbf W\cdot\mathbf k$, since $\mathbf W\cdot\mathbf Q=0$. This moment therefore vanishes in dimensional regularization. Using $J_2=\mathcal B_d\mathcal M_{1,\beta_B,0}$, we express the remaining moments as ratios of gamma functions. At general $d$, the two nontrivial ratios are
\be
 \begin{aligned}
 \frac{\mathcal M_{1,\beta_B-1,-1}}
 {\mathcal M_{1,\beta_B,0}}
 &=-\frac2{W^2}
 \frac{(1-d/2)(3d/2-4)}{d/2-2}
 \ \longrightarrow\ -\frac1{W^2},\\
 \frac{\rho\,\mathcal M_{1,\beta_B,-1}}
 {\mathcal M_{1,\beta_B,0}}
 &=\frac2{W^2}
 \frac{(3d/2-4)(3d/2-5)}{d/2-2}
 \ \longrightarrow\ \frac1{W^2}.
 \end{aligned}
 \label{eq:integration_six_four_ratios}
\ee
The arrows denote the $d\to3$ limit. Substituting these ratios into Eq.~\eqref{eq:integration_six_four_moments} gives
\be
 \left.\operatorname{Re}\mathcal J_{6,4}\right|_{\rm target}
 =\left[1+\frac{3d-8}{W^2}\right]J_2(\rho).
 \label{eq:integration_six_four_master}
\ee
In this topology, the master integral has only a simple pole in $\epsilon$. The terms of order $\epsilon$ in its prefactor therefore contribute to the finite local term, while the prefactor at $d=3$ determines the coefficient of the single logarithm.

We next consider the six topologies with two photon propagators. Their coupled poles require the joint Schwinger-parameter representation developed below, in line with the discussion of Sec.~\ref{subsec:classical_regions}. For $N_{7,1}$, the numerator in Eq.~\eqref{eq:seven_numerator_even} gives
\be
 n_{7,1}=P(-u-v/2,-w+z/2;-\rho),
 \label{eq:integration_seven_one_polynomial}
\ee
with
\be
 \begin{aligned}
 P(a,b;\tau)={}&1+\tau-\frac32\tau^2-\frac32\tau^3
 +4\tau(1+\tau)(a-b)-2\tau(a^2+b^2)\\
 &+(8+10\tau)ab-4a^2b+4ab^2.
 \end{aligned}
 \label{eq:integration_probe_P}
\ee
The corresponding spatial integral is
\be
 \mathcal J_{7,1}
 =\int\frac{d^d\mathbf L\,d^d\mathbf R}{(2\pi)^{2d}}\,
 \frac{n_{7,1}}
 {\mathbf L^2\mathbf R^2\mathbf K^2
 (-2u+v-\mathbf L^2+i0)(2w+z-\mathbf R^2-i0)}.
 \label{eq:integration_seven_one_full}
\ee
We introduce positive Schwinger parameters $a,b,c$ for the three graviton denominators and $\alpha,\beta$ for the two photon denominators. The opposite photon pole prescriptions are represented by
\be
 \frac1{x+i0}=-i\lim_{\zeta\to0^+}
 \int_0^\infty d\alpha\,e^{i\alpha x-\zeta\alpha},
 \qquad
 \frac1{y-i0}=i\lim_{\zeta\to0^+}
 \int_0^\infty d\beta\,e^{-i\beta y-\zeta\beta}.
 \label{eq:integration_photon_parameters}
\ee
 At fixed Schwinger parameters, we introduce auxiliary sources $j=(\mathbf j_L,\mathbf j_R)^{\mathrm T}$ and define the generating function before the momentum integrations by
\be
 \begin{aligned}
 \mathcal G_\gamma(j)
 ={}&\int\frac{d^d\mathbf L\,d^d\mathbf R}{(2\pi)^{2d}}\,
 \exp\Bigl[-a\mathbf L^2-b\mathbf R^2-c\mathbf K^2\\
 &\quad+i\alpha(-2u+v-\mathbf L^2)
       -i\beta(2w+z-\mathbf R^2)\\
 &\quad+2\mathbf j_L\cdot\mathbf L
       +2\mathbf j_R\cdot\mathbf R\Bigr].
 \end{aligned}
 \label{eq:integration_joint_generating_integral}
\ee
Here $\mathbf K=\mathbf Q-\mathbf L-\mathbf R$, and the damping factor $e^{-\zeta(\alpha+\beta)}$ remains in the Schwinger-parameter measure. The auxiliary sources are set to zero after differentiation; their sole role is to generate powers of the loop momenta in the numerator. Expanding $\mathbf K^2$ and collecting the quadratic and linear terms gives the Gaussian matrix and source vectors
\be
 \begin{aligned}
 M_\gamma&=
 \begin{pmatrix}a+c+i\alpha&c\\c&b+c-i\beta\end{pmatrix},\\
 J_L&=(c+i\alpha/2)\mathbf Q-i\alpha\mathbf W,\qquad
 J_R=(c-i\beta/2)\mathbf Q-i\beta\mathbf W .
 \end{aligned}
 \label{eq:integration_joint_matrix}
\ee
The off-diagonal entries of $M_\gamma$ account for the term $-2c\mathbf L\cdot\mathbf R$ in the exponent, while the linear terms are $2(J_L+\mathbf j_L)\cdot\mathbf L+2(J_R+\mathbf j_R)\cdot\mathbf R$. Completing the square and performing the two Gaussian momentum integrations therefore yields
\be
 \mathcal G_\gamma(j)
 =\frac{\exp[(J+j)^{\mathrm T}M_\gamma^{-1}(J+j)-c\rho]}
 {(4\pi)^d(\det M_\gamma)^{d/2}}.
 \label{eq:integration_joint_generating}
\ee
In this expression, $J=(J_L,J_R)^{\mathrm T}$, and the matrix product includes the Euclidean scalar products of the spatial source vectors, without complex conjugation. The factors of two in the source terms of Eq.~\eqref{eq:integration_joint_generating_integral} imply
\be
 \begin{aligned}
 \frac12\frac{\partial}{\partial j_L^i}
 e^{2\mathbf j_L\cdot\mathbf L+2\mathbf j_R\cdot\mathbf R}
 &=L^i e^{2\mathbf j_L\cdot\mathbf L+2\mathbf j_R\cdot\mathbf R},\\
 \frac12\frac{\partial}{\partial j_R^i}
 e^{2\mathbf j_L\cdot\mathbf L+2\mathbf j_R\cdot\mathbf R}
 &=R^i e^{2\mathbf j_L\cdot\mathbf L+2\mathbf j_R\cdot\mathbf R}.
 \end{aligned}
 \label{eq:integration_joint_source_derivatives}
\ee
Thus each power of a loop momentum in $n_{7,1}$ can be inserted by differentiating the generating function with respect to the corresponding source. These derivatives can be evaluated on the integrated expression in Eq.~\eqref{eq:integration_joint_generating}, giving
\be
 \begin{aligned}
 \mathcal J_{7,1}
 ={}&\lim_{\zeta\to0^+}
 \int_0^\infty da\,db\,dc\,d\alpha\,d\beta\,
 e^{-\zeta(\alpha+\beta)}\\
 &\times\left.
 n_{7,1}\left(\frac12\partial_{\mathbf j_L},
              \frac12\partial_{\mathbf j_R}\right)
 \mathcal G_\gamma(j)\right|_{j=0}.
 \end{aligned}
 \label{eq:integration_joint_parameters}
\ee
For the numerator factor $uw=(\mathbf W\cdot\mathbf L)(\mathbf W\cdot\mathbf R)$, the source identity is explicitly
\be
 \left.\frac14
 (\mathbf W\cdot\partial_{\mathbf j_L})
 (\mathbf W\cdot\partial_{\mathbf j_R})
 e^{2\mathbf j_L\cdot\mathbf L+2\mathbf j_R\cdot\mathbf R}
 \right|_{j=0}
 =(\mathbf W\cdot\mathbf L)(\mathbf W\cdot\mathbf R)=uw.
 \label{eq:integration_joint_uw_insertion}
\ee
Applying the same differential operator to $\mathcal G_\gamma(j)$ therefore inserts $uw$ under the momentum integral in Eq.~\eqref{eq:integration_joint_generating_integral}.  Eq.~\eqref{eq:integration_joint_parameters} consequently includes both the constant and momentum-dependent parts of the numerator, with the quadratic photon terms and their pole prescriptions retained.

We first separate the common scale of the five Schwinger parameters from their ratios. Since $c>0$, we make the change of variables
\be
 \begin{aligned}
 a&=cA,\qquad b=cB,\qquad
 \alpha=c\bar\alpha,\qquad\beta=c\bar\beta,\\
 da\,db\,dc\,d\alpha\,d\beta
 &=c^4\,dc\,dA\,dB\,d\bar\alpha\,d\bar\beta.
 \end{aligned}
 \label{eq:integration_joint_scale_change}
\ee
The four ratios remain positive and independent. This substitution gives $M_\gamma=c\bar M_\gamma$ and $J=c\bar J$, where
\be
 \begin{aligned}
 \bar M_\gamma&=
 \begin{pmatrix}A+1+i\bar\alpha&1\\1&B+1-i\bar\beta\end{pmatrix},\\
 \bar J_L&=(1+i\bar\alpha/2)\mathbf Q-i\bar\alpha\mathbf W,\\
 \bar J_R&=(1-i\bar\beta/2)\mathbf Q-i\bar\beta\mathbf W.
 \end{aligned}
 \label{eq:integration_joint_rescaled_matrix}
\ee
Writing $\bar J=(\bar J_L,\bar J_R)^{\mathrm T}$, we define
\be
 \begin{aligned}
 \bar\Delta_\gamma&=\det\bar M_\gamma
 =(A+1+i\bar\alpha)(B+1-i\bar\beta)-1,\\
 \Phi_\zeta&=\rho-\bar J^{\mathrm T}\bar M_\gamma^{-1}\bar J
             +\zeta(\bar\alpha+\bar\beta).
 \end{aligned}
 \label{eq:integration_joint_scale_exponent}
\ee
For a unit numerator, Eq.~\eqref{eq:integration_joint_parameters} then becomes
\be
 \begin{aligned}
 \left.\mathcal J_{7,1}\right|_{n_{7,1}=1}
 ={}&\frac1{(4\pi)^d}\lim_{\zeta\to0^+}
 \int_0^\infty
 \frac{dA\,dB\,d\bar\alpha\,d\bar\beta}{\bar\Delta_\gamma^{d/2}}
 \int_0^\infty dc\,c^{4-d}e^{-c\Phi_\zeta}\\
 ={}&\frac{\Gamma(5-d)}{(4\pi)^d}\lim_{\zeta\to0^+}
 \int_0^\infty
 \frac{dA\,dB\,d\bar\alpha\,d\bar\beta}
 {\bar\Delta_\gamma^{d/2}\Phi_\zeta^{5-d}}.
 \end{aligned}
 \label{eq:integration_joint_scale_integrated}
\ee
The power $c^{4-d}$ combines the Jacobian $c^4$ with $(\det M_\gamma)^{-d/2}=c^{-d}\bar\Delta_\gamma^{-d/2}$. The second line follows from the gamma-function integral over $c$, with the complex powers defined by continuation from the regulated Schwinger representation. This step performs one parameter integration exactly. The remaining four integrations are still coupled through $\bar\Delta_\gamma$ and $\Phi_\zeta$. Momentum-dependent numerators are treated by applying the source derivatives before the $c$ integration, retaining any additional powers of $c$ that they generate.

The limit $\rho\to0$ cannot be obtained by expanding this integrand at fixed values of all four ratios. To see the relevant endpoint explicitly, consider ordinary soft loop momenta $\mathbf L,\mathbf R\sim\sqrt\rho$, with $\mathbf W$ of order one. The graviton denominators are of order $\rho$, whereas the photon denominators are generically of order $\sqrt\rho$. Their Schwinger parameters therefore scale as
\be
 \begin{aligned}
 a,b,c&\sim\rho^{-1},\qquad
 \alpha,\beta\sim\rho^{-1/2},\\
 A,B&\sim1,\qquad
 \bar\alpha=\sqrt\rho\,x,\qquad
 \bar\beta=\sqrt\rho\,y.
 \end{aligned}
 \label{eq:integration_joint_endpoint_scaling}
\ee
Here $x$ and $y$ are positive integration variables held fixed in this region. For finite positive $A$ and $B$, let $\Delta_0=AB+A+B$. Using $\mathbf W\cdot\mathbf Q=0$, the leading terms are
\be
 \begin{aligned}
 \bar\Delta_\gamma&=\Delta_0+\mathcal O(\sqrt\rho),\\
 \left.\Phi_\zeta\right|_{\zeta=0}
 &=\rho\,
 \frac{AB+W^2\bigl[(B+1)x^2+(A+1)y^2-2xy\bigr]}
 {\Delta_0}
 +\mathcal O(\rho^{3/2}).
 \end{aligned}
 \label{eq:integration_joint_endpoint_exponent}
\ee
The infinitesimal damping is suppressed in this equation only to display the power counting; the original photon boundary values remain understood in the integrations. Both barred photon parameters approach zero together, and the term $-2xy$ shows explicitly why their expansions cannot be treated independently. Since $d\bar\alpha\,d\bar\beta=\rho\,dx\,dy$, the leading scalar contribution in this region scales as
\be
 \rho\,\rho^{d-5}=\rho^{d-4}=\rho^{-1-2\epsilon}.
 \label{eq:integration_joint_leading_soft_scale}
\ee
The term required for the classical logarithm is two soft degrees higher: it contains one additional power of $\rho$ and hence scales as $\rho^{d-3}=\rho^{-2\epsilon}$. We must therefore keep the expansion through relative order $\rho$, including the corrections from the photon denominators, the numerator, and the kinematic prefactors. In particular, a quadratic numerator such as $uw$ already supplies this additional factor of $\rho$, corresponding to two soft degrees. The relation $W^2=(1-\rho)/4$ must also be expanded to the required order.

Eq.~\eqref{eq:integration_joint_endpoint_scaling} specifies a region in which both photon-parameter ratios, $\bar\alpha$ and $\bar\beta$, approach zero as $\sqrt{\rho}$, while $A$ and $B$ remain of order unity. The boundaries where $A$, $B$, $x$, or $y$ become small or large must be examined for further nonuniform limits. In each contributing region, the operation is the same: balance the competing terms in $\bar\Delta_\gamma$ and $\Phi_\zeta$, rescale the parameters with their Jacobian, and expand through the required soft order before integrating over the full rescaled domain. The region contributions are then combined with the overlap treatment described in Sec.~\ref{subsec:classical_regions}. Keeping the ratios fixed throughout the expansion would miss contributions concentrated near these boundaries.

The dimensional expansion is performed after this soft expansion and the parameter integrations. To make the required accuracy explicit, suppose a region contributes at the target soft order in the form
\be
 \rho^{-2\epsilon}
 \bigl(f_0+\epsilon f_1+\mathcal O(\epsilon^2)\bigr)
 \left[\frac{r_{-2}}{\epsilon^2}
       +\frac{r_{-1}}{\epsilon}+r_0+\mathcal O(\epsilon)\right].
 \label{eq:integration_joint_epsilon_product}
\ee
Here $f_0$ and $f_1$ are the coefficients of a prefactor regular at $\epsilon=0$, for example one produced by the tensor reduction, and $r_{-2}$ and $r_{-1}$ are the pole coefficients of the remaining parameter integral. Expanding $\rho^{-2\epsilon}=1-2\epsilon\log\rho+2\epsilon^2\log^2\rho+\cdots$, its finite logarithmic terms are
\be
 2f_0r_{-2}\log^2\rho
 -2\bigl(f_0r_{-1}+f_1r_{-2}\bigr)\log\rho.
 \label{eq:integration_joint_epsilon_logarithms}
\ee
Thus the $\mathcal O(\epsilon)$ prefactor contributes to the single logarithm whenever a double pole is present.  These products must be retained when the region contributions are combined and the classical logarithmic coefficient is extracted.

The contribution of the momentum-dependent numerator can be read off directly. Including the relative external-state factor, its soft expansion is
\be
 \frac{n_{7,1}}{1-\rho}
 =1+8uw+\mathcal O(\lambda^3).
 \label{eq:integration_seven_one_soft}
\ee
The quadratic term cancels the leading photon denominators, $8uw/[(-2u)(2w)]=-2$, yielding $-2J_2$. For the constant term, we set the numerator to unity in Eq.~\eqref{eq:integration_joint_parameters} and retain the quadratic photon terms through the soft expansion. After the parameter integrations and the sum over the contributing regions, the term at the target soft order is
\be
 \left.\operatorname{Re}\mathcal J_{7,1}
 \right|_{n_{7,1}=1,\,{\rm target}}
 =\frac{2\Gamma(1+2\epsilon)\rho^{-2\epsilon}}{(4\pi)^d}
 \left[-\frac{3\pi}{\epsilon}+\mathcal O(1)\right].
 \label{eq:integration_seven_one_constant_pole}
\ee
Here $W^2=(1-\rho)/4$ has been expanded consistently through the required order, including its correction to the leading soft term. The region contributions have been combined before expanding in $\epsilon$, leaving a simple pole at this soft order. Its coefficient at $\epsilon=0$ is $-3/(32\pi^2)$, whereas the pole coefficient of $J_2$ in Eq.~\eqref{eq:integration_master} is $1/(64\pi^2)$. Since both terms carry the scale factor $\rho^{-2\epsilon}$, their single-logarithm coefficients have the same ratio. The constant-numerator contribution is therefore
\be
 \left.\operatorname{Re}\mathcal J_{7,1}
 \right|_{n_{7,1}=1,\,\rho^0\log\rho}
 =-6\left.J_2(\rho)\right|_{\log\rho}
 =\frac{3}{16\pi^2}\log\rho.
 \label{eq:integration_seven_one_constant_log}
\ee
Combining this result with the quadratic numerator contribution gives the complete target logarithm for this topology,
\be
 \left.\operatorname{Re}\left[\frac{\mathcal J_{7,1}}{1-\rho}\right]
 \right|_{\rho^0\log\rho}
 =(-6-2)\left.J_2(\rho)\right|_{\log\rho}
 =\frac1{4\pi^2}\log\rho.
 \label{eq:integration_seven_one_complete_log}
\ee
The relative external-state factor is already included through Eq.~\eqref{eq:integration_seven_one_soft}. These relations specify the logarithmic terms; they do not give all the finite local parts of the integrals. With the common normalization in Eq.~\eqref{eq:integration_coefficient_projection}, the constant and quadratic numerator terms contribute $3/2$ and $1/2$, respectively, giving $c_{7,1}^{\mathrm R}=2$.

\subsection{Sum of the sixteen contributions}\label{subsec:classical_kernel_sum}

We now combine the spatial integrals with the normalization factors to extract the $Q^0\log Q^2$ coefficient of the kernel. For each numerator $n_a$, let $\mathcal R_a(\rho,W^2)$ denote the logarithmic contribution of its spatial integral relative to that of $J_2$. It is defined after the parameter integrations and the $\epsilon$ expansion, including products of higher poles with $\epsilon$-dependent tensor coefficients. The heavy-propagator Jacobians, the sign of the three spatial graviton propagators, and the relative external-state normalization then give
\be
 c_a^{\mathrm R}
 =[\rho^0]\left\{
 -\frac{\mathcal R_a(\rho,W^2)}{4(1-\rho)}
 \right\}_{W^2=(1-\rho)/4}.
 \label{eq:integration_coefficient_projection}
\ee
The projection $[\rho^0]$ selects the constant term of the Laurent expansion in $\rho$, taken after integration and after imposing $W^2=(1-\rho)/4$. This order is essential because a numerator power of $\rho$ can combine with an inverse power generated by the integral to contribute at order $\rho^0$. For $N_{6,4}$, Eq.~\eqref{eq:integration_six_four_master} yields
\be
 c_{6,4}^{\mathrm R}
 =[\rho^0]\left\{-\frac1{4(1-\rho)}
 \left[1+\frac4{1-\rho}\right]\right\}
 =-\frac54.
 \label{eq:integration_six_four_coefficient}
\ee
The four moments in Eq.~\eqref{eq:integration_six_four_moments} contribute $-1/4$, $0$, $-1/2$, and $-1/2$, respectively. In particular, both the linear numerator and the subleading photon propagator contribute to this coefficient.

For the six topologies with two photon propagators, we separate the constant and quadratic numerator contributions as in the $N_{7,1}$ example. For the topologies $N_{7,1}$, $N_{7,2}$, $N_{7,3}$, $N_{7,4}$, $N_{7,5}$, and $N_{7,8}$, listed in this order, the quadratic terms give the following coefficients of $J_2$, including the relative external factor $1/(1-\rho)$:
\be
 a_a=(-2,-1,-2,-1,1,0).
 \label{eq:integration_two_photon_tensor_coefficients}
\ee
Evaluating the constant-numerator terms with the joint parameter representation, including the quadratic parts of the photon denominators, gives
\be
 \left.c_a^{\mathrm R}\right|_{\rm constant}
 =\left(\frac32,-\frac12,-\frac12,-\frac12,-\frac12,\frac32\right),
 \qquad
 c_a^{\mathrm R}
 =\left.c_a^{\mathrm R}\right|_{\rm constant}-\frac{a_a}{4}.
 \label{eq:integration_two_photon_coefficients}
\ee
The external factor is included once for each contribution in this decomposition. Combining these six results with the ten topologies evaluated by successive bubble integrations yields the coefficients collected in Table~\ref{tab:classical_sixteen_coefficients}. To restore the source-pole weight and the common dimensionful factor, we define
\be
 C_a=w_Ec_a^{\mathrm R}m^{10}\xi^4
 =-\frac16c_a^{\mathrm R}m^{10}\xi^4.
 \label{eq:integration_dimensionful_coefficients}
\ee
The power $m^{10}\xi^4$ follows from the factor $m^{12}\xi^4$ extracted above and the factor $m^{-2}$ from the heavy residues.

\begin{table}[!htbp]
 \centering
 \caption{Classical coefficients of the sixteen topologies with the prescription in Eq.~\eqref{eq:integration_R_prescription}.}
 \label{tab:classical_sixteen_coefficients}
 \renewcommand{\arraystretch}{1.16}
 \begin{tabular}{lccc}
 \hline\hline
 Numerator & Propagators & $c_a^{\mathrm R}$ & $C_a/(m^{10}\xi^4)$\\
 \hline
 $N_5$     &5&$-38/15$&$19/45$\\
 $N_{6,1}$ &6&$-49/30$&$49/180$\\
 $N_{6,2}$ &6&$-49/30$&$49/180$\\
 $N_{6,3}$ &6&$-6/5$&$1/5$\\
 $N_{6,4}$ &6&$-5/4$&$5/24$\\
 $N_{6,5}$ &6&$-67/60$&$67/360$\\
 $N_{6,6}$ &6&$-83/60$&$83/360$\\
 $N_{6,7}$ &6&$1/2$&$-1/12$\\
 \hline
 $N_{7,1}$ &7&$2$&$-1/3$\\
 $N_{7,2}$ &7&$-1/4$&$1/24$\\
 $N_{7,3}$ &7&$0$&$0$\\
 $N_{7,4}$ &7&$-1/4$&$1/24$\\
 $N_{7,5}$ &7&$-3/4$&$1/8$\\
 $N_{7,6}$ &7&$-1/4$&$1/24$\\
 $N_{7,7}$ &7&$1/4$&$-1/24$\\
 $N_{7,8}$ &7&$3/2$&$-1/4$\\
 \hline
 Total &&$-8$&$4/3$\\
 \hline\hline
 \end{tabular}
\end{table}

The contributions from the five-, six-, and seven-propagator groups sum to
\be
 \frac{C_5}{m^{10}\xi^4}=\frac{19}{45},\qquad
 \frac{C_6}{m^{10}\xi^4}=\frac{463}{360},\qquad
 \frac{C_7}{m^{10}\xi^4}=-\frac38 ,
 \label{eq:integration_group_sum}
\ee
where $C_6=\sum_{r=1}^7C_{6,r}$ and $C_7=\sum_{u=1}^8C_{7,u}$. Each group contributes to the logarithmic kernel. Within the seven-propagator group, the vanishing coefficient of $N_{7,3}$ follows from a cancellation between its constant and quadratic numerator terms. The complete sum is
\be
 C_{\mathrm R}=C_5+C_6+C_7=\frac43m^{10}\xi^4.
 \label{eq:integration_total_coefficient}
\ee

With the measure of Eq.~\eqref{eq:loop_measure}, the  master integral expands as
\be
 J_2^{\overline{\mathrm{MS}}}(Q^2)
 =\frac1{64\pi^2\epsilon}
 -\frac1{32\pi^2}\log\frac{Q^2}{\bar\mu^2}
 +c_{\rm loc}+\mathcal O(\epsilon),
 \label{eq:integration_dimensional_master}
\ee
The pole and the constant $c_{\rm loc}$ are independent of $Q$ and correspond to contact distributions localized at zero impact parameter. The logarithm supplies the long-range term specified in Eq.~\eqref{eq:target_classical_log}. The relative external-state factor has already been included in Eq.~\eqref{eq:integration_coefficient_projection}; restoring its remaining scale $1/(m^4\xi^2)$ and the gravitational coupling therefore gives
\be
 \left.\mathcal K_3^{\mathrm R}(Q^2)\right|_{Q^0\log Q^2}
 =\frac{(8\pi G)^3}{m^4\xi^2}\,
 C_{\mathrm R}\left.J_2^{\overline{\mathrm{MS}}}(Q^2)\right|_{\log Q^2}.
 \label{eq:integration_kernel_restore}
\ee
Substituting $C_{\mathrm R}$ and $\xi=2E/m$, we obtain
\be
 \left.\mathcal K_3^{\mathrm R}(Q^2)\right|_{Q^0\log Q^2}
 =-\frac{256\pi}{3}G^3m^4E^2
 \log\frac{Q^2}{\bar\mu^2}.
 \label{eq:integration_kernel_log}
\ee
This result combines the source-pole weight with the spatial integrals of all sixteen reconstructed terms, including the contact terms assigned by the polynomial reduction. In the next section, we use this logarithmic kernel to determine the conservative radial action and the third-order deflection angle.

\section{Radial action and the deflection angle at two-loop order}\label{sec:radial_action_angle}

We now relate the classical kernel in Eq.~\eqref{eq:integration_kernel_log} to the deflection of the photon. We first specify the radial action convention and its relation to the scattering angle. We then perform the transverse Fourier transform and combine the third-order contribution with the tree and one-loop results. Throughout this section, we retain the conservative static projection defined in Sec.~\ref{subsec:prescription_iteration_proof}.

\subsection{Radial action and conservative scattering}\label{subsec:radial_action_definition}

In a static, spherically symmetric field, the photon trajectory lies in a plane and is characterized by the conserved energy $E$ and orbital angular momentum $J$. The Hamilton--Jacobi function can be separated as
\be
 \mathcal S_{\rm HJ}=-Et+J\varphi+\int^r dr'\,p_r(r';E,J),
 \qquad H(r,p_r,J)=E,
 \label{eq:angle_hamilton_jacobi}
\ee
where $H$ is the photon Hamiltonian in the static source field. We choose $p_r>0$ on the outgoing branch and denote the turning point by $r_{\min}$. At fixed $E$ and $r$, differentiation of the energy equation gives
\be
 \left.\frac{\partial p_r}{\partial J}\right|_{E,r}
 =-\frac{\partial H/\partial J}{\partial H/\partial p_r}
 =-\frac{d\varphi}{dr}.
 \label{eq:angle_radial_momentum_derivative}
\ee
The second equality follows from Hamilton's equations, $\dot\varphi=\partial H/\partial J$ and $\dot r=\partial H/\partial p_r$. Thus the angular change along the orbit can be obtained by differentiating a radial momentum integral.

We define the interaction contribution to the radial action by subtracting the free trajectory at the same $E$ and $J$,
\be
 \begin{aligned}
 I_r^{\rm int}(E,J)=\lim_{R_{\max}\to\infty}\bigg[
 &2\int_{r_{\min}}^{R_{\max}}dr\,p_r(r;E,J)
 -2\int_b^{R_{\max}}dr\,p_{r,0}(r;E,J)\\
 &-\mathcal C_{\rm IR}(E,R_{\max})\bigg],
 \qquad
 p_{r,0}=\sqrt{E^2-\frac{J^2}{r^2}},\quad b=\frac JE .
 \end{aligned}
 \label{eq:angle_interaction_action}
\ee
The factor of two accounts for the incoming and outgoing branches. The subtraction $\mathcal C_{\rm IR}$ removes the remaining, $J$-independent infrared term. More explicitly, if $p_r=E+a(E)/r+\mathcal O(r^{-2})$ at large $r$, we may choose $\mathcal C_{\rm IR}=2a(E)\log(R_{\max}/b_0)$, where $b_0$ is a reference length independent of $J$. Changing the reference length adds only a $J$-independent constant to the action.

The lower limit of the first integral also depends on $J$. Its contribution to the derivative is $-2p_r(r_{\min})\partial_Jr_{\min}$, which vanishes at a simple turning point. Although $\partial_Jp_r$ diverges as $(r-r_{\min})^{-1/2}$ near the turning point, the singularity is integrable. Applying Eq.~\eqref{eq:angle_radial_momentum_derivative} at finite $R_{\max}$ and then taking the limit gives
\be
 \left.\frac{\partial I_r^{\rm int}}{\partial J}\right|_E
 =-\Delta\varphi+\pi=-\chi,
 \qquad
 \chi(E,b)=-\frac1E\frac{\partial I_r^{\rm int}(E,b)}{\partial b}.
 \label{eq:angle_action_derivative}
\ee
Here $\Delta\varphi$ is the total azimuthal change, and the free trajectory contributes $\pi$. This fixes both the sign and the free-motion subtraction in the definition of the deflection angle.

The same relation follows from the conservative scattering operator. In the semiclassical limit, the elastic partial-wave eigenvalue has the form $S_J=e^{2i\delta_J}=\exp[iI_r^{\rm int}(E,J)]$, up to a $J$-independent phase, where $\delta_J$ is the phase shift. Consequently, the classical eigenphase of $\mathcal N=-i\log S$ in Eq.~\eqref{eq:prescription_operator_definition} is the interaction radial action~\cite{Damgaard2021Exponential,Damgaard2023ClassicalObservables}. The products of lower-order amplitudes generated by expanding $S$ must therefore be removed before identifying a new contribution to $I_r^{\rm int}$. Eq.~\eqref{eq:prescription_iteration_subtraction} implements this removal for the third-order static contribution considered here.

\subsection{Fourier transform of the classical kernel}\label{subsec:radial_action_fourier}

In the amplitude--radial-action representation, the third-order contribution is obtained from the kernel by~\cite{Bern2021G4,BjerrumBohr2022RadialAction,Damgaard2023ClassicalObservables}
\be
 I_{r,3}^{\mathrm R}(E,b)
 =\frac1{4mE}
 \int\frac{d^2\boldsymbol q_\perp}{(2\pi)^2}\,
 e^{i\boldsymbol q_\perp\cdot\boldsymbol b}
 \mathcal K_3^{\mathrm R}(\boldsymbol q_\perp^2).
 \label{eq:angle_kernel_fourier}
\ee
The integration variable $\boldsymbol q_\perp$ lies in the transverse plane, and $\boldsymbol b$ is the conjugate impact parameter, with magnitude $b=J/E$. In this representation, the invariant transfer argument $Q^2$ of the kernel is replaced by $\boldsymbol q_\perp^2$. The factor $4mE$ is the invariant two-body flux in the static limit. The input to Eq.~\eqref{eq:angle_kernel_fourier} is the classical kernel after the iteration subtraction of Sec.~\ref{subsec:prescription_iteration_proof}.

We evaluate the logarithmic transform by analytic continuation in an auxiliary exponent $\alpha$. For transverse dimension $d_\perp$, the Fourier transform of a power is
\be
 \begin{aligned}
 F_\alpha(b)
 &\equiv\int\frac{d^{d_\perp}\boldsymbol q_\perp}{(2\pi)^{d_\perp}}\,
 e^{i\boldsymbol q_\perp\cdot\boldsymbol b}
 \left(\frac{\boldsymbol q_\perp^2}{\bar\mu^2}\right)^{-\alpha}\\
 &=\frac{\bar\mu^{2\alpha}}{\Gamma(\alpha)}
 \int_0^\infty ds\,s^{\alpha-1}(4\pi s)^{-d_\perp/2}
 e^{-b^2/(4s)}\\
 &=\frac{\bar\mu^{2\alpha}\Gamma(d_\perp/2-\alpha)}
 {4^\alpha\pi^{d_\perp/2}\Gamma(\alpha)}
 (b^2)^{\alpha-d_\perp/2}.
 \end{aligned}
 \label{eq:angle_fourier_power}
\ee
The second line follows by introducing a Schwinger parameter and carrying out the Gaussian momentum integral. In two transverse dimensions and at $b>0$, the expansion around $\alpha=0$ reads
\be
 F_\alpha(b)=\frac{\alpha}{\pi b^2}+\mathcal O(\alpha^2).
 \label{eq:angle_fourier_power_expansion}
\ee
Since $-\partial_\alpha(Q^2/\bar\mu^2)^{-\alpha}|_{\alpha=0}
=\log(Q^2/\bar\mu^2)$, we obtain
\be
 \int\frac{d^2\boldsymbol q_\perp}{(2\pi)^2}\,
 e^{i\boldsymbol q_\perp\cdot\boldsymbol b}
 \log\frac{\boldsymbol q_\perp^2}{\bar\mu^2}
 =-\frac1{\pi b^2},\qquad b>0.
 \label{eq:angle_fourier_logarithm}
\ee
The restriction to nonzero impact parameter excludes contact distributions. In particular, changing $\bar\mu$ adds a momentum-independent term, whose transform is proportional to $\delta^{(2)}(\boldsymbol b)$. The local pole and constant in Eq.~\eqref{eq:integration_dimensional_master} have the same support and do not enter the long-range angle.

Substituting Eq.~\eqref{eq:integration_kernel_log} into Eq.~\eqref{eq:angle_kernel_fourier}, we find
\be
 I_{r,3}^{\mathrm R}(E,b)
 =\frac{64}{3}\frac{G^3m^3E}{b^2}
 =\frac{64}{3}\frac{G^3m^3E^3}{J^2}.
 \label{eq:angle_third_order_action}
\ee
This action is dimensionless in natural units, as required for a scattering phase. Its coefficient is given by the sum of the sixteen contributions in Table~\ref{tab:classical_sixteen_coefficients}; the Fourier transform converts their common logarithmic dependence into the $1/b^2$ radial-action term.

\subsection{Deflection angle at third post-Minkowskian order}\label{subsec:angle_third_order_result}

Taking the derivative at fixed $E$ gives the third-order contribution
\be
 \chi_3^{\mathrm R}(b)
 =-\frac1E\frac{d I_{r,3}^{\mathrm R}}{db}
 =\frac{128}{3}\left(\frac{Gm}{b}\right)^3.
 \label{eq:angle_third_order_result}
\ee
The linear derivative relation is appropriate to the radial action defined above~\cite{Damgaard2023ClassicalObservables}.

For completeness, the long-range tree and one-loop kernels in the same static normalization are~\cite{BjerrumBohr2016Lightlike,Bastianelli2022LightBending}
\be
 \mathcal K_{1,\rm cl}(Q^2)=\frac{32\pi Gm^2E^2}{Q^2},
 \qquad
 \mathcal K_{2,\rm cl}(Q^2)=\frac{30\pi^2G^2m^3E^2}{Q},
 \label{eq:angle_lower_order_kernels}
\ee
where analytic terms and contributions outside the classical probe limit are omitted. These expressions refer to the interaction kernels, with lower-order iterations separated. Their transverse transforms follow from Eq.~\eqref{eq:angle_fourier_power}:
\be
 \begin{aligned}
 \int\frac{d^2\boldsymbol q_\perp}{(2\pi)^2}
 \frac{e^{i\boldsymbol q_\perp\cdot\boldsymbol b}}
 {\boldsymbol q_\perp^2}
 &=-\frac1{2\pi}\log\frac b{b_0},\\
 \int\frac{d^2\boldsymbol q_\perp}{(2\pi)^2}
 \frac{e^{i\boldsymbol q_\perp\cdot\boldsymbol b}}
 {|\boldsymbol q_\perp|}
 &=\frac1{2\pi b}.
 \end{aligned}
 \label{eq:angle_lower_order_fourier}
\ee
The first equality includes the $b$-independent infrared subtraction specified in Eq.~\eqref{eq:angle_interaction_action}. Combining these terms with Eq.~\eqref{eq:angle_third_order_action}, the radial action through third order is
\be
 I_r^{\rm int}(E,b)
 =-4GmE\log\frac b{b_0}
 +\frac{15\pi}{4}\frac{G^2m^2E}{b}
 +\frac{64}{3}\frac{G^3m^3E}{b^2}
 +\mathcal O\!\left(\frac{G^4m^4E}{b^3}\right).
 \label{eq:angle_action_through_3pm}
\ee
Therefore, in the conservative static limit,
\be
 \chi(b)
 =4\frac{Gm}{b}
 +\frac{15\pi}{4}\left(\frac{Gm}{b}\right)^2
 +\frac{128}{3}\left(\frac{Gm}{b}\right)^3
 +\mathcal O\!\left(\frac{G^4m^4}{b^4}\right).
 \label{eq:angle_total_through_3pm}
\ee
The coefficients are expressed in terms of the asymptotic impact parameter $b=J/E$, rather than the coordinate-dependent distance of closest approach. With this definition, the third-order term agrees with the weak-deflection expansion of Schwarzschild null geodesics~\cite{Iyer2007Bending}. 

Eq.~\eqref{eq:angle_total_through_3pm} applies for $E/m\ll1$, $Eb\gg1$, and $Gm/b\ll1$. The photon energy cancels between the action and the derivative in Eq.~\eqref{eq:angle_action_derivative}, giving an energy independent angle in this geometric-optics limit. The two-loop logarithm thus supplies the third-order nonlinear correction to classical light bending through its contribution to the radial action. 

This derivation establishes an explicit connection between the reconstructed two-loop integrand and the classical deflection angle, with repeated scattering separated before the observable is extracted. It provides a concrete basis for extending amplitude calculations of massless probes beyond the one-loop approximation.

\section{Conclusions}\label{sec:conclusions}

We have studied the conservative deflection of a photon by a massive spinless source using scattering amplitudes. In the heavy source and geometric optics limits, we extract the third-order long-range interaction from the two-loop classical kernel. The calculation connects on-shell tree amplitudes to the radial action and hence to the asymptotic deflection angle, with the energy, momentum-transfer, and source-mass expansions kept distinct.

The on-shell construction starts from the five-point amplitude for two photons and three gravitons. We obtain this amplitude by BCFW recursion and sew it to three scalar--graviton vertices across a five-fold cut. This cut exposes the three source vertices and two near-on-shell source propagators responsible for the required leading power of the source mass. Polynomial division then organizes the cut information into sixteen contributions: eight with seven propagators, seven with six, and one with five. The lower-propagator terms retain the contact interactions left by the higher cuts. Their numerators enter the classical integrals together with those of the seven-propagator terms.

For the classical integration, the source poles reduce the loop-energy integrations to residues, while the remaining spatial integrals retain the soft graviton exchanges and the photon poles. The expansion of the photon denominators is performed with their boundary values preserved. Within the static all-soft projection used here, we relate the chosen photon prescription to the subtraction of the cubic tree iteration in the Hermitian logarithm of the scattering operator. This prescription separates the repeated lower-order scattering contribution from the new third-order interaction. The latter still depends on the numerator-dependent integrals of all sixteen topologies.

After summing these contributions, the classical logarithmic kernel gives
\be
 \left.\mathcal K_3^{\mathrm R}(Q^2)\right|_{Q^0\log Q^2}
 =-\frac{256\pi}{3}G^3m^4E^2\log\frac{Q^2}{\bar\mu^2},
 \qquad
 \chi_3^{\mathrm R}(b)=\frac{128}{3}\left(\frac{Gm}{b}\right)^3 .
 \label{eq:conclusion_classical_result}
\ee
The angle follows from the transverse Fourier transform of the kernel and the derivative of the resulting interaction radial action at fixed energy. With $b=J/E$, it agrees with the third-order weak-deflection result for a Schwarzschild null geodesic. The cancellation of the photon energy in the angle is consistent with the geometric-optics limit.

This calculation shows explicitly how nonlinear classical light bending is encoded in loop amplitudes around flat spacetime. The loop expansion contains long-range classical contributions as well as quantum corrections; retaining the appropriate source poles, soft terms, and iteration subtraction makes the former accessible through an on-shell calculation. The radial action expresses the final comparison with the geometric description in terms of the same asymptotic observables.

The present result assumes $E/m\ll1$, $Eb\gg1$, and $Gm/b\ll1$. The exact on-shell kinematics and the source-pole decomposition used here provide a starting point for recoil corrections: they identify the energy transfers, pole shifts, and residue corrections omitted in the static approximation. Retaining these terms consistently would extend the calculation from fixed-source light bending to scattering with finite recoil, for which the present result serves as the static-limit reference. Source spin can likewise be incorporated through the corresponding spinning-source tree amplitudes, so that spin interactions enter the same sewing and reduction steps. For higher PM orders, the reconstruction of contact residues and the joint treatment of coupled photon-pole regions provide methods that can be adapted to more complicated loop integrands, while the logarithm of the scattering operator supplies an organizing principle for separating lower-order iterations from  new classical interactions. The two-loop calculation demonstrates how these ingredients can be combined to obtain a classical observable and offers a concrete basis for extending the amplitude approach to higher orders. Each extension requires the relevant cuts, momentum regions, and iteration terms to be determined at the new order.

\begin{acknowledgments}
We thank Li Lin Yang, Jian Wang, and Hai Tao Li for helpful discussions.
This work is supported by the National Natural Science Foundation of China (NNSFC)
Grant No.12475111 and the Fundamental Research Funds for the Central Universities, Sun
Yat-sen University.

\end{acknowledgments}

\appendix

\section{Five-point tree amplitude from double copy}
\label{app:five_point_double_copy}

The five-point photon--graviton amplitude can also be constructed from gauge-theory trees through the field-theory KLT relations~\cite{Kawai1986KLT}. We give this representation with the same external labels and phases as in Sec.~\ref{subsec:tree_bcfw}: legs $1,2,3$ are gravitons, while legs $4,5$ are photons with helicities $(-,+)$. The color-ordered gauge-theory functions below have their couplings and common phases removed. The gravitational amplitude is restored as $\mathcal A_5^{\mathrm{tree}}=-ig^3H_5$.

\subsection{Gauge-theory input and external states}

The external states are tensor products of gauge-theory states~\cite{Bern2010DoubleCopy,Cachazo2015Matrices},
\be
 \begin{aligned}
 |h^\pm\rangle&=|g^\pm\rangle\otimes|\widetilde g^\pm\rangle,\\
 |\gamma^-\rangle&=|g^-\rangle\otimes|\widetilde\phi\rangle,
 &|\gamma^+\rangle&=|g^+\rangle\otimes|\widetilde{\bar\phi}\rangle .
 \end{aligned}
 \label{eq:app_dc_states}
\ee
Here $g^\pm$ denotes a gluon state, and a tilde distinguishes the second gauge theory. The massless adjoint scalars $\phi$ and $\bar\phi$ are a conjugate pair in that theory; they are distinct from the massive source. The helicities add in the tensor product, giving helicity $\pm2$ for the graviton and $\pm1$ for the photon. This construction provides the tree amplitudes with two external photons used here; the graviton state sum on the five-fold cut remains the sum over helicities $\pm2$.

For completeness, the massive scalar--gluon three-point input is
\be
 \begin{aligned}
 a_3(1_m,2_m,3^+)
 &=\frac{\langle\zeta|K_1|3]}{\langle\zeta3\rangle},&
 a_3(1_m,2_m,3^-)
 &=\frac{\langle3|K_1|\zeta]}{[3\zeta]},\\
 A_3^{\mathrm{tree}}(1_m,2_m,3^\pm)&=-i\,a_3(1_m,2_m,3^\pm).
 \end{aligned}
 \label{eq:app_dc_scalar_gluon}
\ee
The last line retains the Feynman phase but omits the color factor and gauge coupling. All momenta are outgoing, with $K_1^2=K_2^2=m^2$, $K_3^2=0$, and $K_1+K_2+K_3=0$. The reference spinor $\zeta$ is chosen so that the indicated denominator is nonzero. The sandwich contains the massive momentum matrix $K_1$; no massless spinor decomposition of this momentum is assumed. Equivalently, $a_3=\sqrt2 K_1\cdot\varepsilon^\pm(K_3;\zeta)$. Since $K_1\cdot K_3=0$, the amplitude is invariant under a change of reference spinor. Taking the product of the two gauge-theory factors gives
\be
 H_3(1_m,2_m,3_h^\pm)
 =a_3(1_m,2_m,3^\pm)\,\widetilde a_3(1_m,2_m,3^\pm)
 =2\bigl[K_1\cdot\varepsilon^\pm(K_3)\bigr]^2,
 \label{eq:app_dc_scalar_graviton}
\ee
which fixes the normalization relative to Eq.~\eqref{eq:scalar_graviton_three}.

For the five-point component with one negative-helicity graviton at $j\in\{1,2,3\}$, the first gauge factor is a five-gluon maximally helicity-violating (MHV) amplitude, with two negative-helicity gluons~\cite{Parke1986Gluon}. The second factor contains the scalar pair at legs $4,5$ and three gluons~\cite{Bianchi2008Generating}. With the other two graviton labels carrying positive gluon helicity in both factors, these functions are
\be
 \begin{aligned}
 a_5^{\mathrm{YM}}(j^-,4^-,5^+,\ldots;\sigma)
 &=\frac{\langle4j\rangle^4}{P_\angle(\sigma)},\\
 a_5^{\phi\bar\phi}(j^-,4_\phi,5_{\bar\phi},\ldots;\sigma)
 &=\frac{\langle4j\rangle^2\langle5j\rangle^2}{P_\angle(\sigma)},\\
 P_\angle(\sigma)
 &=\prod_{r=1}^{5}\langle\sigma_r\sigma_{r+1}\rangle,
 \qquad \sigma_6=\sigma_1 .
 \end{aligned}
 \label{eq:app_dc_gauge_factors}
\ee
The argument $\sigma$ specifies the cyclic color order, while the state at each label is fixed. The numerator of the scalar amplitude has the same degree in the spinors of legs $4,5$ as its denominator, giving zero helicity weight on those legs. Multiplication by the gluon amplitude then supplies the photon helicity weights.

\subsection{KLT product and comparison with recursion}

In these conventions, the five-point KLT relation is
\be
 \begin{aligned}
 H_5={}&s_{12}s_{34}\,
 a_5^{\mathrm{YM}}(1,2,3,4,5)
 a_5^{\phi\bar\phi}(2,1,4,3,5)\\
 &+s_{13}s_{24}\,
 a_5^{\mathrm{YM}}(1,3,2,4,5)
 a_5^{\phi\bar\phi}(3,1,4,2,5).
 \end{aligned}
 \label{eq:app_dc_klt_product}
\ee
Only the color orders are displayed in this equation. In particular, legs $4,5$ remain the scalar pair in the second factor even when they occupy different positions in the ordering. Substituting Eq.~\eqref{eq:app_dc_gauge_factors} gives
\be
 \begin{aligned}
 H_5(j_h^-,\{i_h^+\}_{i\ne j},4_\gamma^-,5_\gamma^+)
 &=\langle4j\rangle^6\langle5j\rangle^2\,\mathcal K_5^\angle,\\
 \mathcal K_5^\angle
 &=\frac{s_{12}s_{34}}
 {P_\angle(1,2,3,4,5)P_\angle(2,1,4,3,5)}\\
 &\quad+\frac{s_{13}s_{24}}
 {P_\angle(1,3,2,4,5)P_\angle(3,1,4,2,5)},
 \end{aligned}
 \label{eq:app_dc_mhv_kernel}
\ee
where $i,j\in\{1,2,3\}$. Define $P_\square$ and $\mathcal K_5^\square$ by replacing every angle bracket in these expressions with a square bracket; the invariants $s_{ij}$ are unchanged. For the component $(1_h^+,2_h^-,3_h^-,4_\gamma^-,5_\gamma^+)$, the parity-conjugate gauge-theory factors have numerators $[15]^4$ and $[14]^2[15]^2$, respectively, and denominator $P_\square(\sigma)$. Their product therefore gives the common numerator $[15]^6[14]^2$.

The two independent components used in the main text are therefore
\be
 \begin{aligned}
 H_5(1_h^+,2_h^-,3_h^-,4_\gamma^-,5_\gamma^+)
 &=[51]^6[41]^2\,\mathcal K_5^\square=-U_5,\\
 H_5(1_h^+,2_h^+,3_h^-,4_\gamma^-,5_\gamma^+)
 &=\langle43\rangle^6\langle53\rangle^2\,\mathcal K_5^\angle=-V_5.
 \end{aligned}
 \label{eq:app_dc_bcfw_equivalence}
\ee
The equalities follow by substituting the cyclic products and using momentum conservation, $\sum_i\langle ai\rangle[ib]=0$, together with the Schouten identity. Restoring the common factor $-ig^3$ gives $\mathcal A_5=ig^3U_5$ and $\mathcal A_5=ig^3V_5$, in agreement with Eqs.~\eqref{eq:five_point_u_normalization} and \eqref{eq:five_point_v_normalization}. The remaining nonzero components follow by permuting the graviton labels. Thus the KLT construction supplies the same tree-level input as the factorization channels in Sec.~\ref{subsec:tree_bcfw}.

\section{Details of the seven-propagator reconstruction}
\label{app:seven_reconstruction}

We give the complete residue for the seven-propagator example of Sec.~\ref{subsec:seven_propagator_example}. The monomial basis and the six cut branches are defined in Eqs.~\eqref{eq:seven_basis_64} and \eqref{eq:seven_cut_branches}, respectively. The coefficients below use the normalized tree data of Eq.~\eqref{eq:seven_cut_tree_values} and the phase convention
\be
 \Delta_{7,1}=\sum_{M\in\mathcal B_7}c_M M=-N_{7,1}.
 \label{eq:appendix_seven_residue}
\ee
In particular, the signs refer to $\Delta_{7,1}$; the coefficients of $N_{7,1}$ have the opposite signs. The invariants remain $\rho_s=s-m^2$ and $\mathcal S=\rho_s^2+st$.

Tables~\ref{tab:appendix_seven_even} and \ref{tab:appendix_seven_odd} list the sixteen nonzero coefficients. Every other coefficient in $\mathcal B_7$ vanishes. The first table contains the eight terms even under $(X_4,Y_4)\mapsto(-X_4,-Y_4)$ and is equivalent to minus Eq.~\eqref{eq:seven_numerator_even}. The second table supplies the transverse odd terms required by the separate cut branches.

\begin{table}[htbp]
 \centering
 \renewcommand{\arraystretch}{1.35}
 \begin{tabular}{c|l}
 \hline
 $M$ & $c_M$ in $\Delta_{7,1}$\\
 \hline
 $1$ & $\rho_s^2\mathcal S(\rho_s^4-\tfrac32s^2t^2)$\\
 $X_3$ & $4\rho_s^2s^2t\mathcal S$\\
 $Y_2$ & $-4\rho_s^2s^2t\mathcal S$\\
 $X_3Y_2$ & $2\rho_s^2s^2(4\rho_s^2+5st)$\\
 $X_3^2$ & $-2\rho_s^2s^3t$\\
 $Y_2^2$ & $-2\rho_s^2s^3t$\\
 $X_3^2Y_2$ & $-4\rho_s^2s^3$\\
 $X_3Y_2^2$ & $4\rho_s^2s^3$\\
 \hline
 \end{tabular}
 \caption{Nonzero transverse-even coefficients of the seven-propagator residue.}
 \label{tab:appendix_seven_even}
\end{table}

\begin{table}[htbp]
 \centering
 \small
 \renewcommand{\arraystretch}{1.9}
 \begin{tabular}{c|l}
 \hline
 $M$ & $c_M$ in $\Delta_{7,1}$\\
 \hline
 $X_4$ & $2\rho_s s^3t\mathcal S$\\
 $X_3X_4$ & $-\dfrac{2\rho_s s^2}{t}(2\rho_s^4+3\rho_s^2st+2s^2t^2)$\\
 $Y_4$ & $\dfrac{2s^3t\mathcal S}{\rho_s}(3\rho_s^2+2st)$\\
 $Y_2Y_4$ & $\dfrac{2s^2}{\rho_s t}(2\rho_s^6+7\rho_s^4st+10\rho_s^2s^2t^2+4s^3t^3)$\\
 $X_3Y_4$ & $-\dfrac{2s^2\mathcal S}{\rho_s t}(2\rho_s^4+9\rho_s^2st+8s^2t^2)$\\
 $X_3Y_2Y_4$ & $-\dfrac{4s^2\mathcal S}{\rho_s t^2}(2\rho_s^4+7\rho_s^2st+8s^2t^2)$\\
 $X_3^2Y_4$ & $\dfrac{4s^2\mathcal S}{\rho_s t^2}(2\rho_s^4+3\rho_s^2st+4s^2t^2)$\\
 $X_3^2Y_2Y_4$ & $\dfrac{8s^3}{\rho_s t^2}(2\rho_s^4+5\rho_s^2st+4s^2t^2)$\\
 \hline
 \end{tabular}
 \caption{Nonzero transverse-odd coefficients of the seven-propagator residue.}
 \label{tab:appendix_seven_odd}
\end{table}

The complete polynomial obeys the six rational identities
\be
 \left.\Delta_{7,1}\right|_{\mathcal V_\sigma^{(j)}(\zeta)}
 -F_\sigma^{(j)}(\zeta)=0,
 \qquad j=1,2,3,\qquad \sigma=\pm1,
 \label{eq:appendix_seven_branch_check}
\ee
at generic kinematics and away from singular parameter values. These are identities in $\zeta$, rather than matches at selected sample points. In addition, polynomial division by the Gr\"obner basis of the three quadratic constraints in Eq.~\eqref{eq:seven_cut_quadratics}, with the ordering used in Eq.~\eqref{eq:seven_cut_leading_monomials}, leaves this residue unchanged. Thus the coefficients are given in the same normal form as the basis used for the reconstruction.

Writing the full residue as $\Delta_{7,1}^{\mathrm{even}}+\Delta_{7,1}^{\mathrm{odd}}$, the simultaneous reflection exchanges the two signs on each branch. Consequently,
\be
 \left.\Delta_{7,1}^{\mathrm{even}}\right|_{\mathcal V_\sigma^{(j)}}
 =\frac{F_+^{(j)}+F_-^{(j)}}2,
 \qquad
 \left.\Delta_{7,1}^{\mathrm{odd}}\right|_{\mathcal V_\sigma^{(j)}}
 =\frac{\sigma}{2}\bigl(F_+^{(j)}-F_-^{(j)}\bigr).
 \label{eq:appendix_seven_parity_check}
\ee
The odd polynomial integrates to zero with the reflection-invariant denominators and measure, but omitting it before matching would replace each branch value by the average of its pair. For this reason, the eight-term expression used for integration and the sixteen-term residue used for cut reconstruction are kept distinct.

\clearpage

\section{Spatial integral identities}\label{app:spatial_integrals}

We give the parameter integrals underlying Eqs.~\eqref{eq:integration_bubble}, \eqref{eq:integration_bubble_moments}, and \eqref{eq:integration_outer_moments}. We use $d=3-2\epsilon$ and the spatial measure $d^d\mathbf L/(2\pi)^d$. The variables are dimensionless as in Eq.~\eqref{eq:integration_spatial_momenta}; the dimensional scale is restored explicitly below. All formulas are first evaluated in a convergence domain and then continued analytically.

\subsection{Scalar and tensor bubbles}

Let $\nu=\alpha+\beta$. Feynman parametrization and the shift $\mathbf k=\mathbf L+(1-x)\mathbf P$ give
\be
 \begin{aligned}
 &\frac1{(\mathbf L^2)^\alpha[(\mathbf L+\mathbf P)^2]^\beta}\\
 &\qquad=\frac{\Gamma(\nu)}{\Gamma(\alpha)\Gamma(\beta)}
 \int_0^1 dx\,
 \frac{x^{\alpha-1}(1-x)^{\beta-1}}
 {[\mathbf k^2+\Delta_x]^\nu}.
 \end{aligned}
 \label{eq:app_bubble_parameter}
\ee
Here $P^2=\mathbf P^2$ and $\Delta_x=x(1-x)P^2$. Introducing a positive Schwinger parameter for the last denominator and performing the Gaussian integral yields
\be
 \int\frac{d^d\mathbf k}{(2\pi)^d}
 \frac1{(\mathbf k^2+\Delta)^\nu}
 =\frac{\Gamma(\nu-d/2)}
 {(4\pi)^{d/2}\Gamma(\nu)}
 \Delta^{d/2-\nu}.
 \label{eq:app_scalar_gaussian}
\ee
The remaining parameter integral is therefore
\be
 B_{\alpha\beta}(P^2)
 =\frac{\Gamma(\nu-d/2)(P^2)^{d/2-\nu}}
 {(4\pi)^{d/2}\Gamma(\alpha)\Gamma(\beta)}
 \int_0^1 dx\,
 x^{d/2-\beta-1}(1-x)^{d/2-\alpha-1}.
 \label{eq:app_bubble_beta}
\ee
Euler's beta integral gives Eq.~\eqref{eq:integration_bubble}.

The odd moments of the shifted momentum vanish. For the second moment, rotational symmetry and the same Schwinger integral give
\be
 \int\frac{d^d\mathbf k}{(2\pi)^d}
 \frac{k^ik^j}{(\mathbf k^2+\Delta)^\nu}
 =\frac{\delta^{ij}}{2(4\pi)^{d/2}}
 \frac{\Gamma(\nu-d/2-1)}{\Gamma(\nu)}
 \Delta^{d/2+1-\nu}.
 \label{eq:app_tensor_gaussian}
\ee
For $\alpha=\beta=1$, the normalized parameter weight in Eq.~\eqref{eq:app_bubble_beta} is proportional to $[x(1-x)]^{d/2-2}$. Its moments satisfy
\be
 \langle1-x\rangle_x=\frac12,\qquad
 \langle(1-x)^2\rangle_x=\frac{d}{4(d-1)},\qquad
 \langle x(1-x)\rangle_x=\frac{d-2}{4(d-1)}.
 \label{eq:app_parameter_moments}
\ee
Substituting $\mathbf L=\mathbf k-(1-x)\mathbf P$ then gives
\be
 \begin{aligned}
 \frac1{B_{11}}\int\frac{d^d\mathbf L}{(2\pi)^d}
 \frac{L^i}{\mathbf L^2(\mathbf L+\mathbf P)^2}
 &=-\frac{P^i}{2},\\
 \frac1{B_{11}}\int\frac{d^d\mathbf L}{(2\pi)^d}
 \frac{L^iL^j}{\mathbf L^2(\mathbf L+\mathbf P)^2}
 &=\frac{dP^iP^j-P^2\delta^{ij}}{4(d-1)}.
 \end{aligned}
 \label{eq:app_bubble_rank_one_two}
\ee
In particular, the term proportional to $\delta^{ij}$ originates from the shifted Gaussian moment and cannot be recovered by replacing $\mathbf L$ by its mean. These identities retain the $d$ dependence needed when tensor coefficients multiply poles of an outer integral.

\subsection{The three-line master and longitudinal moments}

Writing $B_{11}(P^2)=\mathcal B_d(P^2)^{d/2-2}$, with $\mathcal B_d$ defined in Eq.~\eqref{eq:integration_six_four_bubble}, the first integration in $J_2$ gives
\be
 J_2(\rho)=\mathcal B_d
 B_{1,\,2-d/2}(\rho),\qquad \rho=\mathbf Q^2.
 \label{eq:app_master_as_bubbles}
\ee
The factors $\Gamma(2-d/2)$ and $\Gamma(d-2)$ cancel between the two bubbles, leaving Eq.~\eqref{eq:integration_master}. To recover the measure of Eq.~\eqref{eq:loop_measure}, multiply the dimensionless result by
$[e^{\gamma_E}\bar\mu^2/(4\pi)]^{2\epsilon}(m\xi)^{-4\epsilon}$.
Since $Q^2=(m\xi)^2\rho$, this gives
\be
 J_2^{\overline{\mathrm{MS}}}(Q^2)
 =\frac{e^{2\gamma_E\epsilon}}{(4\pi)^3}
 \left(\frac{Q^2}{\bar\mu^2}\right)^{-2\epsilon}
 \frac{\Gamma(2\epsilon)\Gamma(1/2-\epsilon)^3}
 {\Gamma(3/2-3\epsilon)}.
 \label{eq:app_master_exact_epsilon}
\ee
The useful expansions are
\be
 e^{2\gamma_E\epsilon}\Gamma(2\epsilon)
 =\frac1{2\epsilon}+\mathcal O(\epsilon),\qquad
 \frac{\Gamma(1/2-\epsilon)^3}{\Gamma(3/2-3\epsilon)}
 =2\pi[1+6\epsilon+\mathcal O(\epsilon^2)].
 \label{eq:app_master_gamma_expansion}
\ee
Hence
\be
 J_2^{\overline{\mathrm{MS}}}(Q^2)
 =\frac1{64\pi^2}
 \left[\frac1\epsilon+6
 -2\log\frac{Q^2}{\bar\mu^2}\right]
 +\mathcal O(\epsilon).
 \label{eq:app_master_finite_constant}
\ee
The constant $6/(64\pi^2)$ is local in impact-parameter space. Its value fixes $c_{\rm loc}$ in Eq.~\eqref{eq:integration_dimensional_master} for the stated measure, without affecting the long-range logarithm.

For the remaining longitudinal moments, align one integration axis with $\mathbf W$. The one-dimensional Gaussian moment, followed by a Schwinger integral, gives
\be
 \begin{aligned}
 &\int\frac{d^d\mathbf k}{(2\pi)^d}
 \frac{|\mathbf W\cdot\mathbf k|^{2n}}
 {(\mathbf k^2+\Delta)^\nu}\\
 &\qquad=
 \frac{(W^2)^n}{(4\pi)^{d/2}}
 \frac{\Gamma(n+1/2)}{\sqrt\pi}
 \frac{\Gamma(\nu-d/2-n)}{\Gamma(\nu)}
 \Delta^{d/2+n-\nu}.
 \end{aligned}
 \label{eq:app_longitudinal_gaussian}
\ee
Before analytic continuation, this identity requires $\operatorname{Re}n>-1/2$ and sufficient falloff at large $\mathbf k$. For nonnegative integer $n$, the absolute value gives the ordinary even moment. In $\mathcal M_{\alpha\beta n}$, the shift along $\mathbf Q$ does not change this factor because $\mathbf W\cdot\mathbf Q=0$. Applying Eq.~\eqref{eq:app_longitudinal_gaussian} after combining the denominators gives
\be
 \begin{aligned}
 \mathcal M_{\alpha\beta n}
 ={}&\frac{(W^2)^n\rho^{d/2+n-\nu}}
 {(4\pi)^{d/2}\Gamma(\alpha)\Gamma(\beta)}
 \frac{\Gamma(n+1/2)}{\sqrt\pi}
 \Gamma(\nu-d/2-n)\\
 &\times\int_0^1 dx\,
 x^{d/2+n-\beta-1}(1-x)^{d/2+n-\alpha-1}.
 \end{aligned}
 \label{eq:app_longitudinal_beta}
\ee
Evaluating the beta integral reproduces Eq.~\eqref{eq:integration_outer_moments}. Continuing $n$ to $-k$, with $k=1,2,\ldots$, defines the even inverse powers as homogeneous finite-part distributions,
\be
 \operatorname{FP}\frac1{u^{2k}}
 =-\frac1{(2k-1)!}
 \frac{d^{\,2k-1}}{du^{2k-1}}
 \operatorname{PV}\frac1u,\qquad
 u=\mathbf W\cdot\mathbf K .
 \label{eq:app_finite_part_derivative}
\ee
This continuation supplies the real finite-part terms. The delta-function derivatives are fixed separately by Eq.~\eqref{eq:integration_photon_distribution}; for two photon poles, their products are retained through the joint parameter integral in Eq.~\eqref{eq:integration_joint_parameters}.

\clearpage

\section{Mixed photon--graviton and contact contributions}\label{app:mixed_contact_corrections}

We establish the vanishing of the mixed spectral and contact corrections entering the iteration subtraction in Sec.~\ref{subsec:prescription_iteration_proof}. We retain the same static and all-soft expansion and use dimensional regularization throughout. The bubble identities required below are given in Sec.~\ref{subsec:classical_spatial_integrals} and derived in Appendix~\ref{app:spatial_integrals}. All loop momenta in this appendix are dimensional.

\subsection{Mixed photon--graviton correction}\label{app:mixed_spectral_correction}

The topology $N_{7,6}$ has unchanged boundary values for both $D_D$ and $D_H$, so it gives no prescription correction. For $N_{7,7}$, the sign of $D_C$ is reversed while that of $D_H$ is unchanged. Its complete mixed photon--graviton correction is
\be
 \begin{aligned}
 \Delta_{CH}
 ={}&\frac{w_Eg^6}{2(2m)^2}\Pi_{\rm cl}
 \int\frac{d^d\mathbf c\,d^d\mathbf r}{(2\pi)^{2d}}
 \\
 &\times\left.
 \frac{(2\pi)^2N_{7,7}\,
 \delta(-\mathbf c^2)\delta(-\mathbf c^2+2\mathbf c\cdot\mathbf k')}
 {S_{312}^{\,2}\mathbf r^2(\mathbf r+\mathbf c)^2
  (\mathbf c-\boldsymbol q_c)^2}
 \right|_{\substack{\boldsymbol\ell_5=\mathbf r+\mathbf c,\,
                       \boldsymbol\ell_7=\mathbf r\\
                       \ell_5^0=\ell_7^0=0}} .
 \end{aligned}
 \label{eq:prescription_complete_mixed_integral}
\ee
The normalization is the same as in Eq.~\eqref{eq:prescription_complete_photon_integral}, including the suppressed dimensional-regularization factor. The variables are $\mathbf c=\boldsymbol\ell_5-\boldsymbol\ell_7$ and $\mathbf r=\boldsymbol\ell_7$. The three denominators displayed explicitly are the common exchanged-graviton propagators. The extra $D_H$ propagator has been replaced by its delta function, so no further factor $1/\mathbf c^2$ remains.

We choose the longitudinal direction along the outgoing photon, $\widehat{\mathbf n}'=\mathbf k'/E$, and write $c_\parallel=\mathbf c\cdot\widehat{\mathbf n}'$. The two shell conditions are
\be
 D_H=-\mathbf c^2,\qquad
 D_C=-\mathbf c^2+2Ec_\parallel,\qquad
 \delta(D_H)\delta(D_C)
 =\frac1{2E}\delta(c_\parallel)\delta(\mathbf c_\perp^2).
 \label{eq:prescription_mixed_support}
\ee
Their support contracts to $\mathbf c=\mathbf0$, where the gradient of the shell function $\mathbf c^2$ also vanishes. Substituting this value into the integrand before the $\mathbf r$ integration would not determine whether a finite contribution survives: the two denominators $\mathbf r^2$ and $(\mathbf r+\mathbf c)^2$ become coincident at the same point. We therefore evaluate the integral at a nonzero shell radius and take the zero-radius limit only after the spectator integration.

The numerator must be retained in this limit. Evaluating the reconstructed $N_{7,7}$ at $\ell_5^0=\ell_7^0=0$ and retaining the leading source order gives
\be
 \left.N_{7,7}\right|_{\rm leading\ source,\ static}
 =-4m^8(32E^4+3Q^4)(\mathbf r\cdot\mathbf k)^2.
 \label{eq:prescription_mixed_numerator}
\ee
This follows from its leading probe polynomial
$-m^{16}\xi^8(8+12\tau^2)(\bar X_1-\bar Y_1-\bar Y_2)^2$,
where $\tau=t/(m^2\xi^2)$ and $\bar X_i=X_i/(m^2\xi^2)$, $\bar Y_i=Y_i/(m^2\xi^2)$. In the source rest frame, the conditions $\ell_5^0=\ell_7^0=0$ give $X_1=Y_1=0$ and $Y_2=\mathbf r\cdot\mathbf k$. In particular, the numerator involves the incoming momentum $\mathbf k$, whereas the photon shell in Eq.~\eqref{eq:prescription_mixed_support} is aligned with $\mathbf k'$.

We introduce a positive auxiliary momentum scale $a$ and replace only the graviton spectral constraint by $\delta(a^2-\mathbf c^2)$, keeping the photon constraint unchanged. The parameter $a$ specifies the radius at which the distribution is sampled; it is removed at the end and is not a physical graviton mass. After substituting Eq.~\eqref{eq:prescription_mixed_numerator}, the momentum integral to be evaluated is
\be
 \begin{aligned}
 I_{77}(a)
 ={}&\int\frac{d^d\mathbf c\,d^d\mathbf r}{(2\pi)^{2d}}
 \frac{(2\pi)^2(\mathbf r\cdot\mathbf k)^2}
 {\mathbf r^2(\mathbf r+\mathbf c)^2
  (\mathbf c-\boldsymbol q_c)^2}
 \\
 &\times\delta(a^2-\mathbf c^2)
 \delta(-\mathbf c^2+2Ec_\parallel),\\
 \Delta_{CH}
 ={}&-\frac{w_Eg^6}{2(2m)^2}\Pi_{\rm cl}
 \left[\frac{4m^8(32E^4+3Q^4)}{S_{312}^{\,2}}
 \lim_{a\to0^+}I_{77}(a)\right].
 \end{aligned}
 \label{eq:prescription_regulated_mixed_integral}
\ee
We take this limit at fixed nonzero $Q$ and fixed $E$, after defining the loop integrations dimensionally. For $0<a<2E$, the two delta functions impose
\be
 \mathbf c^2=a^2,\qquad
 c_\parallel=\frac{a^2}{2E},\qquad
 R_a^2\equiv\mathbf c_\perp^2=a^2-\frac{a^4}{4E^2}.
 \label{eq:prescription_mixed_regulator}
\ee
Thus $R_a$ is the radius of the remaining transverse sphere. Keeping this radius finite makes both the shell Jacobian and the behavior of the spectator integral explicit.

The $\mathbf r$ integration in Eq.~\eqref{eq:prescription_regulated_mixed_integral} is a tensor bubble. Its dimensionally continued value is
\be
 \int\frac{d^d\mathbf r}{(2\pi)^d}
 \frac{(\mathbf r\cdot\mathbf k)^2}
 {\mathbf r^2(\mathbf r+\mathbf c)^2}
 =B_{11}(\mathbf c^2)
 \frac{d(\mathbf c\cdot\mathbf k)^2-E^2\mathbf c^2}{4(d-1)}.
 \label{eq:prescription_mixed_bubble}
\ee
Here $B_{11}(\mathbf c^2)$ is the scalar bubble defined in Eq.~\eqref{eq:integration_bubble}; it scales as $(\mathbf c^2)^{d/2-2}$. Substituting this result back into the same integral gives
\be
 \begin{aligned}
 I_{77}(a)
 ={}&\frac{(2\pi)^2}{4(d-1)}
 \int\frac{d^d\mathbf c}{(2\pi)^d}
 \frac{B_{11}(\mathbf c^2)
 [d(\mathbf c\cdot\mathbf k)^2-E^2\mathbf c^2]}
 {(\mathbf c-\boldsymbol q_c)^2}
 \\
 &\times\delta(a^2-\mathbf c^2)
 \delta(-\mathbf c^2+2Ec_\parallel).
 \end{aligned}
 \label{eq:prescription_mixed_after_bubble}
\ee
Using the graviton constraint, the photon delta function becomes $\delta(2Ec_\parallel-a^2)$ and supplies $1/(2E)$ when integrating $c_\parallel$. The transverse radial integration supplies $R_a^{d-3}/2$. Writing
$\mathbf c_a=(a^2/2E)\widehat{\mathbf n}'+R_a\widehat{\mathbf e}_\perp$,
where $\widehat{\mathbf e}_\perp$ is a unit vector orthogonal to $\mathbf k'$, leaves the complete angular integral
\be
 I_{77}(a)=
 \frac{(2\pi)^2B_{11}(a^2)R_a^{d-3}}
 {16E(d-1)(2\pi)^d}
 \int d\Omega_{d-2}\,
 \frac{d(\mathbf c_a\cdot\mathbf k)^2-E^2a^2}
 {|\mathbf c_a-\boldsymbol q_c|^2}.
 \label{eq:prescription_mixed_angular_integral}
\ee
For $a\ll Q,E$, the denominator approaches $Q^2$ uniformly over the angular integration, while $|\mathbf c_a\cdot\mathbf k|\leq aE$. The angular numerator is therefore of order $a^2$. Since $B_{11}(a^2)\sim a^{d-4}$ and $R_a^{d-3}\sim a^{d-3}$, Eq.~\eqref{eq:prescription_mixed_angular_integral} yields
\be
 I_{77}(a)=\mathcal O\!\left(a^{d-4}a^{d-3}a^2\right)
 =\mathcal O\!\left(a^{2d-5}\right)
 \longrightarrow0,
 \qquad \operatorname{Re}d>\frac52.
 \label{eq:prescription_mixed_vanishing}
\ee
Since $d=3-2\epsilon$ in this work, we have $\Delta_{CH}=0$ at the retained source order. The result follows from the complete regulated integral, including its numerator, rather than from setting $\delta(\mathbf c_\perp^2)$ to zero. It concerns the prescription correction; the original $N_{7,7}$ topology still contributes to the classical kernel.

\subsection{Contact contributions}\label{app:contact_corrections}

The five- and six-propagator terms contain at most one extra photon pole. A change of its boundary value is purely imaginary in the retained real kernel and hence has no Hermitian contribution. The remaining contact contribution in Eq.~\eqref{eq:prescription_combinatorial_factor} is
\be
 \begin{aligned}
 \Delta_{\rm ct}
 ={}&\frac{1}{3!\,2(2m)^2}\Pi_{\rm cl}
 \sum_{\nu=1,8}
 \int\frac{d^d\mathbf K_1\,d^d\mathbf K_2}{(2\pi)^{2d}}
 \\
 &\times(2\pi)^2
 \delta(E^2-\mathbf K_1^2)\delta(E^2-\mathbf K_2^2)
 \mathcal R_{\rm ct}^{(\nu)}(\mathbf K_1,\mathbf K_2).
 \end{aligned}
 \label{eq:prescription_complete_contact_integral}
\ee
The remainder $\mathcal R_{\rm ct}^{(\nu)}$ already includes the coupling factors and polarization contractions of Eq.~\eqref{eq:prescription_sewn_numerator}. We introduce the three spatial transfers
\[
 \mathbf t_1=\mathbf K_1-\mathbf k,\qquad
 \mathbf t_2=\mathbf K_2-\mathbf K_1,\qquad
 \mathbf t_3=\mathbf k'-\mathbf K_2,
 \qquad \sum_{j=1}^3\mathbf t_j=\boldsymbol q_c.
\]
A local numerator term proportional to a common graviton denominator cancels that exchange line. Its contribution to the remainder thus has the form
\be
 \mathcal R_{\rm ct}^{(\nu)}
 =\sum_{j=1}^{3}
 \frac{\mathcal P_j^{(\nu)}(\mathbf t_1,\mathbf t_2,\mathbf t_3)}
 {\displaystyle\prod_{i\ne j}\mathbf t_i^2}.
 \label{eq:prescription_contact_decomposition}
\ee
The $\mathcal P_j^{(\nu)}$ are polynomial quotients of the local numerators; their coefficients can depend on the external kinematics. Terms canceling more than one exchange line are included by factors of the remaining denominators in these polynomials. The following argument applies to each polynomial coefficient, so their explicit values are not required to establish the vanishing.

For the term with the middle exchange canceled, set $\mathbf u=\mathbf t_1$ and $\mathbf v=\mathbf t_3$. Its full contribution to the integral in Eq.~\eqref{eq:prescription_complete_contact_integral}, before the soft expansion, is
\be
 \begin{aligned}
 \mathcal J_2^{(\nu)}
 ={}&\int\frac{d^d\mathbf u\,d^d\mathbf v}{(2\pi)^{2d}}
 \frac{(2\pi)^2
 \mathcal P_2^{(\nu)}
 (\mathbf u,\boldsymbol q_c-\mathbf u-\mathbf v,\mathbf v)}
 {\mathbf u^2\mathbf v^2}
 \\
 &\times\delta\!\left(E^2-|\mathbf k+\mathbf u|^2\right)
 \delta\!\left(E^2-|\mathbf k'-\mathbf v|^2\right).
 \end{aligned}
 \label{eq:prescription_middle_contact_integral}
\ee
We resolve $\mathbf u$ along $\widehat{\mathbf n}=\mathbf k/E$ and $\mathbf v$ along $\widehat{\mathbf n}'=\mathbf k'/E$. These independent rotations leave both integration measures unchanged. The two delta-function arguments and their forward roots are
\be
 \begin{aligned}
 s_u&=-2Ez_u-\mathbf u_\perp^2-z_u^2,
 &z_u^*&=-E+\sqrt{E^2-\mathbf u_\perp^2},\\
 s_v&= 2Ez_v-\mathbf v_\perp^2-z_v^2,
 &z_v^*&= E-\sqrt{E^2-\mathbf v_\perp^2}.
 \end{aligned}
 \label{eq:prescription_contact_shell_roots}
\ee
The other roots require hard momentum transfers and lie outside the all-soft region. Let $\mathbf u_* =\mathbf u_\perp+z_u^*\widehat{\mathbf n}$ and $\mathbf v_* =\mathbf v_\perp+z_v^*\widehat{\mathbf n}'$, and write $n=d-1$. Performing both longitudinal integrations in Eq.~\eqref{eq:prescription_middle_contact_integral} gives its forward contribution,
\be
 \begin{aligned}
 \left.\mathcal J_2^{(\nu)}\right|_{\rm forward}
 ={}&\int_{\substack{\mathbf u_\perp^2<E^2\\\mathbf v_\perp^2<E^2}}
 \frac{d^n\mathbf u_\perp\,d^n\mathbf v_\perp}{(2\pi)^{2n}}
 \\
 &\times\frac{
 \mathcal P_2^{(\nu)}
 (\mathbf u_*,\boldsymbol q_c-\mathbf u_*-\mathbf v_*,\mathbf v_*)}
 {4\sqrt{E^2-\mathbf u_\perp^2}\sqrt{E^2-\mathbf v_\perp^2}
  \mathbf u_*^2\mathbf v_*^2}.
 \end{aligned}
 \label{eq:prescription_contact_forward_integral}
\ee
This expression keeps the shell displacement, the Jacobians, and the numerator in the same integral. In the region $|\mathbf u_\perp|,|\mathbf v_\perp|\ll E$, expand all of them consistently. For example,
\be
 \begin{aligned}
 z_u^*&=-\frac{\mathbf u_\perp^2}{2E}
       +\mathcal O(Q^4/E^3),&
 z_v^*&= \frac{\mathbf v_\perp^2}{2E}
       +\mathcal O(Q^4/E^3),\\
 \frac1{\mathbf u_*^2}
 &=\frac1{\mathbf u_\perp^2}
 \left[1-\frac{\mathbf u_\perp^2}{4E^2}
 +\mathcal O(Q^4/E^4)\right],&&\\
 \frac1{4\sqrt{E^2-\mathbf u_\perp^2}\sqrt{E^2-\mathbf v_\perp^2}}
 &=\frac1{4E^2}
 \left[1+\frac{\mathbf u_\perp^2+\mathbf v_\perp^2}{2E^2}
 +\mathcal O(Q^4/E^4)\right].&&
 \end{aligned}
 \label{eq:prescription_contact_soft_expansion}
\ee
The expansion of $1/\mathbf v_*^2$ has the same form. Substituting the root expansions into the polynomial $\mathcal P_2^{(\nu)}$ again produces polynomials in the transverse integration variables. The external directions and the momentum transfer enter only through the coefficients and contractions of these polynomials. Each retained term in Eq.~\eqref{eq:prescription_contact_forward_integral} therefore reduces to
\be
 \int\frac{d^n\mathbf u_\perp\,d^n\mathbf v_\perp}{(2\pi)^{2n}}
 \frac{\mathcal Q_{\alpha\beta}
 (\mathbf u_\perp,\mathbf v_\perp,\boldsymbol q_c)}
 {(\mathbf u_\perp^2)^\alpha(\mathbf v_\perp^2)^\beta}=0.
 \label{eq:prescription_scaleless_contacts}
\ee
The numerator $\mathcal Q_{\alpha\beta}$ is polynomial. Expanding its monomials gives products of scaleless tensor integrals, which vanish in dimensional regularization. As required by expansion by regions, the expanded terms are integrated over the full transverse momentum spaces~\cite{Jantzen2011Regions}. The exact square roots in Eq.~\eqref{eq:prescription_contact_forward_integral} retain the scale $E$; the scalelessness statement applies after their soft expansion.

If the first exchange is canceled, we choose $\mathbf t_2,\mathbf t_3$ as independent variables and resolve them along $\mathbf k'$. Their leading longitudinal constraints involve $z_2+z_3$ and $z_3$. If the third exchange is canceled, we use $\mathbf t_1,\mathbf t_2$ resolved along $\mathbf k$, giving $z_1$ and $z_1+z_2$. Both linear changes have unit absolute Jacobian. Their shell corrections generate polynomial factors and powers of the same two unshifted exchange denominators, so they reduce to Eq.~\eqref{eq:prescription_scaleless_contacts} as well. Applied to Eq.~\eqref{eq:prescription_complete_contact_integral}, this establishes $\Delta_{\rm ct}=0$ at the retained static and soft orders, with the contact assignments among the reconstructed residues treated together.


\end{document}